\documentclass[a4paper,11pt]{article}

\usepackage{jheppub} 
\usepackage{graphicx,color}
\usepackage{amsmath}
\usepackage{slashed}
\usepackage{multirow}
\usepackage{autobreak}
\usepackage[normalem]{ulem}
\usepackage{hyperref}
\usepackage{cleveref}
\usepackage{slashed}
\usepackage{amssymb}
\usepackage{epsfig}
\usepackage{array}
\usepackage{mathtools}
\usepackage{caption}
\usepackage[utf8]{inputenc}
\usepackage{mathtools, nccmath}
\usepackage{float}
\usepackage[section]{placeins}
\usepackage[utf8]{inputenc}
    \usepackage[T1]{fontenc}
    \usepackage{lmodern}
    \usepackage[x11names]{xcolor}
    \usepackage{framed}
\colorlet{shadecolor}{blue!10}

\SetSymbolFont{largesymbols}{bold}{OMX}{txex}{b}{n}

\DeclareMathAlphabet{\mathpzc}{OT1}{pzc}{m}{it}

\newcommand{\be}{\begin{eqnarray*}}
\newcommand{\ee}{\end{eqnarray*}}

\renewcommand{\vec}{\bf}

\usepackage{parskip}

\def\qr {{\mathfrak{q}}}

\newcommand{\ba}{\begin{array}}
\newcommand{\ea}{\end{array}}
\newcommand{\bd}{\begin{displaymath}}
\newcommand{\ed}{\end{displaymath}}
\newcommand{\besub}{\begin{subequations}}
\newcommand{\eesub}{\end{subequations}}

\def\m{\mu}
\def\n{\nu}

\def\q2 {q^2}

\def\mchi {m_\chi}
\def\ca{c_{\ell_1}}

\def\bt{\begin{table}}
\def\et{\end{table}}

\usepackage{bbm}  
\usepackage{amsmath}                                                

\newcommand{\nc}{\newcommand}
\nc{\beq}{\begin{equation}}  \nc{\eeq}{\end{equation}}
\nc{\bea}{\begin{eqnarray}}  \nc{\eea}{\end{eqnarray}}
\nc{\baa}{\begin{array}}     \nc{\eaa}{\end{array}}
\nc{\bit}{\begin{itemize}}   \nc{\eit}{\end{itemize}}
\nc{\ben}{\begin{enumerate}} \nc{\een}{\end{enumerate}}
\nc{\bce}{\begin{center}}    \nc{\ece}{\end{center}}
\nc{\bpm}{\begin{pmatrix}}   \nc{\epm}{\end{pmatrix}}
\nc{\bvt}{\begin{verbatim}}  \nc{\evt}{\end{verbatim}}
\nc{\bal}{\begin{align}}
\def\mcr{\nonumber\\[6pt]}

\def\to{\rightarrow}

\def\boldoverdot{\,{\raise6pt\hbox{\bf.}\!\!\!\!\>}}

\def\gcal{{\cal G}}

\def\lcal{{\cal L}}

\def\ee{{\bf e}}

\usepackage{bm}
\def\diag{\hbox{\diag}}

\def\m{\hbox{m}}

\def\vevof#1{\left\langle #1 \right\rangle}

\def\doubleundertext#1{
{\undertext{\vphantom{y}#1}}\par\nobreak\vskip-\the\baselineskip\vskip4pt%
\undertext{\hbox to 2in{}}}
\def\inbox#1{\vbox{\hrule\hbox{\vrule\kern5pt
     \vbox{\kern5pt#1\kern5pt}\kern5pt\vrule}\hrule}}
\def\sqr#1#2{{\vcenter{\hrule height.#2pt
      \hbox{\vrule width.#2pt height#1pt \kern#1pt
         \vrule width.#2pt}
      \hrule height.#2pt}}}

\def\today{\ifcase\month\or
  January\or February\or March\or April\or May\or June\or
  July\or August\or September\or October\or November\or December\fi
  \space\number\day, \number\year}
\def\pmb#1{\setbox0=\hbox{#1}%
  \kern-.025em\copy0\kern-\wd0
  \kern.05em\copy0\kern-\wd0
  \kern-.025em\raise.0433em\box0 }

\def\pmbb#1{\setbox0=\hbox{#1}%
  \kern-.02em\copy0\kern-\wd0
  \kern.04em\copy0\kern-\wd0
  \kern-.02em\raise.03464em\box0 }
\def\sumprime_#1{\setbox0=\hbox{$\scriptstyle{#1}$}
  \setbox2=\hbox{$\displaystyle{\sum}$}
  \setbox4=\hbox{${}'\mathsurround=0pt$}
  \dimen0=.5\wd0 \advance\dimen0 by-.5\wd2
  \ifdim\dimen0>0pt
  \ifdim\dimen0>\wd4 \kern\wd4 \else\kern\dimen0\fi\fi
\mathop{{\sum}'}_{\kern-\wd4 #1}}
	\title{\boldmath Effective Leptophilic WIMPs at the $e^+e^-$ collider}

\author[a]{Basabendu Barman,}
\author[b]{Subhaditya Bhattacharya,}
\author[c]{Sudhakantha Girmohanta,}
\author[b]{Sahabub Jahedi}

\affiliation[a]{Centro de Investigaciones, Universidad Antonio Nariño\\
Carrera 3 este \# 47A-15, Bogotá, Colombia}
\affiliation[b]{Department of Physics, Indian Institute of Technology Guwahati, Assam 781039, India}
\affiliation[c]{\ C. N. Yang Institute for Theoretical Physics and
Department of Physics and Astronomy, \\
Stony Brook University, Stony Brook, New York 11794, USA }

\emailAdd{basabendu88barman@gmail.com}
\emailAdd{subhab@iitg.ac.in}
\emailAdd{sudhakantha.girmohanta@stonybrook.edu}
\emailAdd{sahabub@iitg.ac.in}

\abstract{We consider higher-dimensional effective (EFT) operators consisting of fermion dark matter (DM) connecting to Standard Model (SM) leptons 
upto dimension six. Considering all operators together and assuming the DM to undergo thermal freeze-out, we find out relic density allowed 
parameter space in terms of DM mass ($m_\chi$) and New Physics (NP) scale ($\Lambda$) with one loop direct search constraints from 
XENON1T experiment. Allowed parameter space of the model is probed at the proposed International Linear Collider (ILC) via monophoton 
signal for both Dirac and Majorana cases, limited by the centre-of-mass energy $\sqrt s=$1 TeV, where DM mass can be probed within 
$m_\chi<\frac{\sqrt{s}}{2}$ for the pair production to occur and $\Lambda>\sqrt s$ for the validity of EFT framework.

}


\begin{document} 
\begin{flushright}
 PI/UAN-2021-700FT
\end{flushright}

\maketitle
\flushbottom

\section{Introduction}
\label{sec:intro}

The existence of dark matter (DM) is motivated from several astrophysical~\cite{Zwicky:1933gu, Zwicky:1937zza, Rubin:1970zza, Clowe:2006eq} and cosmological~\cite{Hu:2001bc, Aghanim:2018eyx} evidences (for a review, 
see, e.g. Refs.~\cite{Jungman:1995df, Bertone:2004pz, Feng:2010gw}), although a laboratory discovery is still awaited. Excepting some broad characteristics like electromagnetic charge neutrality, stability over the Universe's 
life time etc., other properties like mass, spin, interactions (other than gravitational) are still unknown. Anisotropy in the cosmic microwave background (CMB) radiation provides the most precise measurement of the DM relic 
density, usually expressed as $\Omega_{\tt DM}h^2 \simeq 0.12$~\cite{Aghanim:2018eyx}, where $\Omega$ refers to cosmological density and $h$ is the reduced Hubble constant in the unit of 100 km/sec/Mpc. Since 
the Standard Model (SM) of particle physics fails to offer a viable particle DM, one has to explore beyond the realms of the SM. 

The weakly interacting massive particle (WIMP)~\cite{Jungman:1995df, Kolb:1990vq} by far is the most popular DM candidate. WIMPs are assumed to be in thermal and chemical equilibrium in the early 
universe due to sizeable DM-SM coupling and undergoes a thermal freeze-out once the interaction rate falls below the Hubble expansion rate ($\mathcal{H}$) of the universe. The interaction cross-section that gives rise to the observed relic abundance of DM turns out to be of the order of the weak interaction ($\sim 10^{-10}~\rm{GeV}^{-2}$), suggesting the name WIMP. This very fact also opens up the possibility for WIMPs to be probed in a plethora of experimental frontiers like direct DM search, collider search and indirect searches (for a review, see~\cite{Arcadi:2017jqd,Roszkowski:2017nbc}). 
However, other possibilities like Feebly Interacting Massive Particle, FIMP~\cite{Hall:2009bx,Bernal:2017kxu} or Strongly Interacting Massive Particle, SIMP~\cite{Hochberg:2014dra} etc.
are getting more attention due to null experimental observation so far, but the quest for hunting WIMP-like DM is still on. 

Theoretical and phenomenological studies of particle DM have been done mainly in two ways: constructing a UV complete model as an extension of the SM or by constructing DM-SM effective (EFT) operators, where the later is the focus of the present paper. DM EFT operators can be cooked up in a standard prescription with the Lorentz invariant contact interactions of the form $\sim\mathcal{O}_\text{DM}\,\mathcal{O}_\text{SM}$ 
(suppressed by appropriate powers of the new physics (NP) scale), where $\mathcal{O}_\text{DM}$ consists of dark sector fields and obey some dark symmetry, while $\mathcal{O}_\text{SM}$ 
contains SM fields and follow SM gauge symmetry (for a review see~\cite{Bhattacharya:2021edh}). Such EFT constructions can describe DM relic density, direct search
~\cite{Fitzpatrick:2012ib, Fitzpatrick:2012ix, Cirelli:2013ufw, SuperCDMS:2015lcz, Bishara:2016hek, DeSimone:2016fbz, DEramo:2016gos, Brod:2017bsw, Bishara:2017pfq, Li:2018orw, Criado:2021trs}, 
indirect search~\cite{Beltran:2008xg, Cao:2009uw, Cheung:2010gj, Goodman:2010qn, Cheung:2010ua, Blumenthal:2014cwa,Klasen:2015uma, Bell:2016uhg, Bell:2019pyc},
and production at collider in terms of just three parameters of the theory, namely, the NP scale $\Lambda$, the DM mass $m_\chi$ and the coupling $c_i$ in a
 model-independent way\footnote{However, these effective operators can be associated to an UV complete description involving all the relevant fields and symmetry at an energy scale larger than the NP scale of interaction.}.

Considering collider signal, the DM is missed in the detector, but its production in association with any initial state radiation (like a photon or a jet) gives rise to mono-$X$ plus missing energy signal, a typical one in the context of DM EFT frameworks\footnote{Similar searches are also done in `{\it simplified model}' with $s/t$-channel mediators (see~\cite{Abdallah:2015ter, Abercrombie:2015wmb}).}. Mono-$X$ signal is searched 
extensively at LHC (for a review, see~\cite{Kahlhoefer:2017dnp, Penning:2017tmb, Boveia:2018yeb}) and the absence of an excess has provided bounds on the DM parameter 
space\footnote{Higgs and Z-invisible decays~\cite{10.1093/ptep/ptaa104} also provide a collider probe for DM having mass less than $\frac{m_{h/Z}}{2}$.}. Similar analysis have also been 
done in context of lepton colliders~\cite{Dreiner:2012xm, Chae:2012bq, deBlas:2018mhx, Habermehl:2020njb}. A crucial point is to associate the collider signal 
with the DM bounds {\it viz.,} relic density, direct detection and indirect searches, which has not been strictly followed in several existing analyses. 




The study of the complementarity between (in)direct and collider searches adopting the EFT approach have also been done extensively in context of LHC
~\cite{Goodman:2010yf, Goodman:2010ku, Bai:2010hh, Rajaraman:2011wf,Buckley:2011kk,Fox:2011pm, Dreiner:2013vla, Buchmueller:2013dya, Petrov:2013nia, Chang:2013oia, Altmannshofer:2014cla,
Bell:2015sza,Belyaev:2016pxe, Capdevilla:2017doz, Belyaev:2018pqr} as well as in context of lepton collider ($e^+e^-$) or proposed International Linear Collider (ILC)
~\cite{Fox:2011fx,Yu:2013aca,Essig:2013vha,Kadota:2014mea,Yu:2014ula,Freitas:2014jla,Dutta:2017ljq,Liu:2019ogn,
Choudhury:2019sxt,Bharadwaj:2020aal,Kundu:2021cmo} with possible UV completions~\cite{Dutta:2017ljq, Liu:2019ogn,Bharadwaj:2020aal}. More exhaustive connections are also thought of, 
for example in~\cite{Bertuzzo:2017lwt}, where the bounds on MeV-scale DM are concocted with CMB, BBN, LHC, LEP, direct detection experiments and meson decays. 

We note further that EFT approach primarily dictates to consider all the operators having same mass dimension on equal footing absent a hint of specific NP
\footnote{With an exception for potential tree generated (PTG) or loop generated (LG) operators~\cite{GonzalezMacias:2015rxl}.}, which we do here, 
although many of the analyses have been projected by taking one operator at a time. This generic technique also provides us with an opportunity to address different non-zero combinations and signs 
of the corresponding Wilson coefficients to subsume specific UV complete set-ups, as well as drastically distinct phenomenology at collider, as we elaborate.
Importantly, in collider study of DM EFT one needs to ensure that the center-of-mass (CM) energy of the reaction lies below the mass of the NP scale 
($\Lambda$). In a hadron collider like the LHC, the partonic CM energy ($\sqrt{\hat{s}}$) is unknown, hence it is not possible to guarantee that 
$\sqrt{\hat{s}}<\Lambda$ holds, particularly for DM pair production whose invariant mass can not be constructed\footnote{In~\cite{Busoni:2013lha, Busoni:2014haa, Busoni:2014sya} the validity of EFT in context of DM 
collider search has been studied in detail in presence of $s$ or $t$ channel mediators.}. This is in contrast to DM direct searches, where the momentum transfer involved in the scattering of DM particles with heavy nuclei are of the order of tens of keV, way below the NP scale, making EFT description more reliable.  As a consequence, even though hadron colliders have a larger reach, applicability of DM EFT is questionable. On the other hand, in leptonic colliders like ILC,  knowledge of CM energy of the reaction, together with symmetric beams and the possibility of polarizing the beams to reduce SM background contribution, make DM EFT studies much more concrete. Motivated from these, in this work, we explore the DM EFT in $e^+\,e^-$ collider, assuming the DM to be a Dirac or Majorana fermion, where it interacts {\it preferentially} to the SM leptons. By considering all the operators of dimension six together, we explore the resulting parameter space for the DM abiding bounds from Planck observed relic abundance and also limits from spin-independent direct search experiments.
 

The paper is organized as follows: DM-SM effective operators are noted in Section \ref{sec:dmeffop}, while the DM constraints are described in Section \ref{sec:constraints}, followed by collider prospects of leptophilic DM at ILC in Section \ref{sec:collider}. We remark on the UV completion of 
our effective description in Section \ref{sec:uv} and finally summarize in Section \ref{sec:sum}. Appendices~\ref{app:cross} contain all the relevant annihilation 
cross-section formulae;~\ref{sec:negative} contain DM constraints for some special choices of the Wilson coefficients.

\section{DM-SM Effective Operators}
\label{sec:dmeffop}
We take up EFT approach to study DM physics, which is also motivated from the absence of a specific hint of dark sector particles from ongoing experiments and
assume effective DM-SM interactions of the following form :

\beq
\lcal_{\tt int}=\frac{c}{\Lambda^{d+d'-4}}\mathcal{O}^{(d)}_\text{SM}\mathcal{O}^{(d')}_\text{DM}\,,
\label{eq:non-renorm}
\eeq

\noindent where $\Lambda$ denotes NP scale, $\mathcal{O}^{(d)}_\text{SM}$ consists of SM fields (having mass dimension $d$) and are invariant under $ \gcal_{\tt SM} $ (SM gauge symmetry), 
while $\mathcal{O}^{(d')}_\text{DM}$ consists of DM fields (having mass dimension $d'$) and is invariant under dark symmetry $ \gcal_{\tt DM} $; necessitated by the stability of DM upto a scale of universe life time and
$c$ denotes dimensionless couplings, also called the Wilson coefficients. We also assume $\mathcal{O}_\text{DM}$ is singlet under $ \gcal_{\tt SM} $ and $\mathcal{O}_\text{SM}$ is singlet under 
$ \gcal_{\tt DM}$\footnote{Exceptions to this simplification have been studied in context of many UV complete models, where dark sector particles transform nontrivially under $ \gcal_{\tt SM} $, see for example,~\cite{Diaz-Cruz:2010czr}.}.
Also note that $ \mathcal{O}_\text{DM}$ must have at least two dark-sector fields, since all dark fields transform non-trivially under $ \gcal_{\tt DM} $. For simplicity, we assume here $\gcal_{\tt DM}=Z_2$. 
Eq.~\eqref{eq:non-renorm} dictates the freeze-out when DM is assumed to be WIMP or freeze-in when it is assumed to be FIMP, as well as its interaction in direct, indirect and collider search experiments.  

DM EFT operators involving a scalar ($\Phi$), fermion (Dirac or Majorana) ($\chi$) or vector bosons ($X$) as DM, have been constructed in several works 
(see, for example, \cite{Matsumoto:2014rxa, Duch:2014yma,Duch:2014xda,Macias:2015cna}). Let us consider the DM to be a fermion $\chi$ which transforms under a 
dark symmetry $ \gcal_{\tt DM} = Z_2$, and write all possible EFT operators in Eq.~\eqref{eq:op5} (dimension five) and Eq.~\eqref{eq:op6} (dimension six), 
suppressed by appropriate powers of NP scale $\Lambda$. 

\beq
\begin{array}{ll}
\mathcal{O}_{D1}^5 = \frac{g}{8\pi^2\Lambda}\left(\overline{\chi}\sigma_{\mu\nu}\chi\right)B^{\mu\nu}\,, &~\mathcal{O}_{D2}^5 = \frac{g}{8\pi^2\Lambda}\left(\overline{\chi}i\sigma_{\mu\nu}\gamma^5\chi\right)B^{\mu\nu}\,, \cr
\mathcal{O}_3^5= \frac{c_{H_1}}{\Lambda}\left(\overline{\chi}\chi\right)\left(H^\dagger H\right)\,, &~\mathcal{O}_4^5=\frac{c_{H_2}}{\Lambda}\left(\overline{\chi}i\gamma^5\chi\right)\left(H^\dagger H\right).
\end{array}
\label{eq:op5}
\eeq



\beq
\begin{array}{ll}
\mathcal{O}_{DQ}^6 = \frac{c_{q_1}}{\Lambda^2}\left(\overline{\chi}\gamma_\mu\chi\right)\left(\overline{\qr}\gamma^\mu \qr\right)\,, &\mathcal{O}_{DL}^6 = \frac{c_{\ell_1}}{\Lambda^2}\left(\overline{\chi}\gamma_\mu\chi\right)\left(\overline{\ell}\gamma^\mu \ell\right)\, , \cr
\mathcal{O}_{Q}^6 = \frac{c_{q_2}}{\Lambda^2}\left(\overline{\chi}\gamma_\mu\gamma^5\chi\right)\left(\overline{\qr}\gamma^\mu \qr\right)\,, &\mathcal{O}_{L}^6 = \frac{c_{\ell_2}}{\Lambda^2}\left(\overline{\chi}\gamma_\mu\gamma^5\chi\right)\left(\overline{\ell}\gamma^\mu \ell\right)\,, \cr
\mathcal{O}_{Q1}^6 = \frac{c_{q_3}}{\Lambda^2}\left(\overline{\chi}\gamma_\mu\gamma^5\chi\right)\left(\overline{\qr}\gamma^\mu \gamma_5 \qr\right) \,, &\mathcal{O}_{L1}^6 = \frac{c_{\ell_3}}{\Lambda^2}\left(\overline{\chi}\gamma_\mu\gamma^5\chi\right)\left(\overline{\ell}\gamma^\mu \gamma_5 \ell\right)\,, \cr
\mathcal{O}_{DQ1}^6 = \frac{c_{q_4}}{\Lambda^2}\left(\overline{\chi}\gamma_\mu\chi\right)\left(\overline{\qr}\gamma^\mu \gamma_5 \qr\right)\,,& \mathcal{O}_{DL1}^6 = \frac{c_{\ell_4}}{\Lambda^2}\left(\overline{\chi}\gamma_\mu\chi\right)\left(\overline{\ell}\gamma^\mu \gamma_5 \ell\right)\,, \cr
\mathcal{O}_{DH}^6=\frac{c_{H_3}}{\Lambda^2}\left(\overline{\chi}\gamma^\mu\chi\right)(H^\dagger i\overset\leftrightarrow{D}_\mu H)\,, & \mathcal{O}_{H}^6=\frac{c_{H_4}}{\Lambda^2}\left(\overline{\chi}\gamma^\mu\gamma^5\chi\right)(H^\dagger i\overset\leftrightarrow{D}_\mu H). 
\end{array}
\label{eq:op6}
\eeq

\noindent Here $H$ stands for the SM Higgs isodoublet, $\qr$ stands for either left handed (LH) doublet or right handed (RH) singlet SM quarks (of all flavors), 
and $\ell$ stands for SM lepton, LH doublet or RH singlet. SM gauge invariance ensures only LH doublets or RH singlets to appear in the SM fermion 
current. For brevity we omit flavour indices and for simplicity assume the interactions to be flavour diagonal\footnote{Flavour non-diagonal DM interactions have been studied in context of lepton flavour universality violation, as in~\cite{DAmbrosio:2002vsn}.}. 
The covariant derivative $D_\mu$ is defined as
$$D_\mu = \partial_\mu - i g \frac{\sigma^i}{2}W_\mu^i - i\frac{g'}{2} B_\mu~,$$ where $g, g'$ stands for $SU\left(2\right)_L$ and $U\left(1\right)_Y$ 
gauge coupling strengths, $\sigma^i$ are the Pauli spin matrices with $i\in 1,2,3$, and $W_\mu^i$ and $B_\mu$ are the $SU\left(2\right)_L$ and 
$U\left(1\right)_Y$ gauge bosons. $B^{\mu\nu}=\partial^\mu B^\nu-\partial^\nu B^\mu$ represents the $U(1)_Y$ gauge field strength tensor, which is itself SM 
gauge invariant. The hermitian conjugate of covariant derivatives is defined as $\overset\leftrightarrow{D}_\mu \equiv D_\mu -\overset\leftarrow{D}_\mu^\dagger$. 
We also assume the operators to have different couplings to SM leptons ($c_{\ell_i}$), quarks ($c_{q_i}$) and SM Higgs ($c_{H_i}$). Note further, that
the operators with tag $D$ (for example, $\mathcal{O}_{DQ}^6, \mathcal{O}_{DH}^6$ etc.) are present only when DM $\chi$ is a Dirac fermion, which naturally indicates 
all the $D$-tagged operators to be dropped for Majorana $\chi$. In the following, we consider both these possibilities and indicate the distinction they provide in subsequent 
phenomenology. We further point out that any operator having an interaction with gauge field strength tensor (for example, $B^{\mu\nu}$) are generated at least in a one-loop level 
via NP in a perturbative UV theory and hence classified as loop generated operators (LG) having additional suppression factor $\sim 1/(16 \pi^2)$ (see, for example,~\cite{Macias:2015cna}). 
Therefore, $\mathcal{O}_{D{1,2}}^5$ will have smaller contribution and their presence can be ignored compared to those other operators, which are potential tree-generated (PTG).

A priori for a fermion DM, dimension five operators (Eq.~\eqref{eq:op5}) naturally dominate over others and 
subsequent phenomenology is that of Higgs portal interaction, which have been studied extensively~\cite{Fedderke:2014wda,Matsumoto:2014rxa,Bishara:2015cha}. 
However, one may think of NP scenarios which gives rise to predominantly DM-SM fermion interactions, where the Higgs portal interactions can be neglected and the four fermion 
operators (as in Eq.~\eqref{eq:op6}) having the following form start playing a crucial role: 
\beq
\frac{c_{f}}{\Lambda^2}\Bigl(\overline{\chi}\Gamma^\mu_\chi\chi\Bigr)\Bigl(\overline{f}\Gamma_{\mu,f} f\Bigr)\,; ~~ \Gamma^\mu_{\chi/f}: \{\gamma^\mu, \gamma^\mu\gamma^5\}\,; ~~f=\{\qr, \ell\}\,.
\label{op-fermion}
\eeq
Further classification may also emerge if the DM couple preferentially to SM leptons or quarks
(due to some specific NP scenarios as we will highlight later) as: 
\begin{itemize}
 \item Leptophilic: $c_{\ell_i}=1, c_{H_i}=c_{q_i}=0$ ;
 \item Hadrophilic: $c_{q_i}=1, c_{H_i}=c_{\ell_i}=0$ .
\end{itemize}
It is easy to see that Leptophilic or Hadrophilic operators provide completely different phenomenology; 
for example, Hadrophilic DM operators are more prone to direct search constraints and can be produced at the LHC. On the other hand, 
Leptophilic DM is interesting for at least two reasons; they can hide from direct search constraints to a great extent and can be probed at ILC 
in its effective limit. We note that the non-zero values of the Wilson coefficients chosen here is representative and can be 
compensated by a new choice of $\Lambda$\footnote{However, when several operators are considered together, as we do here, 
different relative coupling strengths can't be reproduced by naive scaling of $\Lambda$.}. By Leptophilic DM ($c_{\ell_i}=1$), we usually refer to the situation 
where all the operators ($i=\{1,..4\}$) are assumed to have equal coupling strength, unless explicitly specified to the cases where some of them are considered 
absent. We will also see that the sign of the couplings will play an important part, particularly in collider phenomenology. We comment on some possible UV completion of leptophilic models considered here in section~\ref{sec:uv}.


\section{DM constraints on Leptophilic Operators}
\label{sec:constraints}

The first exercise is to find constraints on the operators from existing experimental limits from DM searches. They mainly include relic density, direct search and 
indirect searches, which we are going to discuss in the following subsections, highlighting the processes which contribute to these observables and finally the 
allowed parameter space available to such operators after these constraints.

\subsection{Relic Density}
\label{sec:relic}

DM Relic density provides the most important constraints on the parameter space of the EFT model. Here we assume $\chi$ to be a 
WIMP-like fermion DM in thermal and chemical equilibrium with SM particles in the early universe, which decouples at some later epoch
as the universe expands and cools down. The the relic density within the WIMP scenario (see~\cite{Kolb:1990vq} for details) 
is obtained from the solution to the Boltzmann equation:

\beq
\frac{d Y_\chi}{d x} = -1.32\,\sqrt{g_*}~M_{\tt pl}~ \frac{\mchi }{x^2}~\vevof{\sigma v}_{2_{\rm DM}\to 2_{\rm SM}} ~\left( {Y_\chi}^2-Y_{\tt eq}^{2} \right)\,, 
\label{eq:BEQ2to2}
\eeq

\noindent where $M_{\tt pl} $ denotes the reduced Planck mass, $Y_\chi = n/s$ refers to DM yield ($n$ is the DM density, $s$ is the total entropy density) and 
$x= \mchi /T$ ($T$ is the temperature). $Y_{\tt eq}$ denotes the value of $Y_\chi$ in thermal equilibrium given by Maxwell Boltzmann distribution for 
non-relativistic species:
\beq
 Y_{\tt eq}(x)=0.145~\frac{g_{\tt DM}}{g_*} x^{3/2}e^{-x}\,.
 \label{eq:yeq}
\eeq
In the above, $g_{\tt DM}$ refers to the number of DM internal states; $g_*$ denotes the effective relativistic degrees of freedom
\beq
g_*=\sum_{i=\rm bosons}g_i \theta(T-m_i)+\frac{7}{8}\sum_{i=\rm fermions}g_i \theta(T-m_i)\,,
\label{eq:DOFeqn}
\eeq

\noindent where $ g_i $ are the internal degrees of freedom of particle $i$ with mass $ m_i $. The key in Eq.~\eqref{eq:BEQ2to2} is to assume that $Y_\chi=Y_{\tt eq}$ at $x \to 0$, ensuring DM to be in thermal bath. Finally, in Eq.~\eqref{eq:BEQ2to2}, $\vevof{\sigma v}_{2_{\rm DM}\to 2_{\rm SM}} $ denotes the thermal average of the DM annihilation cross-section$\times$ velocity for the process $ \chi\bar\chi\to \ell\bar\ell $ mediated by operators with $c_{\ell_i}=1$. It is straightforward to compute the 
corresponding annihilation cross-sections for all these operators. For example, the operator $\mathcal{O}_{DL}^6$ provides the following annihilation 
cross-section for a Dirac DM $\chi$ 
\bal
\langle{\sigma v}\rangle_{2_{\rm DM}\to 2_{\rm SM}} &=  \frac{\mchi ^2}{2\pi\Lambda^4} \sum_\ell \left(c_{\ell_1} \right)^2   \left\{ 2+\frac{m_\ell^2}{\mchi ^2} +\left[ \frac{8\mchi ^4 - 4 m_\ell^2 \mchi ^2 + 5 m_\ell^4 }{24 \mchi ^2 (\mchi ^2 - m_\ell^2)} \right]v^2  \right\} \sqrt{1- \frac{m_\ell^2}{\mchi ^2}}\,, \mcr
&= a + b v^2\,,
\label{ann-cross-sec}
\end{align}
where $m_\ell$ is the lepton mass, $v$ is the M\"oller velocity $ v= \sqrt{(p_\chi.p_{\bar{\chi}})^2-\mchi^4}/(E_\chi E_{\bar{\chi}})$~\cite{Gondolo:1990dk, Edsjo:1997bg} and we have ignored higher powers of $v$. 
This interaction then gives rise to s-wave ($\propto a$) and p-wave ($\propto b$) contributions~\cite{Kolb:1990vq, Bauer:2017qwy}. Annihilation 
cross-section for other operators are furnished in Appendix~\ref{app:cross}.

\begin{figure}[htb!]
$$
\includegraphics[scale=0.35]{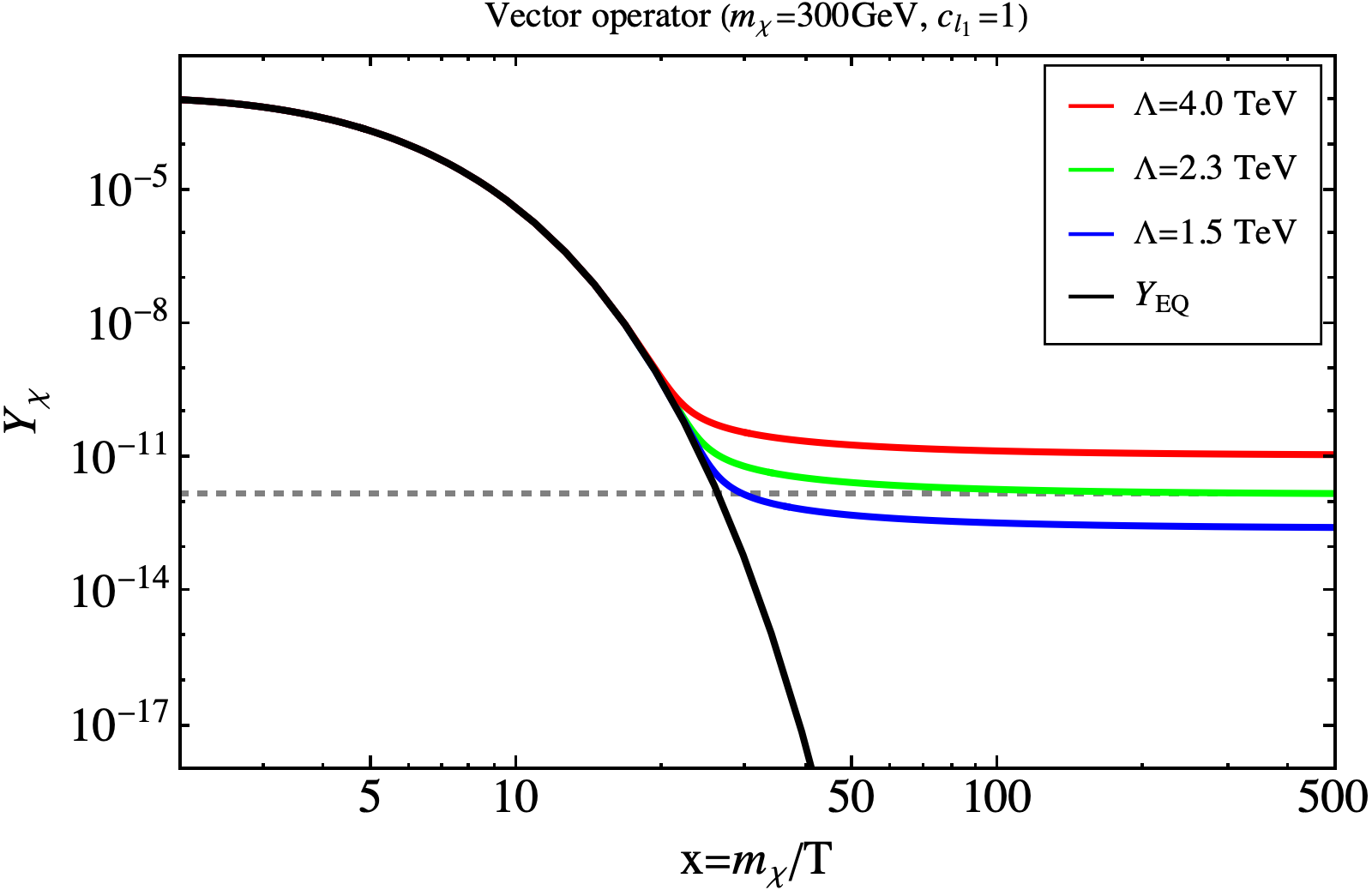}~~
\includegraphics[scale=0.35]{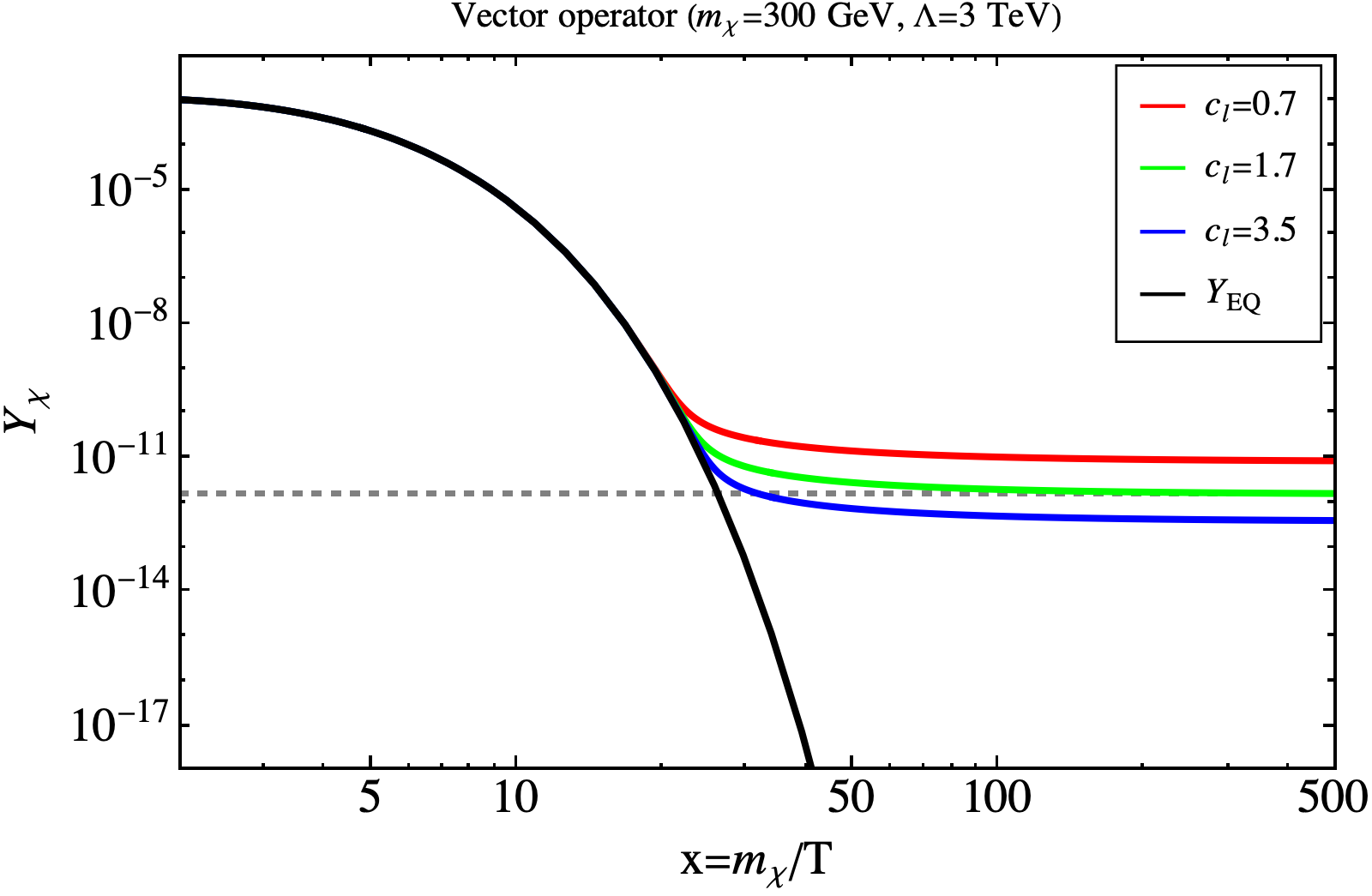}
$$
\caption{Left: Evolution of DM (Dirac) yield $Y_\chi$ with $x=m_\chi/T$ for operator $\mathcal{O}_{DL}^6$ (see Eq.~\eqref{eq:op6}), 
for three different choices of the EFT scale $\Lambda=\{4.0,2.3,1.5\}$ TeV shown respectively in red, blue and green keeping $\ca=1$ and 
$\mchi=$ 300 GeV fixed. Right: Same as left, but for a fixed $\Lambda=3$ TeV with three choices of $\ca=\{0.7,1.7,3.5\}$ shown in red, 
blue and green respectively. In both the plots the thick black curve shows the equilibrium DM yield ($Y_{\tt eq}$) and the grey dashed 
straight line indicates the yield that produces correct relic abundance (see Eq.~\eqref{relic-value}).}
\label{fig:beq}
\end{figure}

The relic abundance of DM after thermal freeze-out has an approximate analytical form

\beq
    \Omega_{\tt DM}\,{\sf h}^2 \simeq \frac{(1.04 \times 10^9\,\text{GeV})x_f}{M_{\tt pl}\sqrt{g_*}(a + 3b/x_f)}\,, 
    \label{theo.relic}
\eeq

\noindent with $\Omega_{\tt DM} = \rho_{\tt DM}/\rho_c$, where $\rho_c, \rho_{\tt DM} $ denote the critical and DM densities respectively and  $x_f=m_\chi /{T_f}$, with $ T_f$ the freeze-out temperature when the DM decouples from thermal bath. In presence of all the operators with $c_{\ell_i}=1$, 
all the individual amplitudes add together to the total annihilation cross-section, with 
 \beq
\langle{\sigma v}\rangle_{\rm tot}\sim \Big \vert \sum_{\mathcal {O}_i} \mathcal {M}_{\mathcal{O}_i}\Big\vert^2 \,,
\eeq
where $\mathcal {M}_{\mathcal{O}_i}= \frac{c_{\ell_i}}{\Lambda^2}(\bar{\chi}\Gamma\chi)(\bar{\ell}\Gamma\ell); ~\rm{with}~\Gamma=\{\gamma^\mu, \gamma^\mu\gamma^5\}$ denote the matrix elements for DM pair annihilation to SM via respective operators as in Eq.~\eqref{eq:op6}. 
The relic density obtained from the model is then compared to the current Planck data~\cite{Aghanim:2018eyx}

\beq
\Omega_{\tt DM} {\sf h}^2 = 0.11933\pm 0.00091 \,,
\label{relic-value}
\eeq

\noindent that finds relic density allowed parameter space of the model. In Fig.~\ref{fig:beq}, we show the thermal freeze-out of the DM following Eq.~\eqref{eq:BEQ2to2} in $Y_\chi-x$ plane, for operator $\mathcal{O}_{DL}^6$ (see Eq.~\eqref{eq:op6}) assuming a Dirac DM $\chi$. The equilibrium distribution $Y_{\tt eq}$ is shown by black thick line, where we choose $\mchi=300$ GeV. Horizontal grey dashed line corresponds to $Y_\chi$ value for correct DM relic density following Eq.~\eqref{relic-value}, which is approximately given by the condition $m_\chi Y_\chi \approx 0.44$ eV. 
In the left panel, we vary the EFT scale $\Lambda=\{4.0, 2.3, 1.5\}$ TeV, shown respectively by red, green and blue curves assuming $\ca=1$ and we see that the green curve with $\Lambda=2.3$ TeV provides the correct relic after freeze-out, whereas the case with larger (smaller) $\Lambda$, corresponding to the red (blue) curve in Fig.~\ref{fig:beq} provides over (under) abundance following the inverse dependence of relic density to annihilation cross-section as in Eq.~\eqref{theo.relic}. In the right panel of Fig.~\ref{fig:beq}, we show the thermal freeze-out for three different choices of $\ca=\{0.7, 1.7, 3.5\}$ in red, green and blue respectively for a fixed $\Lambda=3$ TeV. Again, we see that the case with $\ca=1.7$ satisfy correct relic while the smaller (larger) $\ca$ provides over (under) abundance attributed to Eq.~\eqref{theo.relic}. The relic density allowed parameter space of the model can then be found in terms of $\{\Lambda,\mchi\}$ assuming $c_{\ell_i}=1$. We use the numerical tool {\tt MicrOmegas}~\cite{Belanger:2010pz} to compute relic density for the model for both Dirac and Majorana cases. The results, together with direct search constraints, will be discussed in the next section.


\subsection{Direct detection}\label{sec:loopDD}

\begin{figure}[htb!]
$$
\includegraphics[width=0.5\textwidth]{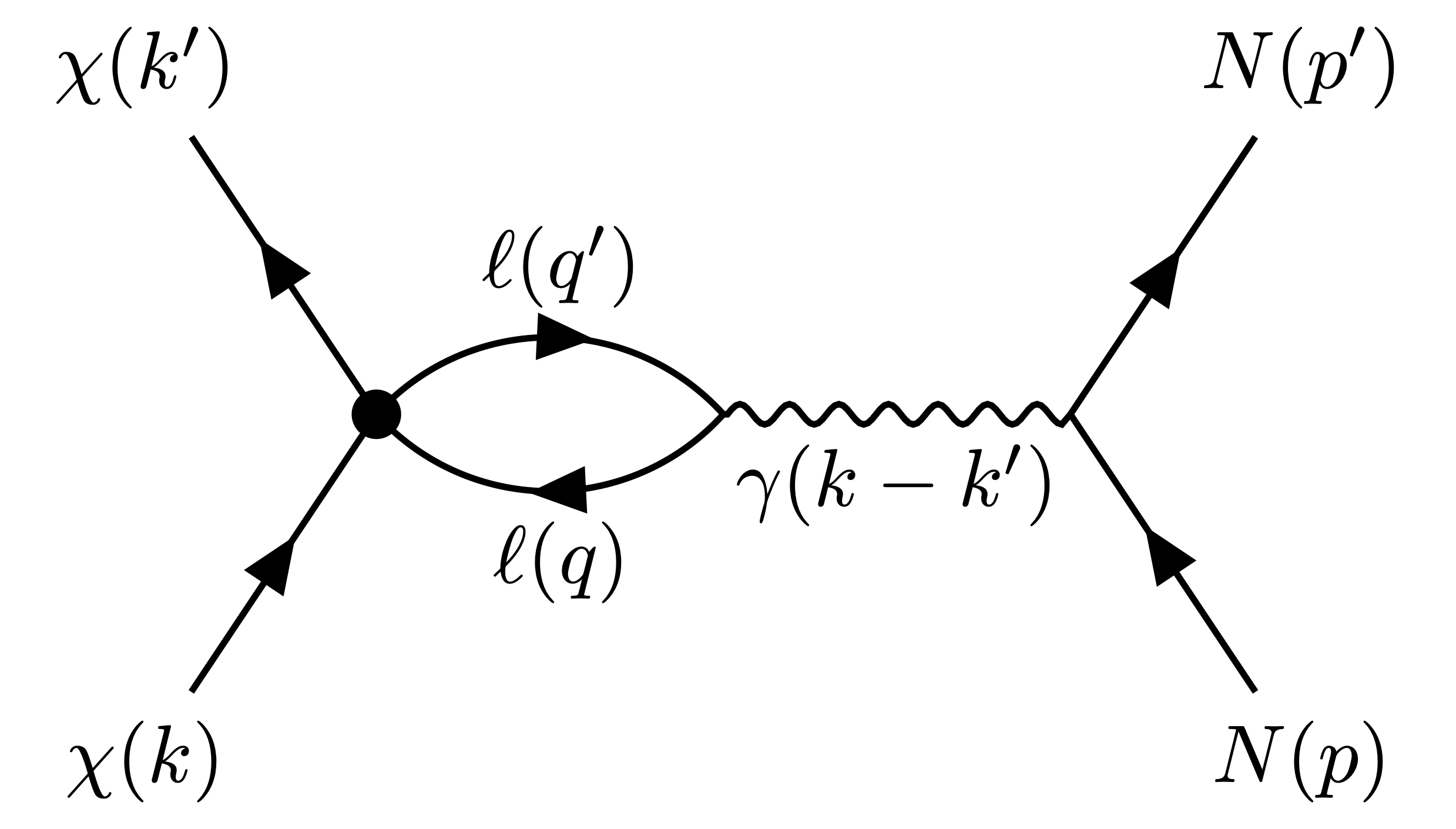}
$$
\caption{Loop-induced DM-nucleon direct detection mediated by a virtual photon. Here $q$ is the loop momentum, and $q'=k-k'+q$. The blob represents the effective nature of the vertex.
}\label{fig:1-loop-dd}
\end{figure}

Direct DM search relies on the DM scattering with earthly detectors and observation of nuclear recoil to confirm the presence of DM interaction. Unfortunately no DM signal has been confirmed yet, resulting a strong bound on the DM-SM interaction cross-section, the latest from XENON1T~\cite{XENON:2018voc}.
Leptophilic DM as defined, do not have a direct coupling to detector nucleus. But, it is possible to induce couplings to quarks via photon exchange 
at one loop level as in Fig.~\ref{fig:1-loop-dd}, where in the loop one may have any charged lepton to which the DM couples to\footnote{Similar diagrams with a Z, or Higgs-boson propagator are suppressed in comparison with the virtual photon mediation by a factor $\simeq (k-k')^2/m_{Z,H}^2$~\cite{PhysRevD.80.083502}.}. Instead of a nuclear recoil, it is also possible to knock off an electron from the atomic orbital of the detector material directly via the effective DM interaction with the lepton. However, as it has been shown in~\cite{PhysRevD.80.083502}, the loop-induced DM-nucleon scattering always dominates over the DM-electron scattering 
cross-section, where the latter is suppressed by the momentum wave function. Therefore, we ignore the DM-electron scattering here and focus on the 
one loop DM-nucleus interaction via one loop. Now, following the most general 4-fermion interaction as in Eq.~\eqref{op-fermion} \cite{Kopp:2009et}, 
the one loop contribution to DM-nucleon interaction involves the following integration:

%

\bea
\int\frac{d^4 q}{\left(4\pi\right)^4} \text{Tr}\Biggl[\Gamma_\ell\frac{\slashed{q}'+m_\ell}{q^{'2}-m_\ell^2}\gamma^\mu\frac{\slashed{q}+m_\ell}{q^{2}-m_\ell^2}\Biggr] \,, 
\label{eq:loop} 
\eea

\noindent where $q$ and $q'=k-k'+q$ denote the loop momentum, with $k,k'$ as the momenta carried by the incoming and outgoing DM particles respectively. 
Note that, the loop contribution is non-zero only for vector and tensor lepton currents $\Gamma^\mu_\ell=\gamma^\mu,\sigma^{\mu\nu}$. 
For scalar, pseudoscalar and axial-vector currents $\Gamma^\mu_\ell=1,\gamma_5,\gamma^\mu\gamma_5$ this contribution in Eq.~\eqref{eq:loop} identically 
vanish. Therefore, non-zero direct search contribution for leptophilic DM arises only for operators $\mathcal{O}_{DL}^6, \mathcal{O}_{L}^6$. However, the 
contribution from $\mathcal{O}_{L}^6$ to the direct detection cross-section vanishes in the non-relativistic limit and $\mathcal{O}_{DL}^6$ doesn't apply to Majorana 
DM. Therefore, direct detection constraints do not yield any bound on the parameter space for leptophilic (or hadrophilic) Majorana DM. 
Subsequently, we focus on the vector type interaction, namely $\mathcal{O}_{DL}^6 = \frac{c_{\ell_1}}{\Lambda^2} (\bar\chi \gamma^\mu \chi) (\bar\ell \gamma_\mu \ell)$, and calculate the loop-induced matrix element relevant for nuclear recoil experiments.

%
%
The matrix element for $\chi N\to\chi N$ scattering (where $N$ denotes detector nucleus) via one-loop interaction, modulo implicit sum over the light 
quark generations with charge $Q_j$ is thus given by:
\bea\begin{split}
&\mathcal{M} = L^{(1)}\Bigl(\overline{u}_\chi\,\gamma^\mu\,u_\chi\Bigr)\,\Bigl\langle N\Bigl|Q_j\,\Bigl(\overline{\qr}_j\gamma_\mu \qr_j\Bigr)\Bigr|N\Bigr\rangle= L^{(1)} \Bigl(\overline{u}_\chi\,\gamma^\mu\, u_\chi\Bigr) Z\,F\left(q\right)\, \Bigl(\overline{u}_N\,\gamma_\mu\,u_N\Bigr)  \,,
    \end{split}\label{eq:amp-dd}
\eea
where $F\left(q\right)$ denotes the nuclear form factor, $Z$ denotes the Atomic number of the nucleus and the loop factor $L^{(1)}$ is given by
\bea
 L^{(1)} =  \frac{2\alpha_\text{em}}{\pi}\frac{c_{\ell_1}}{\Lambda^2} \int_0^1\,dx\, x\left(1-x\right)\ln \Biggl[\frac{-x\left(1-x\right)\mathcal{Q}^2+m_\ell^2}{\mu^2}\Biggr]\,.
\eea

\noindent $\mathcal{Q}$ in the above expression momentum transfer ($\mathcal{Q}=k-k'$) in the process and $\mu$ is the renormalization scale. 
For $m_\ell\gg\left|\vec{\mathcal{Q}}\right|$ one may neglect the momentum transfer and obtain

\bea
L_{LL}^{(1)} \approx \frac{2\alpha_\text{em}}{\pi}\frac{c_{\ell_1}}{\Lambda^2} \underbrace{\int_0^1 \,dx\, x\left(1-x\right)\ln \Biggl[\frac{m_\ell^2}{\mu^2}\Biggr]}_{=\frac{1}{6}\ln \Biggl[\frac{m_\ell^2}{\mu^2}\Biggr]} = \frac{\alpha_\text{em}}{3\pi}\frac{c_{\ell_1}}{\Lambda^2}\ln\Biggl[\frac{m_\ell^2}{\mu^2}\Biggr]+\mathcal{O}\Bigl(\left|\vec{\mathcal{Q}}\right|/m_\ell\Bigr)\,,
\eea
\noindent where $LL$ stands for the ``leading log'' contribution, neglecting the remaining logarithmic dependence on momentum transfer. With the leading log approximation the matrix element takes the form

\bea\begin{split}
& \mathcal{M}=\Biggl(\frac{\alpha_\text{em}}{3\pi}\frac{c_{\ell_1}}{\Lambda^2}\log\Biggl[\frac{m_\ell^2}{\mu^2}\Biggr] ZF\left(q\right)\Biggr)\Bigl(\overline{u}_\chi\,\gamma^\mu\, u_\chi\Bigr)\Bigl(\overline{u}_N\,\gamma_\mu\, u_N\Bigr).     
\label{eq:me-dir} 
    \end{split}
\eea

For a single operator it is possible to compute the direct search cross-section analytically as illustrated~\cite{Kopp:2009et}. 
However, since we are considering all operators together, it is difficult to obtain an analytical expression 
for the loop-induced direct detection cross-section. More importantly, the Wilson coefficients ($c_{\ell_i}$) can vary substantially
from the typical energies for obtaining DM relic density (where they are defined) to the typical energies of DM direct search experiments 
($\sim$ GeV). We use the {\tt runDM}~\cite{DEramo:2016gos} package to obtain couplings at the energy scale of direct detection
 $\sim 2~\rm GeV$ starting from  energy scale of $\sim$ TeV where the Wilson coefficients are assumed to be $c_{\ell_i}=1$. 
 The renormalization group evolution (RGE) typically introduces mixing between different DM-SM interactions, affecting the size of couplings, 
 and even inducing new couplings which do not appear in a naive comparison. The relic density calculation however remains unaffected with or without the running. 

Note that the matrix element in Eq.~\eqref{eq:me-dir} depend on the renormalization scale $\mu$. To extract physical quantities that does not depend on $\mu$, 
we need to define the renormalization condition. In our case this condition is that at scale $\Lambda \sim 1$ TeV, the coefficient ($c_{q_{1}}^{(\qr_j)}$) 
of the {\it effective} DM-quark operator $(\bar \chi \gamma^\mu \chi) (\bar{\qr}_j\gamma_\mu\qr_j)$ is zero. As an 
illustrative example with only one generation of lepton, in the leading-log approximation (following Eq.~\eqref{eq:me-dir}), 
at the $\overline{\text{MS}}$ scale $\mu$, the coefficient $c_{q_{1}}^{(\qr_j)}$ reads
\begin{equation}
	c_{q_{1}}^{(\qr_j)} (\mu)\approx Q_j \frac{\alpha_{\rm em}}{3 \pi}  \frac{c_{\ell_1}}{\Lambda^2}  \ln \bigg( \frac{m_\ell^2}{\mu^2}\bigg) \,.
	\label{cq_VV_eqn}
\end{equation}
Therefore, our {\it renormalization condition} reads
\bea
 	c_{q_{1}}^{(\qr_j)} (\Lambda) \approx  Q_j \frac{\alpha_{\rm em}}{3 \pi} \frac{c_{\ell_1}}{\Lambda^2} \ln \bigg( \frac{m_\ell^2}{\Lambda^2}\bigg) = 0 \, ,
 	 \label{eq:rencond}
\eea
\noindent where we set $\mu\equiv\Lambda$~\cite{Kopp:2009et}. Using Eq.~\eqref{eq:rencond}, it is easy to infer the coefficient at any energy 
scale $\Lambda_f$ ($ \le \Lambda$) as
\begin{equation}
	c_{q_{1}}^{(\qr_j)} (\Lambda_f) \approx  Q_j \frac{\alpha_{\rm em}}{3 \pi}  \frac{c_{\ell_1}}{\Lambda^2}  \ln \bigg( \frac{\Lambda^2}{\Lambda_f^2}\bigg)\ .
	\label{eq:cq}
\end{equation}
 \begin{table}[htb!]
        \centering
        \resizebox{\textwidth}{!}{%
        \begin{tabular}{|c|c|c|c|c|c|c|c|c|}
        \hline
        \hline
         Relevant Scales & $c_{q_{1}}^{(u)}$ & $c_{q_{1}}^{(d)}$ & $c_{q_{1}}^{(c)}$ & $c_{q_{1}}^{(s)}$ & $c_{q_{1}}^{(b)}$ & $c_{\ell_{1}}^{(e)}$ & $c_{\ell_{1}}^{(\mu)}$ & $c_{\ell_{1}}^{(\tau)}$ \\
         \hline 
         $\Lambda=3$ TeV & 0 & 0 & 0 & 0 & 0 & 1 & 1 & 1 \\
         \hline
         $\Lambda_{\rm NR}=2$ GeV & 0.024 & -0.012 & 0.024 & -0.012 & -0.012 & 0.965 & 0.965 & 0.965 \\
         \hline
         \hline
        \end{tabular}
        }
        \caption{Running of the coupling coefficients $c_{q_1}^{(\qr)}$, and $c_{\ell_1}^{(\ell)}$ of the operators ${\cal O}_{DQ}^6$ and ${\cal O}_{DL}^6$ respectively (see Eq. (\ref{eq:op6})) using {\tt runDM} package, from an initial scale $\Lambda=3$ TeV to the non-relativistic scale $\Lambda_{\rm NR}=2$ GeV relevant for DM-nucleon scattering. For Leptophilic case, only non-zero coefficient of the operator ${\cal O}_{DL}^6$ is assumed with $c_{\ell_1}^{(\ell)}=1$, at the scale $\Lambda$. Note that this is only relevant for Dirac DM, as the vector DM current is zero for the Majorana case.}
        \label{running_table}
    \end{table}

%
Notice in Eq.~\eqref{eq:cq}, that $c_{q_{1}}^{(\qr_j)}$ is zero at the scale $\Lambda_f =  \Lambda$, as per {\it renormalization condition}. 
The DM-nucleon cross section at  direct detection scale $\Lambda_\text{NR}$, is then given by
\bea
\sigma_{\rm SI} \approx \frac{\mu_p^2 c_{\ell_1}^2}{9\pi A^2}\Biggl(\frac{\alpha_\text{em}Z}{\pi \Lambda^2}\Biggr)^2\Biggl[\ln\Biggl(\frac{\Lambda^2}{\Lambda_\text{NR}^2}\Biggr)\Biggr]^2  \,,
\label{dd_px_eqn}
\eea
\noindent where we have ignored the term proportional to DM velocity $v\sim 10^{-6}$, used $F(q)=1$ for brevity and $\mu_p=m_p\,m_\chi/\left(m_p+m_\chi\right)$ 
as the reduced mass of the DM-nucleon system. Its worth reminding that in deriving Eq.~\eqref{dd_px_eqn}, we have assumed the presence of only the vector operator 
$\mathcal{O}_{DL}^6 = \frac{c_{\ell_1}}{\Lambda^2} (\bar\chi \gamma^\mu \chi) (\bar\ell \gamma_\mu \ell)$ at the effective scale $\Lambda$. The running of 
the couplings to a lower scale ($\Lambda_{\rm{NR}}$) generates the DM-quark interaction via operator mixing, thereby producing the spin-independent direct detection 
cross-section as in Eq.~\eqref{dd_px_eqn}. In Table \ref{running_table}, we show the values of the running coupling coefficients $c_{q_1}^{(\qr)}$, and $c_{\ell_1}^{(\ell)}$ 
at direct search experiment scale $\Lambda_\text{NR} = 2$ GeV evaluated using {\tt runDM} package, while the input values are assumed at NP scale $\Lambda=3$ TeV. 
Using them, together with $m_\chi =500$ GeV, we find $ \sigma_{\rm SI} \approx 3 \times 10^{-47}~\text{cm}^2$, 
close to the sensitivity of existing direct search cross-section. It is clear, that the loop-induced WIMP-nucleon spin-independent scattering cross-section 
can provide a very strong bound on the leptophilic DM parameter space, which will be shown together with the 
relic density allowed parameter space in Subsection.~\ref{sec:paramspace2}.

\subsection{Indirect detection}
\label{sec:indirect}

Although DM occasionally annihilate after thermal-freeze-out, it can still occur in regions with very large DM density, for example, at the centre of galaxies. The annihilations produce a bunch of SM particles in pair of various kinds, such as:

\bea
\nonumber
\chi + \bar{\chi} \to f \bar{f},~ \gamma\gamma,...~~f: ~\text{SM ~fermions};
\eea 

\noindent where the presence of anti-particle ($\bar{f}$) and photons in particular, play crucial role in elucidating the presence of DM, if an excess is found.   
Various satellite and ground based telescopes (like the Fermi Large Area Telescope (FermiLAT)~\cite{Fermi-LAT:2016afa} or Cherenkov Telescope Array (CTA)~\cite{CTAConsortium:2012fwj}) 
have searched for excess in anti-particle or photons beyond astrophysical processes which may 
arise from the DM annihilation. For example, the production rate ($\Gamma$) of SM particles from DM 
annihilation is crucially dependent on parameters like DM density ($\rho_{\tt DM}$) and strength of annihilation cross-section $(\sigma v)$~\cite{Cirelli:2010xx, Slatyer:2017sev, Hooper:2018kfv}

\bea
\Gamma \sim \int_V \left(\frac{\rho^2_{\tt DM}}{\mchi^2}\right) dV \times (\sigma v) (N_\text {SM}) ;
\label{eq:ind}
\eea 

\noindent where $N_\text{SM}$ represents number of SM particles per annihilation event. Note that annihilation cross-section present in the equation above is also crucial to produce thermal freeze-out for WIMP like DM to produce correct relic density. In context of the effective leptophilic DM operators that we are concerned, we can therefore calculate annihilation cross-section to produce current indirect search observation and relate to the annihilation cross-section to produce correct relic density and 
thus produce a bound on the model. We further note that significant constraint can only be obtained for leptophilic Dirac DM as Majorana DM would produce velocity dependent annihilation cross-section, which results in a suppressed annihilation rate, and hence does not provide any reasonable bound from indirect detection. A detailed analysis of indirect detection bound on the DM parameter space is beyond the scope of this paper. We simply calculate the thermal averaged DM pair annihilation cross-section into different SM leptonic final states as shown in table \ref{indir_table}, and project the bound from the non-observation of gamma-ray signals from DM 
annihilation in dwarf satellite galaxies~\cite{MAGIC:2016xys}. In the absence of any boost factor {\it viz.,} Sommerfeld enhancement~\cite{Feng:2010zp}, the bounds from indirect search turns out to be much 
milder than that due to direct detection, and constraints the relic density allowed parameter space for the DM only at very low $m_\chi$. 

\begin{table}[htb!]
        \centering
        \begin{tabular}{|c|c|c|}
        \hline
          $\langle \sigma v \rangle_{\chi \bar\chi \to e^+e^-}$ $({\rm cm}^3/{\rm s})$  & $\langle \sigma v \rangle_{\chi \bar\chi \to \mu^+\mu^-}$ $({\rm cm}^3/{\rm s})$ & $\langle \sigma v \rangle_{\chi \bar\chi \to \tau^+\tau^-}$ $({\rm cm}^3/{\rm s})$ \\
          \hline
          $7.5 \times 10^{-27}$  & $7.5 \times 10^{-27}$ & $7.5 \times 10^{-27}$ \\
          \hline
          \end{tabular}
        \caption{Velocity averaged annihilation cross-section $\langle \sigma v \rangle$ (s-wave) of dominating channels for the relic density satisfying points. We have chosen illustrative values $c_{\ell_i}=1$ (Dirac DM). The annihilation to quark final states are negligibly small in comparison to the leptonic final states. Note that the annihilation to neutrino final states are identically zero due to interference effects (e.g., see Eq.~\eqref{ceq1_cases}).}
        \label{indir_table}
    \end{table}

\subsection{Viable parameter space}
\label{sec:paramspace2}

\begin{figure}[htb!]
$$
\includegraphics[scale=0.4]{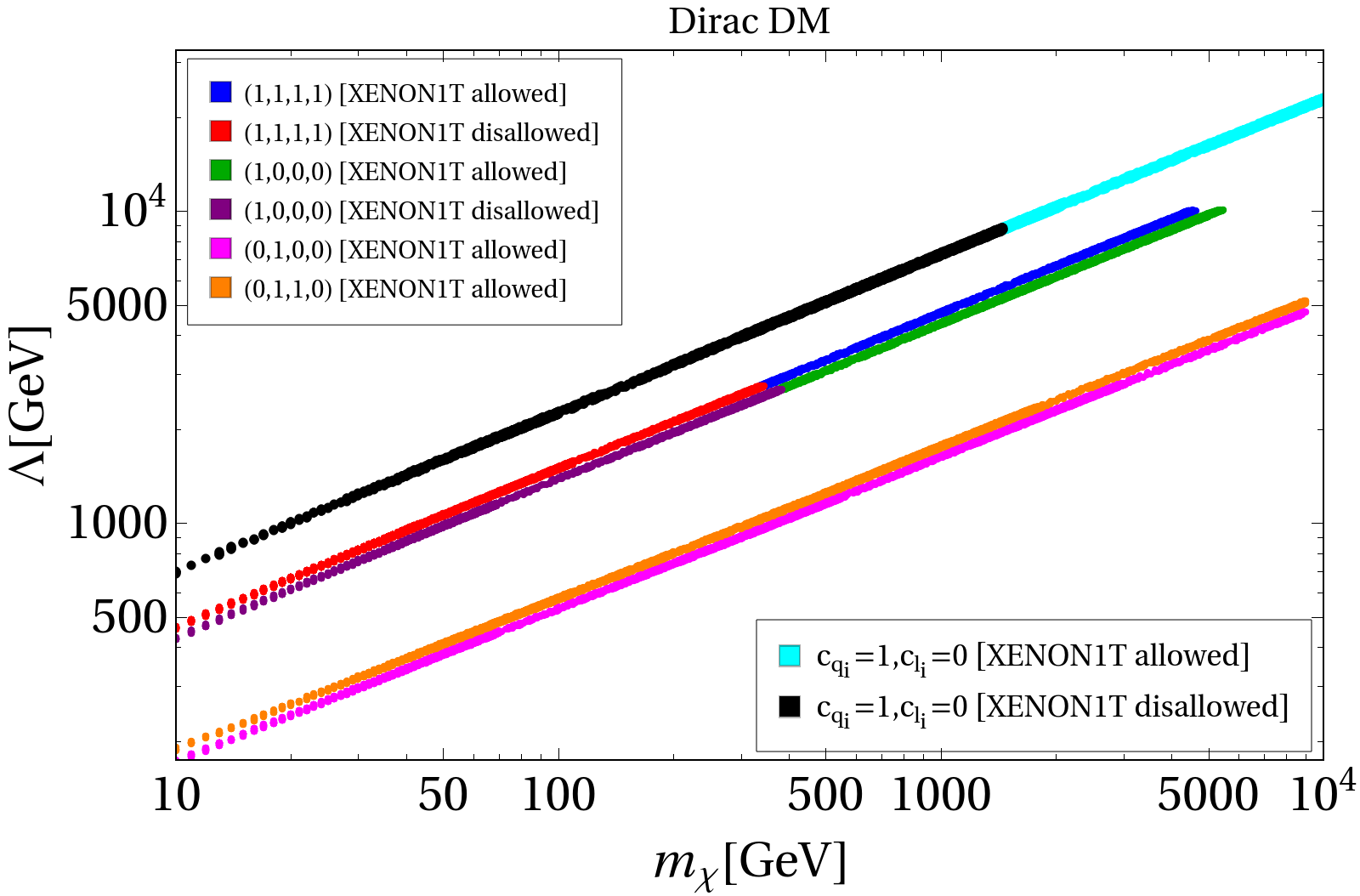}~~
\includegraphics[scale=0.4]{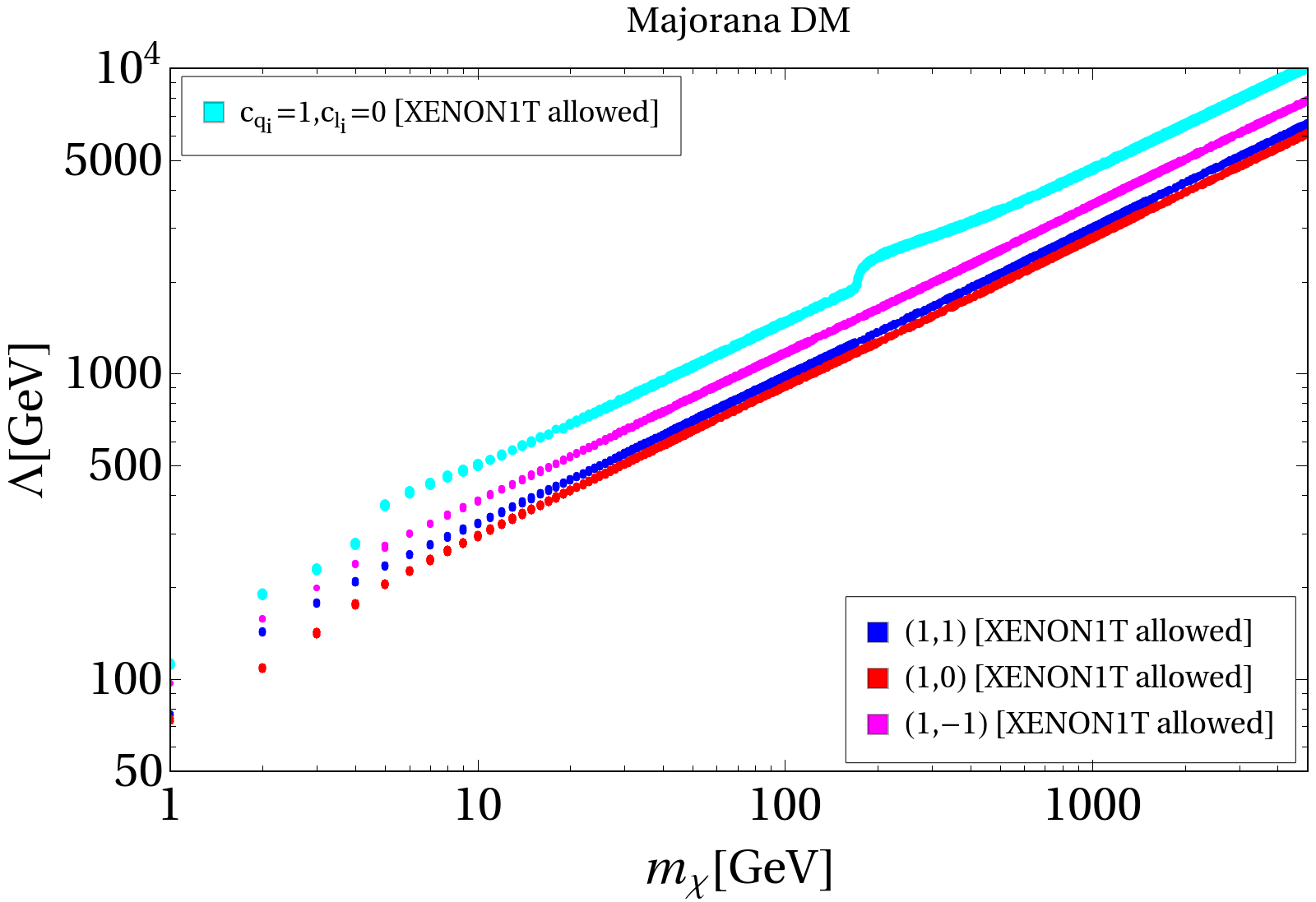}
$$
$$
\includegraphics[scale=0.4]{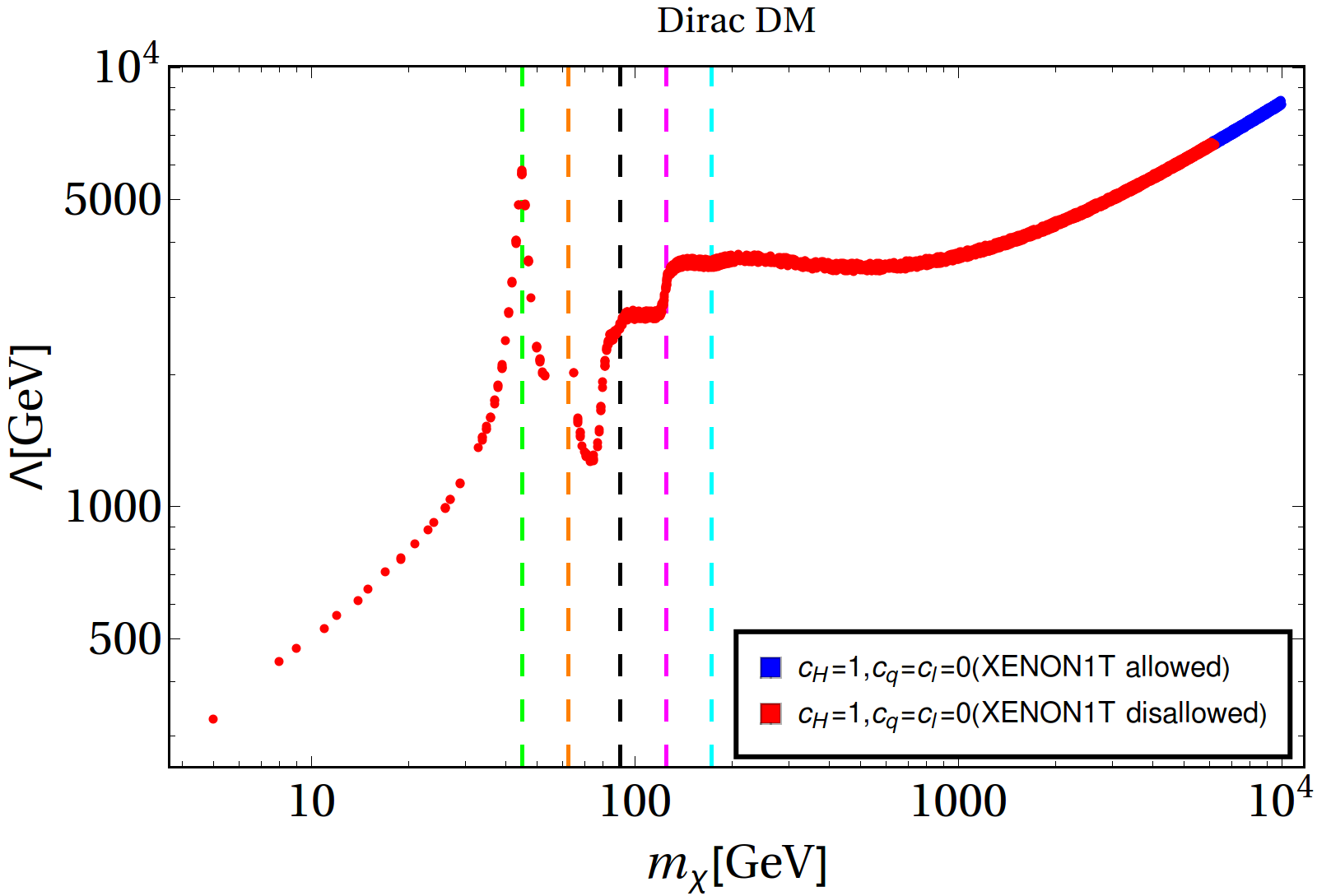}
$$
\caption{Relic density and direct search (XENON1T) allowed parameter space for hadrophilic ($c_{q_i}=1, c_{\ell_i}=0$), 
and leptophilic ($c_{q_i}=0, c_{\ell_i}=1$) cases in the top panel and Higgs portal case ($c_{H_i}=1,c_{\ell_i}=c_{q_i}=0$) in the bottom panel. 
Top Left (right) figure describes Dirac (Majorana) DM. The coupling values $(c_{\ell_1}, c_{\ell_2}, c_{\ell_3},  c_{\ell_4})$ is denoted as an ordered set for Dirac DM and $(c_{\ell_2}, c_{\ell_3})$ for Majorana DM. XENON1T allowed region and disallowed regions are marked in different colours (see figure inset and text for details); for example, in hadrophilic Dirac DM case (top left), XENON1T allowed region is shown in cyan, while the disallowed region is shown in black. Note, there is no spin-independent direct detection bound on Majorana DM. }
\label{fig:paramspace}
\end{figure}

After discussing the major constraints on DM model coming from relic density, direct and indirect search experiments, we summarise here the available DM 
parameter space for leptophilic DM operators from all of these constraints, in terms of $\mchi-\Lambda$. In Fig.~\ref{fig:paramspace} we show the Planck observed relic density allowed parameter space for 
Dirac (top left) and Majorana (top right) DM in $\Lambda-m_\chi$ plane by considering all relevant operators together for both hadrophilic  ($c_{\ell_i}=0,c_{q_i}=1$) and leptophilic ($c_{\ell_i}=1,c_{q_i}=0$) cases.  
We vary the two free parameters of the theory, namely the cut-off scale $\Lambda$ and the DM mass $m_\chi$, in the following range $$\Lambda: \{1-10^4\}~\text{GeV};~ m_\chi:\{1-10^4\}~\text{GeV}$$
 while considering $\Lambda\gtrsim m_\chi/2\pi$ in order to make sure that the effective theory description remains valid. First of all, notice that in all plots, with the increase in DM mass, $\Lambda$ 
 also rises to satisfy the relic abundance, following that the relic density of thermal WIMP goes as $\Omega_\chi h^2\sim 1/\langle\sigma v\rangle\sim \Lambda^4/m_\chi^2$. 
 XENON1T allowed region and disallowed regions are marked in different colours. 
 
 In the top left plot of Fig.~\ref{fig:paramspace}, for hadrophilic case, the relic density allowed region is significantly ruled out from 
 spin-independent direct detection bound upto $m_\chi\lesssim 6.4~\rm TeV$ for $\Lambda\lesssim 18~\rm TeV$ (shown by black points) making it difficult for collider probe. 
In the same plot, we also show the leptophilic cases where coupling values $(c_{\ell_1}, c_{\ell_2}, c_{\ell_3},  c_{\ell_4})$ is denoted as an ordered set for 
Dirac DM.  Amongst different possibilities considered, we see that vector coupling plays a crucial role as those cases with $c_{\ell_1}=1$ almost overlap and segregates 
from those with $c_{\ell_1}=0$. As explained earlier in Sec.~\ref{sec:loopDD}, the loop-induced direct search plays an important role as this rules out leptophilic DM mass $m_\chi$ and $\Lambda$ significantly. 
For example, (1,1,1,1) case is ruled out upto $(m_\chi,\Lambda)\sim$(345, 2780) GeV, and (1,0,0,0) case is ruled out upto $(m_\chi,\Lambda)\sim$(380, 2700) GeV. 
Other choices of model parameters, such as (1,1,0,0), (1,1,1,0) etc. are very close to the (1,0,0,0) case and hence we did not show them here.
Note that for the cases with $c_{\ell_{1}}=0$, the spin-independent direct search cross-section vanishes, and there is no bound on $\Lambda-m_\chi$ parameter space from direct detection. 
Also notice that for the same $m_\chi$, the leptophilic cases requires a lower $\Lambda$ than the hadrophilic case to satisfy relic abundance. 
This stems from larger annihilation cross-section due to the color factor to quark final states, which therefore requires a comparatively larger $\Lambda$ 
to tame down the annihilation cross-section to satisfy the DM relic. 

The same exercise is done for Majorana DM in the top right panel of Fig.~\ref{fig:paramspace} for both leptophilic (red, blue, and magenta) and hadrophilic (cyan) case. Couplings $(c_{\ell_2}, c_{\ell_3})$ are used as an 
ordered pair for defining Majorana DM models. We again see here, that the hadrophilic case requires larger $\Lambda$ than the leptophilic cases following the same line of arguments as before. 
Here one should notice the presence of two prominent bumps corresponding to the opening of $b\bar{b}$ and $t\bar{t}$ final states for the hadrophilic case. For  Majorana DM, there is no significant 
direct detection bound because of the absence of DM vector current, and due to the fact that the contribution of ${\cal O}_{L}^6$ (see Eq. (\ref{eq:op6})) to the direct detection cross-section vanishes in 
the non-relativistic limit. This makes all of the relic density allowed parameter space available starting from DM mass as low as 1 GeV with $\Lambda\sim 10~\rm GeV$. 

We have also shown the allowed parameter space corresponding to the Higgs portal operators with $(c_{H_i}=1,c_{q_i}=c_{\ell_i}=0)$ case for comparison in the bottom panel of Fig.~\ref{fig:paramspace}. 
The vertical black dashed lines correspond to (from left to right) resonances due to $Z$ and Higgs boson, opening of $Z$, Higgs and top quark final states respectively. Here we see both relic density and 
direct detection bounds are satisfied for $m_\chi\gtrsim 6.2~\rm TeV$ and $\Lambda\gtrsim 7~\rm TeV$, again making this model difficult to probe in ongoing collider experiment. 
We carry out the analysis of viable parameter space with negative Wilson coefficients in Appendix~\ref{sec:negative}.

\begin{figure}[htb!]
$$
\includegraphics[scale=0.5]{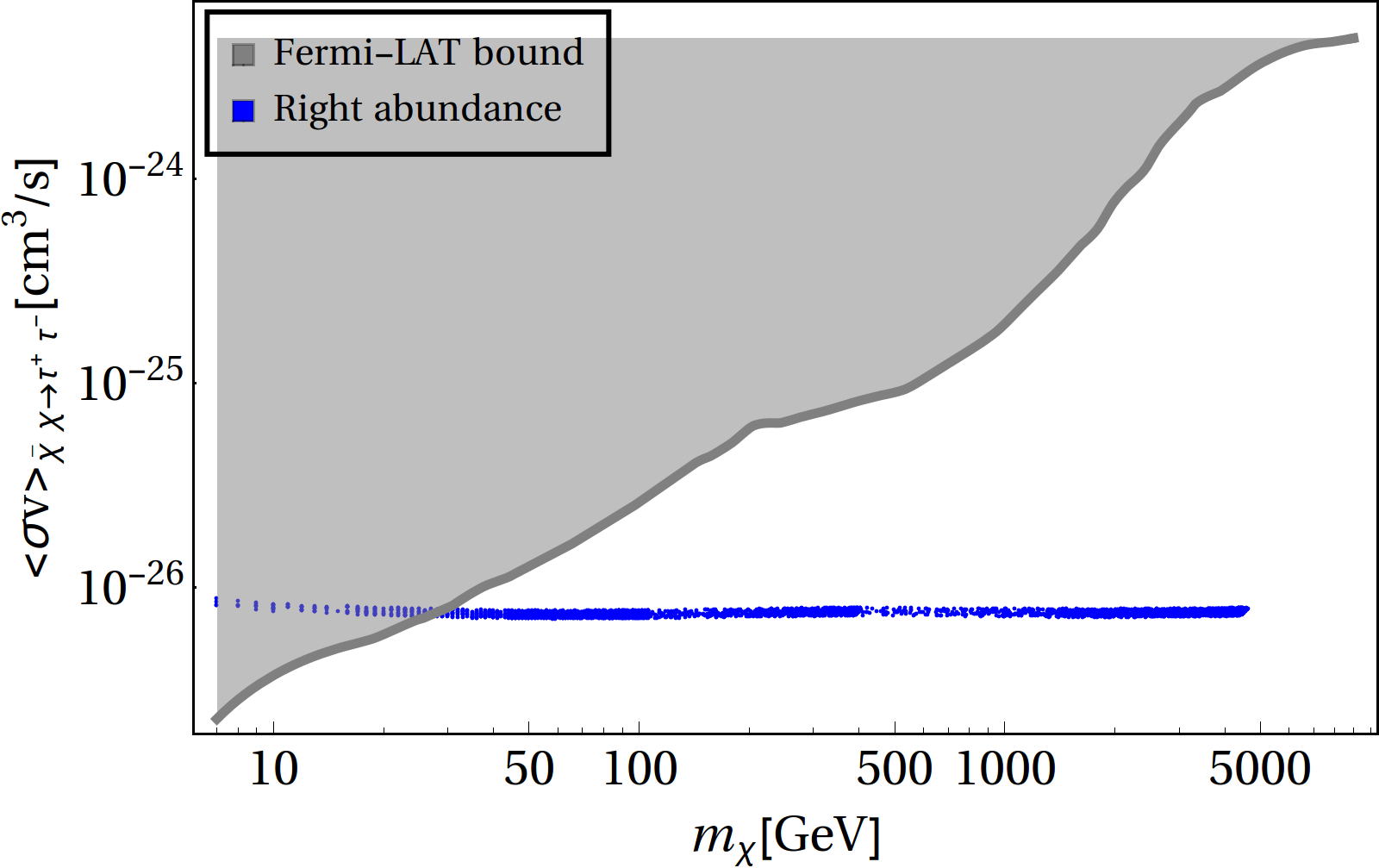}
$$
\caption{Indirect detection bound on Dirac DM annihilation cross-section. The gray region is excluded from the combined data analysis of MAGIC cherenkov telescopes and Fermi large area telescope (FermiLAT) that searched for the gamma-ray signals from DM annihilation \cite{MAGIC:2016xys}. The blue points represent thermally averaged annihilation cross-section ($\langle \sigma v \rangle)$ for DM annihilating into $\tau^+ \tau^-$ pair.  As can be seen from the figure, this bound rules out DM having mass below $30$ GeV. }
\label{indirect:dirac}
\end{figure}

\begin{figure}[htb!]
    \centering
    $$
    \includegraphics[width=0.6\textwidth]{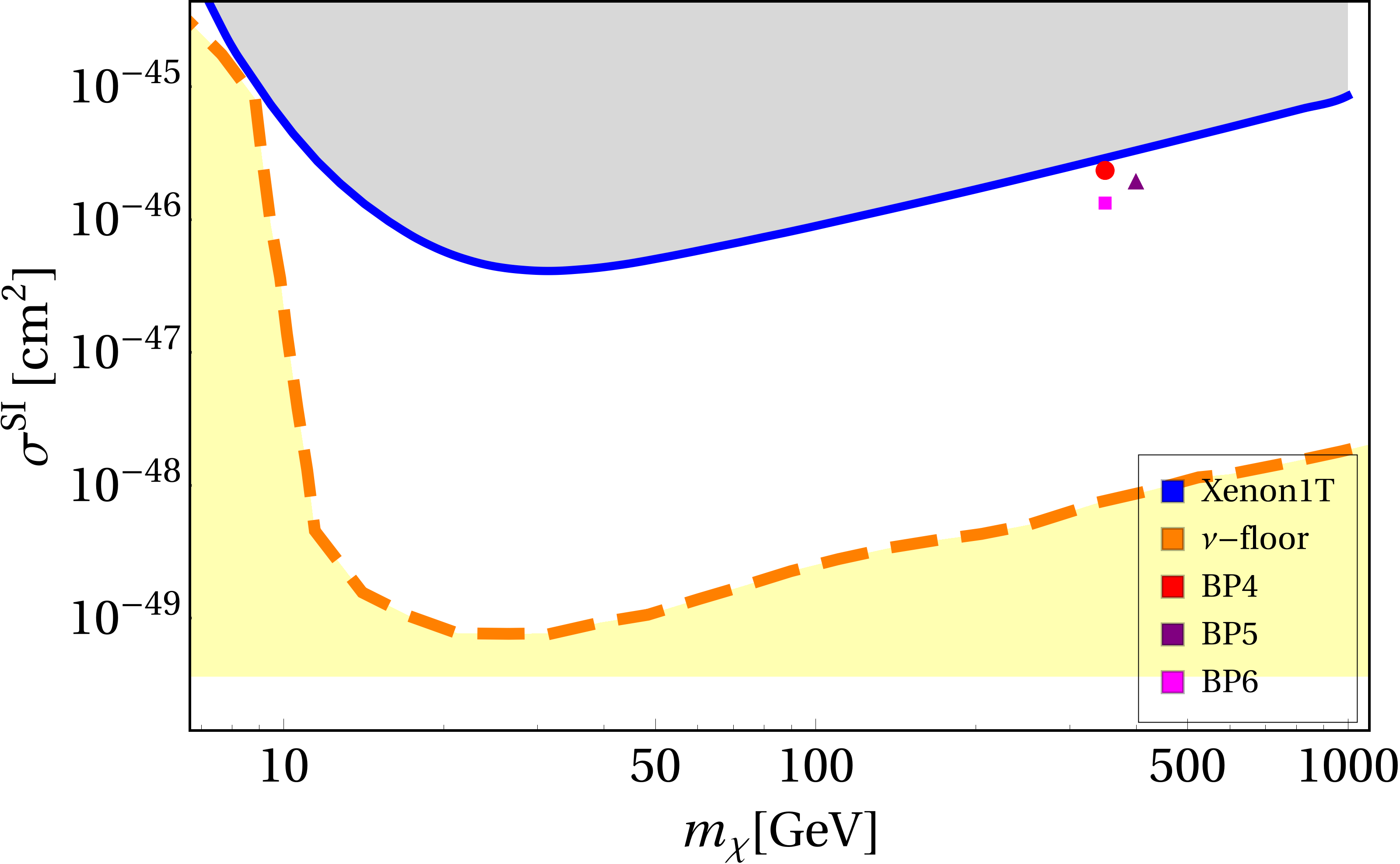}
    $$
    \caption{ Benchmark points (BP4 in red, BP5 in purple and BP6 in magenta: see table~\ref{tab:BP}) for Dirac DM confronted with XENON1T bound shown by 
    gray excluded region in $\mchi$ (in GeV) versus the spin-independent direct detection cross-section $\sigma_{\rm SI}$ (in cm$^2$) plane. The Yellow region at the bottom corresponds to neutrino floor. Note that
    Majorana DM do not face any direct search constraints in leptophilic case as considered.}
    \label{DD_fig}
\end{figure}

Let us now turn to indirect search bounds. The constraint on thermally-averaged cross-section for $\tau^+ \tau^-$ final state in our model yields a bound on 
Dirac DM mass, namely $m_\chi \lesssim 30$ GeV (Fig.~\ref{indirect:dirac}). This bound is much less stringent than the bound from direct detection (as in Fig.~\ref{fig:paramspace}). 
In further analysis for collider search, we therefore address benchmark points where DM constraints are coming mainly from relic density 
and direct search\footnote{A recent analysis concerning bound on leptophilic DM from AMS-02 data can be found in~\cite{John:2021ugy}.}. A few such benchmark points 
are tabulated in table~\ref{tab:BP} where we specify the DM mass ($\mchi$), corresponding NP scale ($\Lambda$) and Wilson coefficients as input model parameter in an 
ordered set $(c_{\ell_1}, c_{\ell_2}, c_{\ell_3}, c_{\ell_4})$ for Dirac case and $(c_{\ell_2}, c_{\ell_3})$ for Majorana DM cases. 
The benchmark points for Dirac DM are also represented in the spin-independent direct detection cross-section ($\sigma_{\rm SI}$ (in cm$^2$) plane) versus DM mass plane in Fig.~\ref{DD_fig}
with XENON1T bound. As Majorana DM do not face any bound from direct search, corresponding benchmark points can't be shown in this plane.

\begin{table}[htb!]
	\centering
	\begin{tabular}{|l|c|c|c|c|c|c|c|c|c|r|}
		\hline
		DM scenario & Benchmark Points & Model  & $m_\chi$ (GeV) & $\Lambda$ (GeV) \\
		\hline
		 & BP1 & $(1, 1)$ & 70 & 816 \\
		 Majorana & BP2 & $(1, 1)$ & 112 & 1025 \\
		 & BP3 & $(1, -1)$  & 70 & 990 \\
		 \hline
		& BP4 & $(1, 1, 1, 1)$  & 350 & 2800 \\
		Dirac & BP5 & $(1, 1, 1, 1)$  & 400 & 3000 \\
		& BP6 & $(1, 1, -1, -1)$  & 350 & 3300 \\
		\hline
	\end{tabular}
	\caption{Benchmarks points satisfying DM relic density and direct detection constraints for Majorana (BP1, BP2, BP3) and Dirac (BP4, BP5, BP6) DM scenario considering all leptophilic operators together. 
	Different choices for the Wilson coefficients are denoted as different models and represented as an ordered set of allowed $\{c_{\ell_i}\}$. For example, $(1, 1, -1, -1)$ in BP6 refers to the choice 
	$(c_{\ell_1}=1, c_{\ell_2}=1, c_{\ell_3}=-1, c_{\ell_4}=-1)$, whereas $(1, -1)$ in BP3 denotes $(c_{\ell_2}=1, c_{\ell_3}=-1)$.}
	\label{tab:BP}
\end{table}

Before closing this section, we would like to mention that the recent result from PandaX-4T~\cite{PandaX-4T:2021bab} puts a stringent bound on the spin-independent DM-nucleon cross-section, 
for example, $3.3\times 10^{-47}~\text{cm}^2 (90\%$ C.L.) for $\m_\chi \sim$ 30 GeV. Satisfying this exclusion limit follows a similar exercise as above and requires an even larger mass ($m_\chi$) and 
effective scale ($\Lambda$) for Dirac DM (when $c_{\ell_1} \neq 0$), like $\{m_\chi,\Lambda\}:\{430,3150\}~\rm GeV$. Note that this can still be probed at a 1 TeV lepton collider. However, 
for further analysis we content ourselves with the already published limit from XENON1T data~\cite{XENON:2018voc}.

\section{Collider search prospects of Leptophilic DM}
\label{sec:collider}

It is obvious that DM operators that connects to SM leptons have very suppressed production (via loop) at the currently running LHC and given the SM background contribution, it is difficult to find them at LHC.
\begin{figure}[htb!]
$$
\includegraphics[scale=0.15]{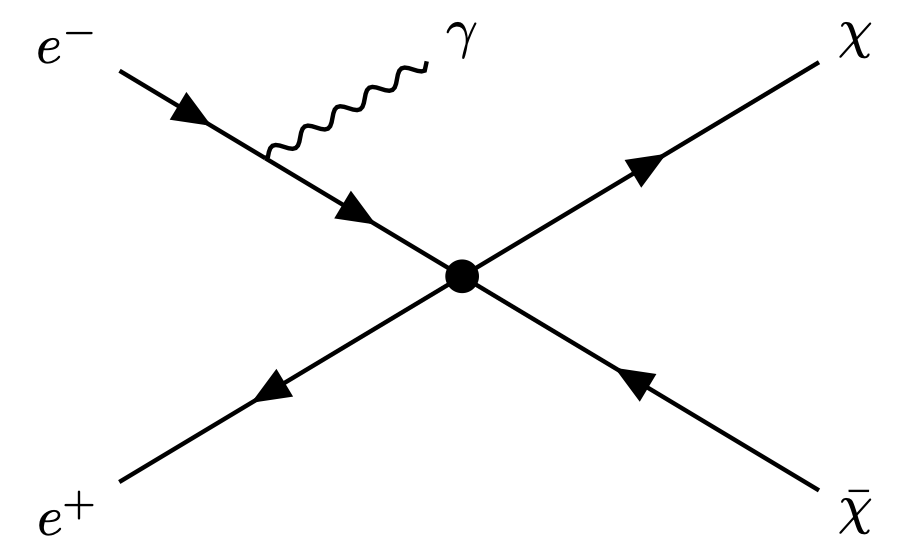}~~
\includegraphics[scale=0.15]{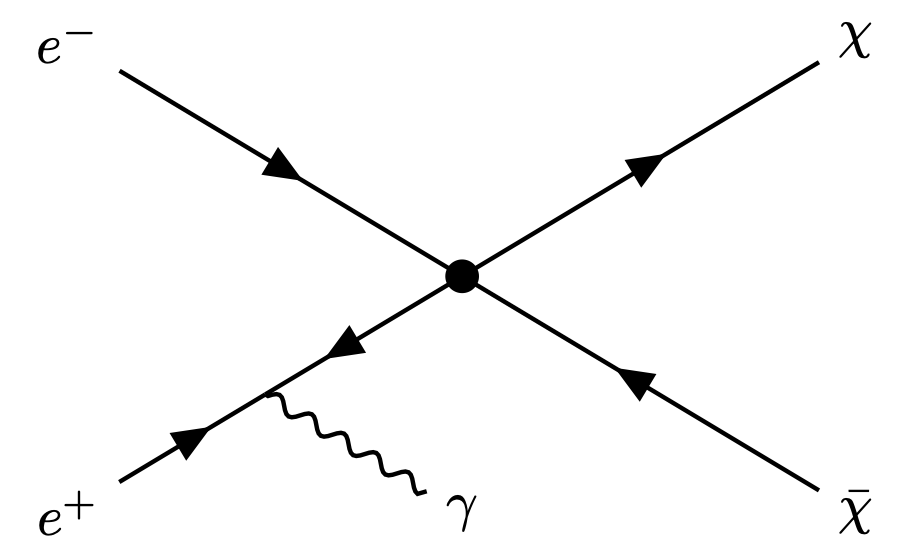}
$$
\caption{Mono-photon with missing energy $\gamma+\slashed{E_T}$ signature at the collider where the photon is irradiated from the initial state charged leptons. The blob represents the effective nature of the vertex.}\label{fig:sig-feyn}
\end{figure}

On the other hand, such operators can be probed at $e^+e^-$ machine. The main production 
of DM occurs via the contact interaction dictated by the EFT operators. However, DM being invisible at the detector a pair production of DM  would not produce 
a signal at the detector, so one needs to rely upon one or more initial state radiation (ISR) photon. Therefore, the signal in such a case is predominantly 
given by 
\bea
e^+e^- \to \chi \bar{\chi} + \gamma \rightarrow \text{ME}+ \gamma  \,; 
\eea 
where ME refers to missing energy with monophoton. The Feynman graph is shown in Fig.~\ref{fig:sig-feyn}. 
Note that higher multiplicity of radiated photon will diminish the cross-section of the process further. Therefore, mono-photon events with missing energy is a vanilla EFT signature of DM production at colliders that have been extensively studied in 
the literature in context with both hadron and lepton colliders. 

As has been emphasized already, in a hadron collider, since the partonic CM frame is not the same as the hadronic CM frame, it is rather inconsistent to apply effective theory. 
The leptonic collider, on the other hand, provides a simple kinematics where a higher NP scale ($\Lambda$) compared to the known CM energy ($\sqrt{s}$), with $\Lambda > \sqrt{s}$ makes the EFT limit validated. 
As we have seen from the previous section, for a hadrophilic Dirac DM, the spin-independent direct search limit on the relic density allowed parameter space becomes so severe that it only allows DM as heavy as about 
$\mchi \sim$ 6 TeV with the NP scale as large as $\Lambda \sim$ 18 TeV, that lies almost out of LHC sensitivities. For a leptophilic Dirac DM, on the other hand, it is possible to bring down the allowed DM mass to about 
345 GeV with $\Lambda\sim 2.8~\rm TeV$. Such a DM can not certainly be pair-produced at a lepton collider with $\sqrt{s}=500~\rm GeV$, but can be probed with CM energy of 1 TeV. 
For leptophilic Majorana DM since all of its relic density satisfying parameter space is also allowed by direct search limits, this scenario is testable at a lepton collider with CM energy of 500 GeV or even below. 
In the following we perform event-level collider simulation for both Dirac and Majorana DM at those select benchmark points mentioned in Table~\ref{tab:BP}.

It is worth mentioning here that constraints on leptophilic DM from mono-photon searches at LEP have been analyzed in~\cite{Fox:2011fx} using effective interactions and investigated with experimental data 
from DELPHI~\cite{DELPHI:2003dlq, DELPHI:2008uka}. This results in a bound on the effective DM-electron coupling scale, $\Lambda:250-500$ GeV for DM mass $m_\chi\lesssim 80$ GeV, 
depending on the operator structure. The most stringent bound on $\Lambda$ arises from vector current (i.e., $O^6_{DL}$ in our case) which disallows $\Lambda<500$ GeV at 90\% C.L. for $m_\chi\lesssim 60$ GeV. 
For DM mass $m_\chi\gtrsim 100$ GeV, LEP is completely insensitive for kinematic reasons. On the other hand, LHC bound on leptophilic DM has been discussed in~\cite{Bell:2014tta}, 
where the DM communicates with the SM via leptophilic $Z'$ and the interaction with quarks (partons) take place via 1-loop. However, the bounds are model-dependent, and analyzed for $4\ell+\slashed{E_T}$ 
final states, taking each lepton flavour in turn. Hence, this is not directly applicable for our framework. In Table~\ref{tab:BP}, 
all the benchmark points are well above the LEP exclusion limit. 


\subsection{DM Production cross-section and beam polarization}

The central point of the collider analysis depends on the DM production cross-section at $e^+e^-$ collider. Here we elaborate on that. 
Since the electroweak part of the SM is chiral, appropriate beam polarization may be helpful to reduce SM backgrounds and increase NP signal, so
we focus on polarized beams. The general expression for differential cross-section of the process $e^+e^- \rightarrow \chi \bar \chi$ with partial beam polarization 
($-1< P_{e^\pm} < 1$)\footnote{Note that $P_{e^\pm}=\frac{n_R-n_L}{n_R+n_L}$ refers to longitudinal beam polarisation containing a mixture of $n_L$ left handed and $n_R$ right handed electrons or positrons.} 
is given by~\cite{Fujii:2018mli,Bambade:2019fyw}

\begin{equation}\begin{split}
\frac{d \sigma (P_{e^-}, P_{e^+})}{d \Omega} &=\frac{1}{4}\Biggl\{\Bigl(1+P_{e^-}\Bigr)\Bigl(1+P_{e^+}\Bigr) \bigg( \frac{d \sigma}{d \Omega}\bigg)_{\rm RR}+\Bigl(1-P_{e^-}\Bigr)\Bigl(1-P_{e^+}\Bigr) \bigg( \frac{d \sigma}{d \Omega}\bigg)_{\rm LL}\\&+\Bigl(1+P_{e^-}\Bigr)\Bigl(1-P_{e^+}\Bigr) \bigg( \frac{d \sigma}{d \Omega}\bigg)_{\rm RL}+\Bigl(1-P_{e^-}\Bigr)\Bigl(1+P_{e^+}\Bigr) \bigg( \frac{d \sigma}{d \Omega}\bigg)_{\rm LR}\Biggr\}  \, ,               
                \end{split}\label{eq:pol-sig}
\end{equation}
\noindent where $\sigma_{ij}$ with $i,j\in L,R$ is the cross section for a given process with completely polarized beams of the four possible orientations. The final state of 
our interest are fermion pairs and the operators that mediate the DM-SM interactions are either vector or axial-vector operators. Hence, spin conservation indicates 
that the total spin of initial state should either be $\pm 1$, indicating that $\sigma_{\rm RR}=\sigma_{\rm LL}=0$ always, reducing the cross-section to the form:

    \begin{align}
        \nonumber
        \frac{d \sigma (P_{e^-}, P_{e^+})}{d \Omega} &= \frac{(1+ P_{e^-}) (1-P_{e^+})}{4} \bigg( \frac{d \sigma}{d \Omega}\bigg)_{\rm RL}  \\\nonumber 
        &  + \frac{(1- P_{e^-}) (1+P_{e^+})}{4} \bigg( \frac{d \sigma}{d \Omega}\bigg)_{\rm LR}  \\
       \label{sigpol}
    \end{align}
    where, for Dirac DM,

    \begin{align}
    \nonumber
    \bigg( \frac{d \sigma}{d \Omega}\bigg)^{\tt dirac}_{\rm RL} & = \frac{s \sqrt{1-\beta_{\chi}^2}}{32 \pi^2 \Lambda^4} \bigg[ \Big\{\ (1+\beta^2_{\chi})(c_{\ell_1}+c_{\ell_4})^2+(1-\beta^2_{\chi})(c_{\ell_2}+c_{\ell_3})^2 \Big\}  +4\sqrt{1-\beta^2_{\chi}} \\  &  \quad
    \Big\{(c_{\ell_1}+c_{\ell_4})(c_{\ell_2}+c_{\ell_3})\Big\}\cos\theta 
    +(1-\beta^2_{\chi})\Big\{(c_{\ell_1}+c_{\ell_4})^2+(c_{\ell_2}+c_{\ell_3})^2\Big\} \cos^2 \theta \bigg] \ ,
    \label{RLeqn}
    \end{align}

    \begin{align}
    \nonumber
    \bigg( \frac{d \sigma}{d \Omega}\bigg)^{\tt dirac}_{\rm LR} & = \frac{s \sqrt{1-\beta_{\chi}^2}}{32 \pi^2 \Lambda^4} \bigg[ \Big\{\ (1+\beta^2_{\chi})(c_{\ell_1}-c_{\ell_4})^2+(1-\beta^2_{\chi})(c_{\ell_2}-c_{\ell_3})^2 \Big\}  +4\sqrt{1-\beta^2_{\chi}} \\  &  \quad
    \Big\{(c_{\ell_4}-c_{\ell_1})(c_{\ell_2}-c_{\ell_3})\Big\}\cos\theta 
    +(1-\beta^2_{\chi})\Big\{(c_{\ell_1}-c_{\ell_4})^2+(c_{\ell_2}-c_{\ell_3})^2\Big\}\cos^2\theta \bigg] \ ,
    \label{LReqn}
    \end{align}
where, $\beta_{\chi}=\frac{2 m_{\chi}}{\sqrt{s}}$ , and $\theta$ is the DM scattering angle in the center of mass frame.  

On the other hand, for Majorana DM, we can use Eq.~\eqref{sigpol} for generic differential cross-section along with: 
    \begin{align}
        \bigg( \frac{d \sigma}{d \Omega}\bigg)^{\tt majorana}_{\rm RL} & = \frac{s}{16 \pi^2 \Lambda^4} \Big[ \Big( c_{\ell_2} + c_{\ell_3} \Big)^2 \Big( 1- \beta_{\chi}^2 \Big)^{3/2}(1+\cos^2\theta) \Big] \ ,
  \label{RLeqn_maj}
    \end{align}
and 
    \begin{align}
        \bigg( \frac{d \sigma}{d \Omega}\bigg)^{\tt majorana}_{\rm LR} & = \frac{s}{16 \pi^2 \Lambda^4} \Big[ \Big( c_{\ell_2} - c_{\ell_3} \Big)^2 \Big( 1- \beta_{\chi}^2 \Big)^{3/2}(1+\cos^2\theta) \Big] \ .
  \label{LReqn_maj}
    \end{align}

Furthermore, it is evident from Eq.~\eqref{RLeqn} and Eq.~\eqref{LReqn} that the combination $\{c_{\ell_2}, c_{\ell_3}\}$, and $\{c_{\ell_1}, c_{\ell_4}\}$ breaks 
the Parity symmetry, as these specific combination of coefficients gives rise to $V \pm A$ type interactions. 
It is then straightforward to find the analytical expression of unpolarized cross-sections ($\sigma_0$) for leptophilic Majorana and Dirac DM production via 
designated operators as given by,

\bea
     \sigma^{\tt dirac}_{0} &=& \frac{s}{12 \pi \Lambda^4} \sqrt{1-\beta_{\chi}^2} \bigg[ \Big(c_{\ell_1}^2 + c_{\ell_4}^2 \Big) \Big(1+  \frac{\beta_{\chi}^2}{2} \Big) + \Big(c_{\ell_2}^2 + c_{\ell_3}^2 \Big) \Big(1-\beta_{\chi}^2 \Big)\bigg]\,, \nonumber \\ 
     &=& \frac{s}{3 \pi \Lambda^4} \left(1- \frac{m_\chi^2}{s}\right) \sqrt{1-\frac{4 m_\chi^2}{s}}\,; ~{\rm where} ~c_{\ell_i}=1~(i=1{\text -}4)\,,
     \label{eq:dirac}
\eea

and

\bea
\sigma^{\tt majorana}_{0}  &=& \frac{s}{6 \pi \Lambda^4}\Big(c_{\ell_2}^2 + c_{\ell_3}^2 \Big) \left(1-\beta_{\chi}^2\right)^{3/2}\,, \nonumber \\ 
&=& \frac{s}{3 \pi \Lambda^4} \left(1-\frac{4 m_\chi^2}{s}\right)^{3/2} \,; ~~~{\rm where} ~c_{\ell_i}=1~(i=2,3)\,.
\label{eq:majorana}
\eea

It is worthy mentioning that polarised cross-section can also be written as \cite{Bambade:2019fyw}
\bea
&&\sigma\Bigl(P_{e^-},P_{e^+}\Bigr)=2\sigma_0\,\Bigl(\mathcal{L}_\text{eff}/\mathcal{L}\Bigr)\,\Bigl[1-\mathcal{A}_{LR} \, P_\text{eff}\Bigr]  \,;\\
&&\sigma_0 =  \frac{1}{4}\Biggl[\sigma_\text{RR}+\sigma_\text{LL}+\sigma_\text{RL}+\sigma_\text{LR}\Biggr]\,; ~~\mathcal{L}_\text{eff} = \frac{1}{2}\Bigl(1-P_{e^-}\,P_{e^+}\Bigr)\mathcal{L} \,;
\eea
\noindent where $P_{\rm eff} = \frac{P_{e^-}-P_{e^+}}{1-P_{e^-}P_{e^+}}$, and $\mathcal{L}_\text{eff}$ refers to effective Luminosity of ILC.

We further note from the Eqs.~(\ref{RLeqn}), (\ref{LReqn}), (\ref{LReqn_maj}) and (\ref{RLeqn_maj}) that in the limit of all $c_{\ell_i}=1$:

\begin{equation}
\sigma_{LR}=0;~~\mathcal{A}_{LR}=\frac{\sigma_\text{LR}-\sigma_\text{RL}}{\sigma_\text{LR}+\sigma_\text{RL}} = -1.
\label{eq:asym}
\end{equation}

The above feature can also be understood from the spinor structure for $e^+ e^-$. Consider for example, 
the electron current $\bar v(p')\gamma^\mu\gamma_5 u(p)$ corresponding to 
Majorana DM operator $\mathcal{O}_{L1}^6$. Let us now consider right-polarised electron and replace the current by 
$\bar v(p') \gamma^\mu\gamma_5 \mathcal{P}_R u(p)$, where $\mathcal{P}_R = \frac{1+ \gamma_5}{2}$ represents right chiral projection operator. Clearly this is same as 
$(\mathcal{P}_R v(p'))^\dagger \gamma^0\gamma^\mu\gamma_5 u(p)$, which is non-zero only for left-handed positrons, i.e. with $RL$ combination. 
Trivially, this can be seen to be true for the opposite helicity combination by replacing $\mathcal{P}_R \to \mathcal{P}_L$, i.e. with $LR$ combination. 
When we take all operators together (for both Dirac and Majorana cases), we practically sum the corresponding amplitudes. 
In the limit of $c_{\ell_i}=1$, this results in 
\bea
\mathcal{L}_{\rm eff}=\sum_i \mathcal{O}_i \sim \begin{cases}
    \left(\bar{\chi}\gamma^\mu\gamma^5\chi\right)\bar{e}\gamma_\mu\left(1+\gamma^5\right)e & \text{Majorana DM} \\
    \left(\bar{\chi}\gamma^\mu\gamma^5\chi + \bar{\chi}\gamma^\mu\chi\right)\bar{e}\gamma_\mu\left(1+\gamma^5\right)e  & \text{Dirac DM}
                                            \end{cases}.
\label{ceq1_cases}
\eea
\noindent This expression  is non-vanishing only when ${\mathcal{P}_R\, e }\ne 0$, i.e., the electron is {\it right-handed}. This automatically implies that the positron is 
{\it left-handed}, resulting in a non-vanishing $\sigma_{RL}$ combination. In other words, we infer that the signal cross-section 
is maximum for fully right handed electron and left handed positron ($\{+,-\}$ combination) of the $e^-e^+$ beam. This is evident from Table~\ref{tab:bp-pol}, 
where we show the signal production cross-section for the benchmark points for different choices of the beam polarization. Note here following ILC design, the 
maximum right polarised electron and left polarised positron is possible upto $\{P_{e^-}:P_{e^+}\}=\{+80\%:-30\%\}$, which we abide by.

A different choice for the signs of the Wilson coefficients will however alter this behaviour. Let us consider a distinct set of values for the Wilson coefficients, namely, $c_{\ell_i}\equiv\{1,1,-1,-1\}$ for Dirac DM and 
$c_{\ell_i}\equiv\{1, -1\}$ for Majorana DM, which yields effective Lagrangian of the form
\bea
\mathcal{L}_{\rm eff}= \begin{cases} \left(\bar{\chi}\gamma^\mu\gamma^5\chi + \bar{\chi}\gamma^\mu\chi\right)\, \bar{e}\gamma_\mu\left(1-\gamma^5\right)e & \text{Dirac DM} \\
\left(\bar{\chi}\gamma^\mu\gamma^5\chi\right) \, \bar{e}\gamma_\mu\left(1-\gamma^5\right)e & \text{Majorana DM}
\end{cases}
\label{eq:11m1m1}
\eea
\noindent which clearly shows that the interaction vanishes for $\mathcal{P}_L\,e=0$ and results in $\sigma_\text{RL}=0$ as can also be inferred from Eq.~\eqref{RLeqn}. 
As a result, for this case, ${\mathcal{A}_{LR}}=+1$, and one has to choose  left-polarized electron and right-polarized positron beam to maximize the signal production cross-section. 
This choice of the beam polarization, in context with the ILC, is maximally possible with $\{P_{e^-}:P_{e^+}\}=\{-80\%:+30\%\}$. 
However, this choice will result in a suppressed signal cross-section compared to the SM background, as seen from Table~\ref{tab:bp-pol} and Table~\ref{tab:bckg-pol}. 
Consequently, the signal significance will deteriorate, making this Dirac DM scenario more difficult to probe at the ILC. 
Therefore, we do not consider $\{1,1,-1,-1\}$ combination of the Wilson coefficients for event simulation.  
Worthy to mention here that, in analyzing the signal and the background we have adopted the ILC specifications regarding beam bunch length, 
bunch population, horizontal and vertical beam size etc following~\cite{Behnke:2013xla}. 

 \begin{table}[htb!]
 	\centering
 	\begin{tabular}{|c|c|>{\centering\arraybackslash}p{0.1\textwidth}|>{\centering\arraybackslash}p{0.1\textwidth}|>{\centering\arraybackslash}p{0.1\textwidth}|>{\centering\arraybackslash}p{0.1\textwidth}|>{\centering\arraybackslash}p{0.1\textwidth}|>{\centering\arraybackslash}p{0.1\textwidth}|}
 		\hline
 		\multicolumn{2}{|c|}{Beam}&
 		\multicolumn{6}{|c|}{Production cross-section ($\sigma_{e^+e^-\rightarrow\chi \bar \chi}$ ) (pb)}\\
 		\cline{3-8}
 		\multicolumn{2}{|c|}{polarization}&
 		\multicolumn{3}{|c|}{Majorana DM ($\sqrt{s}=$ 250 GeV)}&
 		\multicolumn{3}{|c|}{Dirac DM ($\sqrt{s}=$ 1 TeV)}\\
 		\hline 
 		\multicolumn{1}{|c|}{$P_{e^-}$} & \multicolumn{1}{|c|}{$P_{e^+}$}  & \multicolumn{1}{|c|}{BP1} & \multicolumn{1}{|c|}{BP2} &
 		\multicolumn{1}{|c|}{BP3} & \multicolumn{1}{|c|}{BP4} &
 		\multicolumn{1}{|c|}{BP5} &
 		\multicolumn{1}{|c|}{BP6} \\
 		\hline
 		-0.8 & +0.3 & 0.40 & 0.02 & 0.78 & 0.05 & 0.03 & 0.42 \\
 		+0.8 &  -0.3 & 6.73 & 0.34 & 0.05 & 0.81 & 0.47 & 0.03 \\
 		0.0 & 0.0  &  2.87 & 0.14 & 0.33 & 0.35 & 0.20 & 0.18 \\
 		\hline
 	\end{tabular}
 	\caption{Production cross-section for the benchmark points in table~\ref{tab:BP} for different choices of the beam polarization.}
 	\label{tab:bp-pol}
 \end{table}
\begin{figure}[htb!]
$$
\includegraphics[scale=0.30]{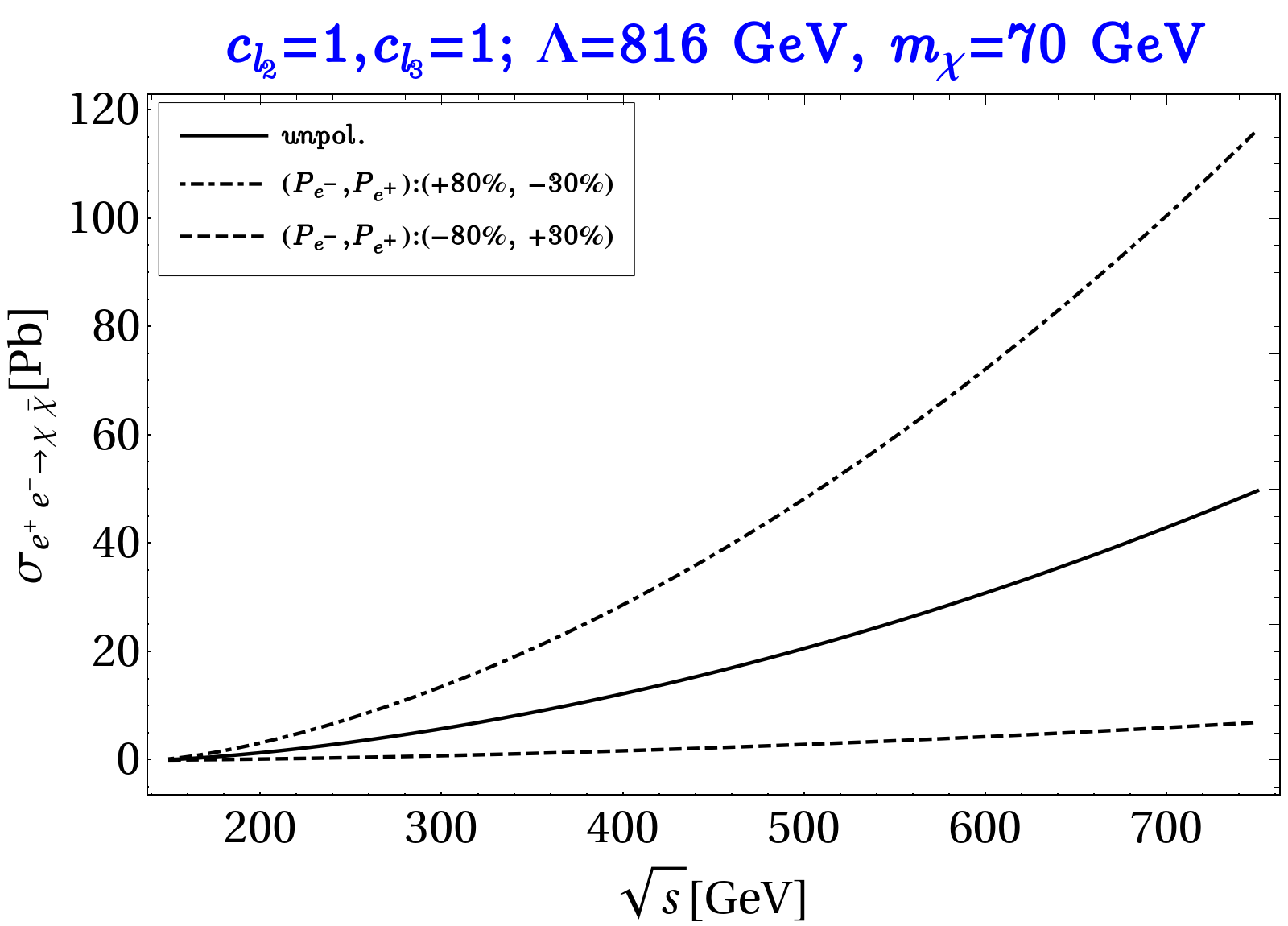}~~~~
\includegraphics[scale=0.29]{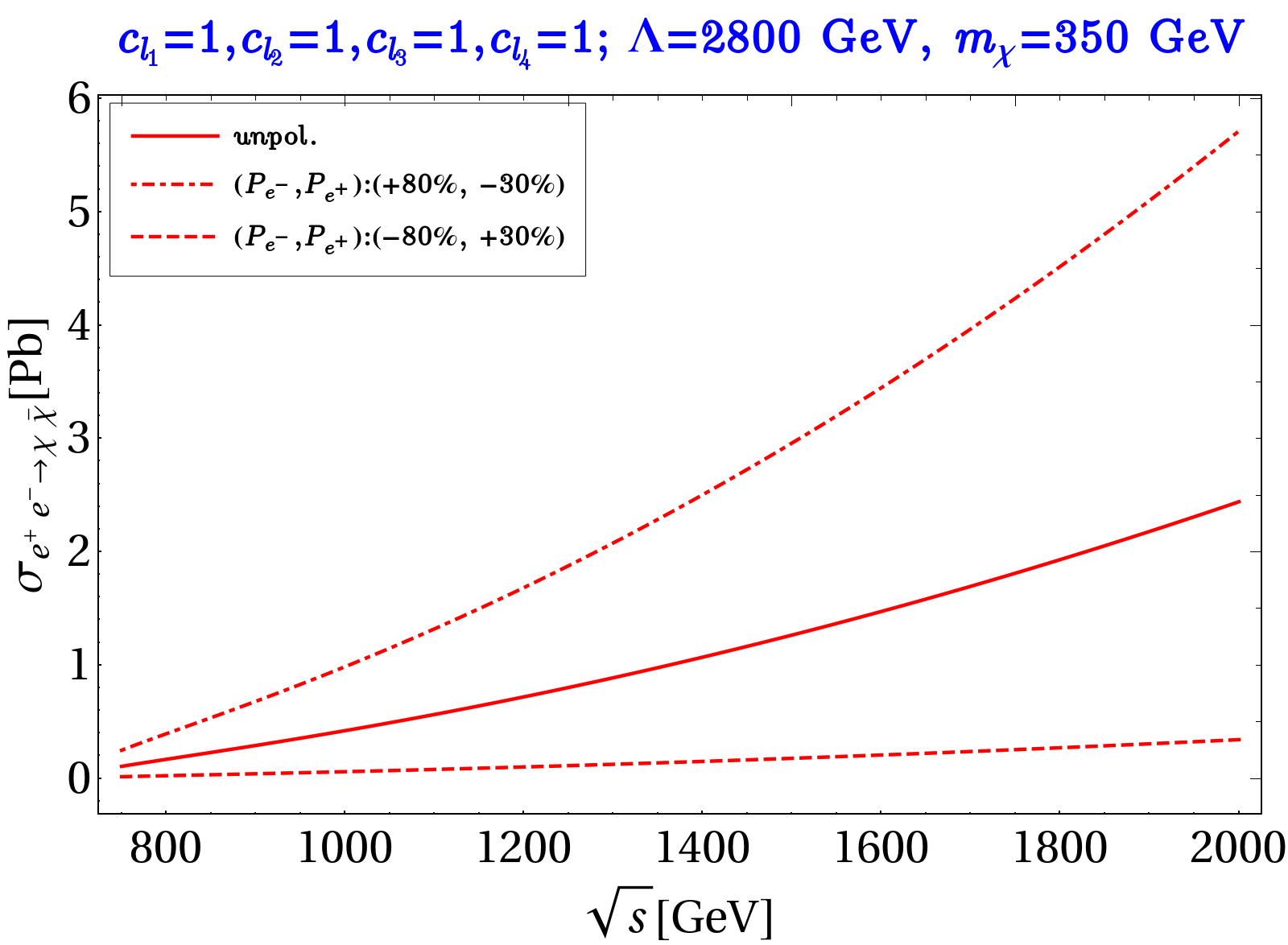}
$$
$$
\includegraphics[scale=0.30]{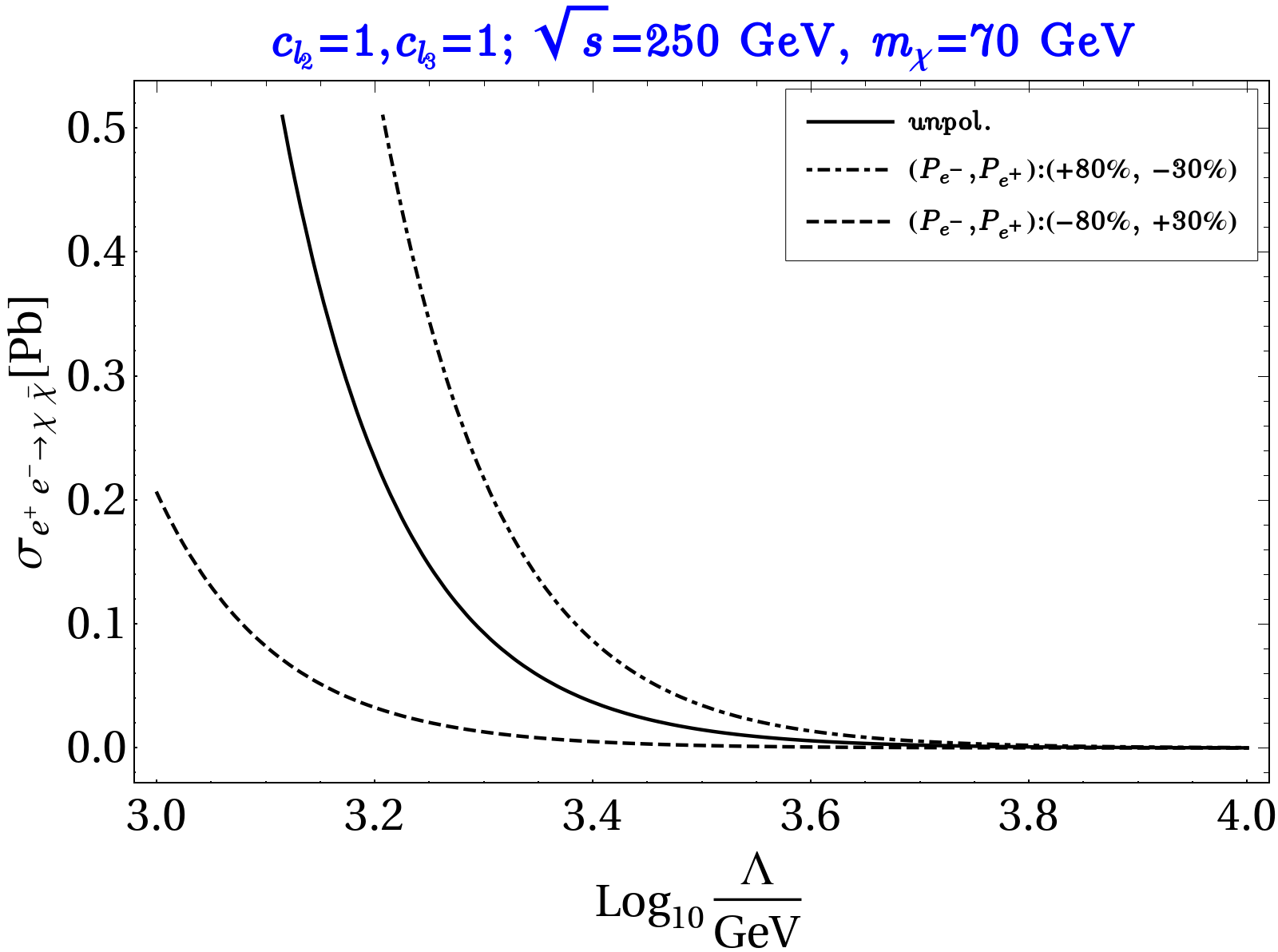}~~~~
\includegraphics[scale=0.29]{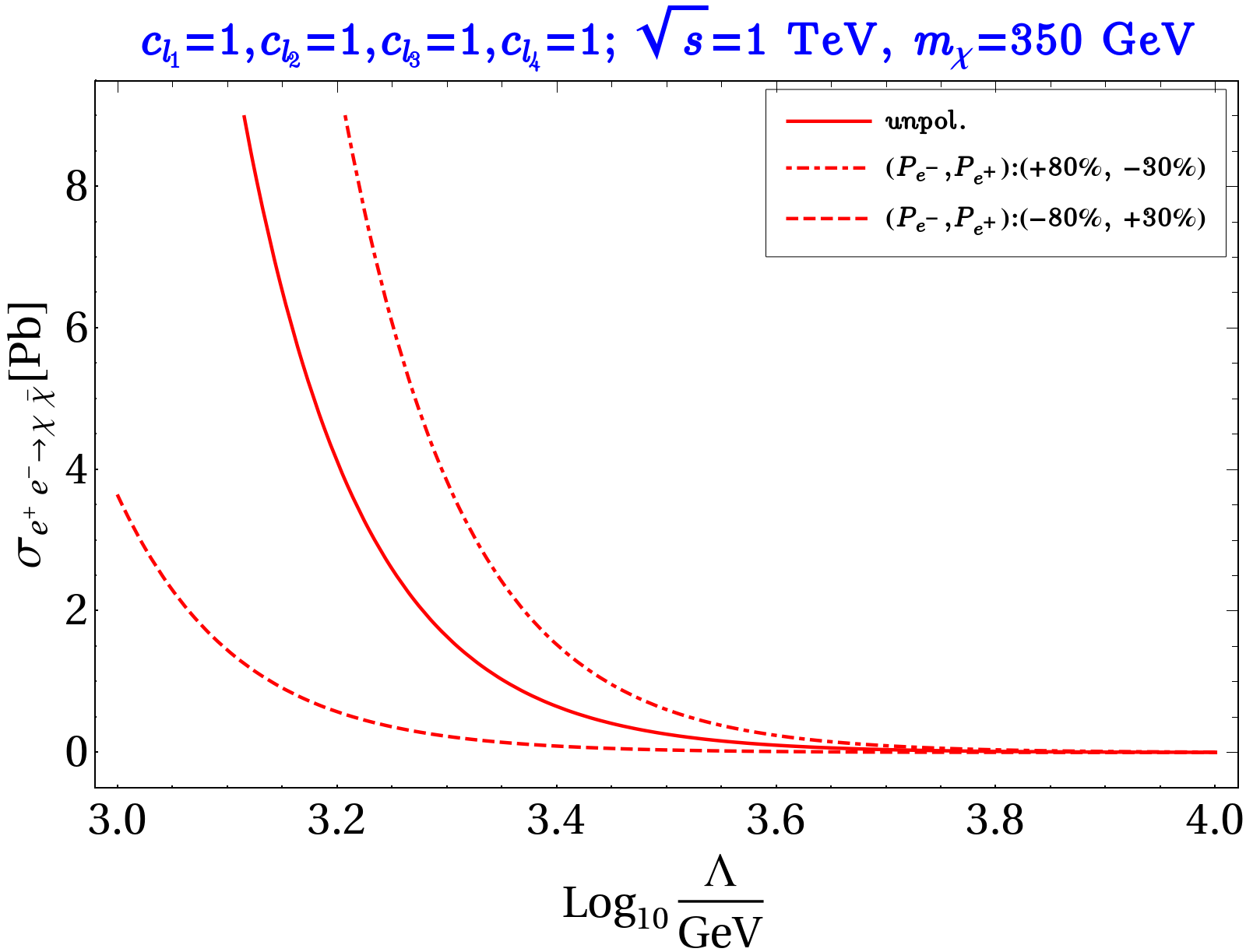}
$$
$$
\includegraphics[scale=0.30]{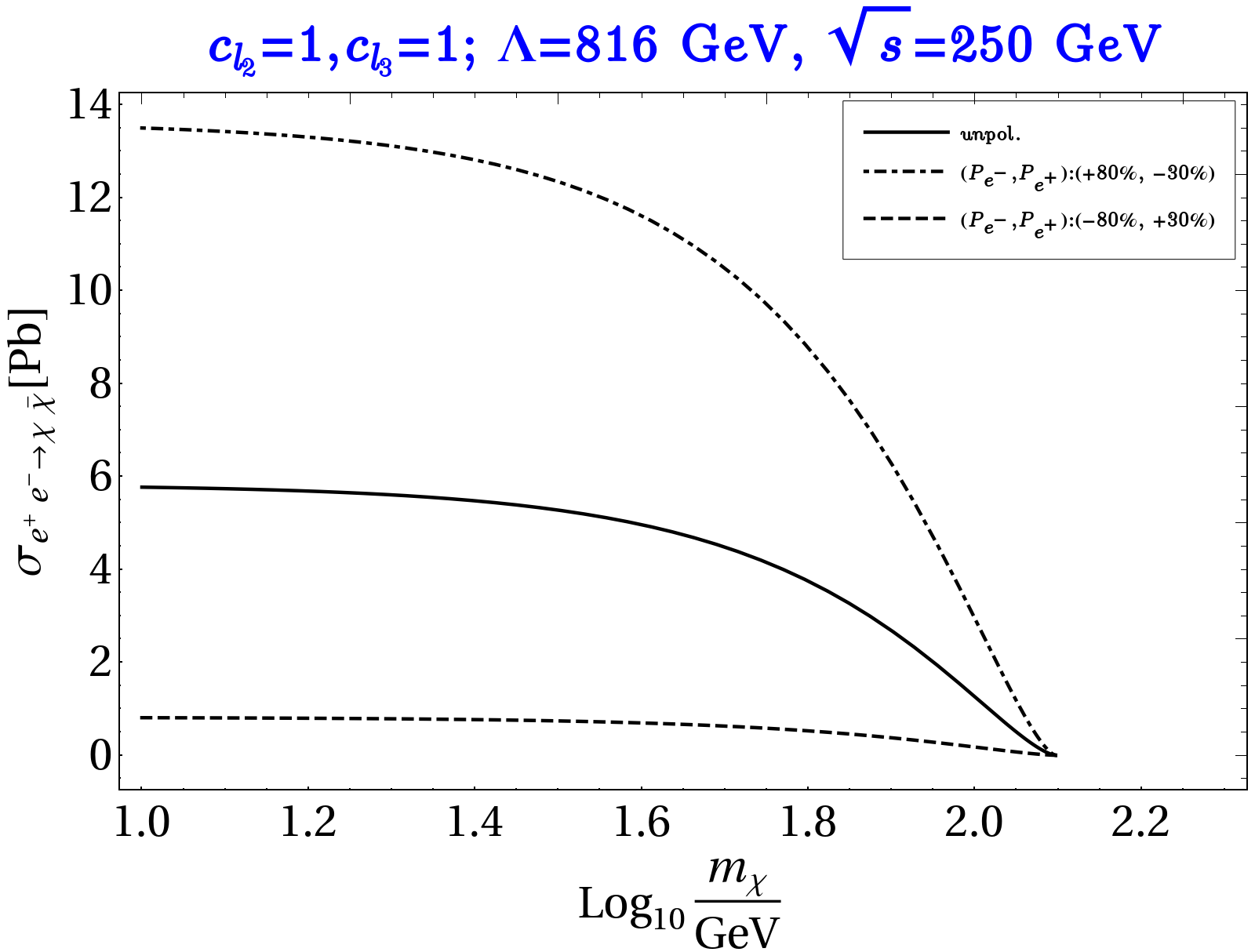}~~~~
\includegraphics[scale=0.30]{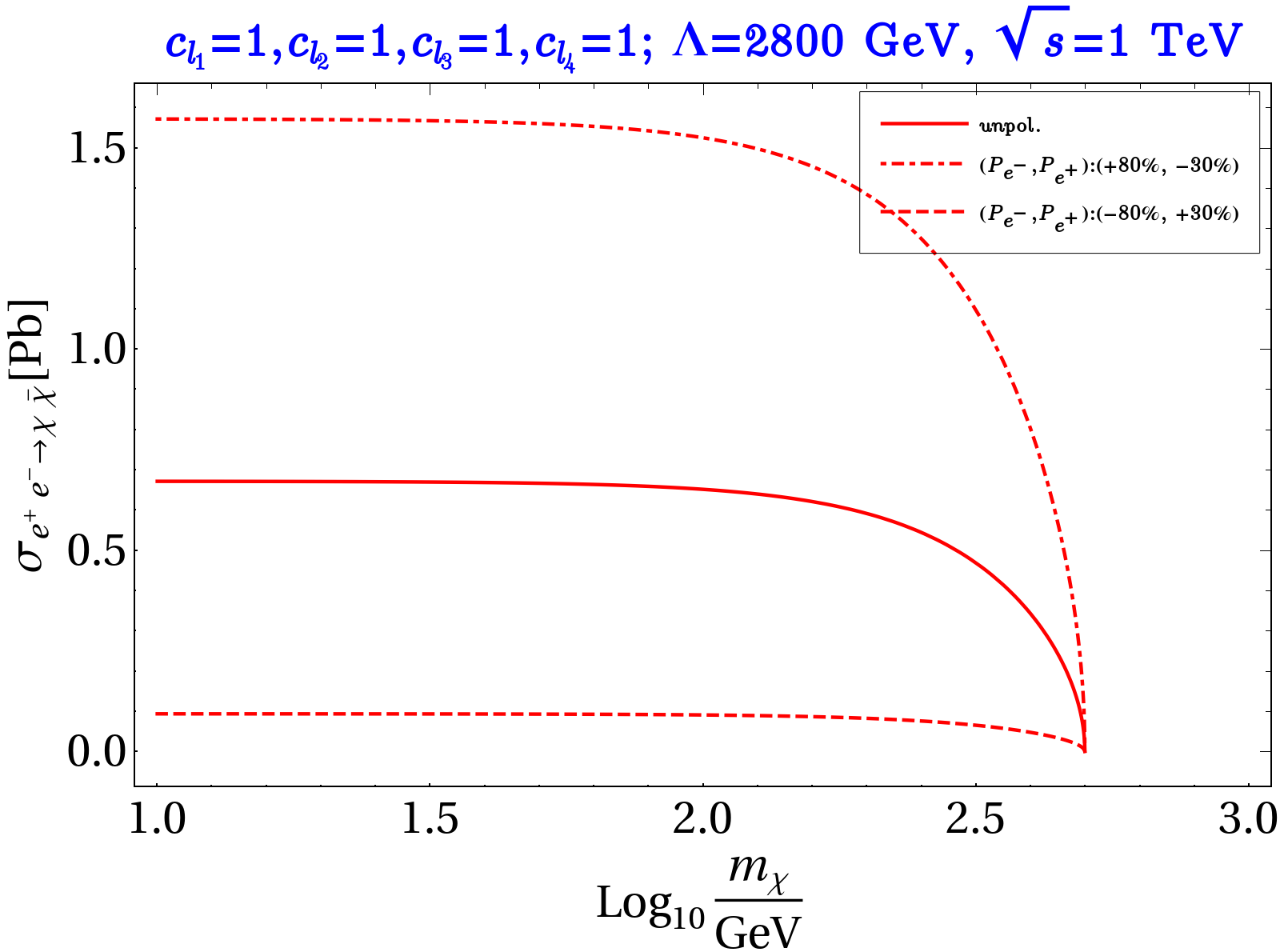}
$$
\caption{$\sigma_{e^+ e^- \to \chi \bar \chi}$ as function of $\sqrt{s}$ (top panel), $\Lambda$ (middle panel) and $\mchi$ (bottom panel) for Majorana DM (BP1) in 
the left column and Dirac DM (BP4) in the right column for three choices of polarisation: $\{P_{e^-}:P_{e^+}\}=\{0,0\};\{+0.8,-0.3\};\{-0.8,+0.3\}$ as mentioned in 
the figure inset. The other parameters kept fixed are mentioned in Figure heading.}
\label{fig:dm-prod}
\end{figure}

We plot the variation of DM production cross-section ($\sigma_{e^+ e^- \to \chi \bar \chi}$) for Majorana DM (BP1) in the left column and Dirac DM 
(BP4) in the right column of Fig.~\ref{fig:dm-prod}. In the top, middle and bottom panel we show the variation with respect to $\sqrt{s}$, $\Lambda$ and $\mchi$ 
respectively each for three different choices of the beam polarization: $\{P_{e^-}:P_{e^+}\}=\{0,0\};\{+0.8,-0.3\};\{-0.8,\\ +0.3\}$ shown by solid, 
dashed-dot and dashed black (red) curves for Majorana (Dirac) DM cases with all accessible operators put together. The other parameters kept fixed 
are mentioned in the Figure headings. We find the signal cross-section to be maximum for the $\{+80\%,-30\%\}$ case as elaborated above, 
with all Wilson coefficients chosen positive. The shape of the plot in the left panel (for Majorana DM) and in the right panel (for Dirac DM) can be 
verified from the Eq.~(\ref{eq:majorana}) and Eq.~(\ref{eq:dirac}) respectively where we find that total cross-section to increase with the CM energy ($\sqrt{s}$) 
upto effective limit $\Lambda>\sqrt{s}$ (top panel). In the middle panel, the cross-section falls with larger $\Lambda$ as $\sigma_{e\bar e \to \chi\bar \chi}\sim 1/\Lambda^4$ 
for a fixed DM mass. The cross-section slowly falls with $\mchi$ and vanishes for $\sqrt{s}\lesssim 2m_\chi$, as expected from the phase-space 
dependence shown in the bottom panel of Fig.~\ref{fig:dm-prod}. 

\begin{figure}[htb!]
$$
\includegraphics[scale=0.18]{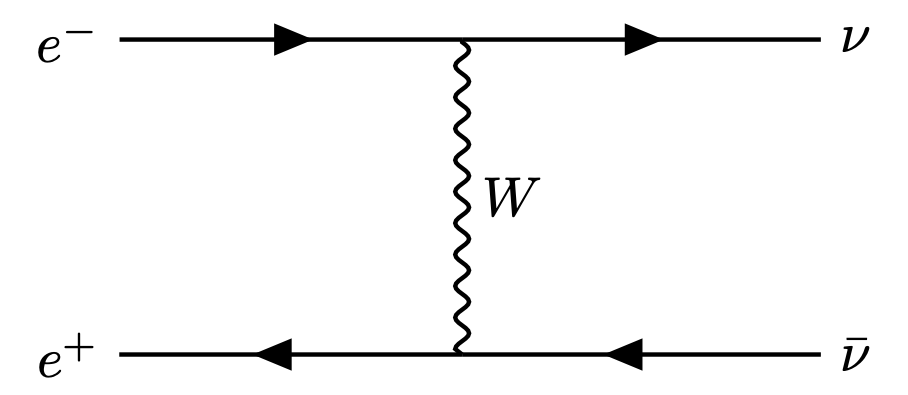}~~
\includegraphics[scale=0.18]{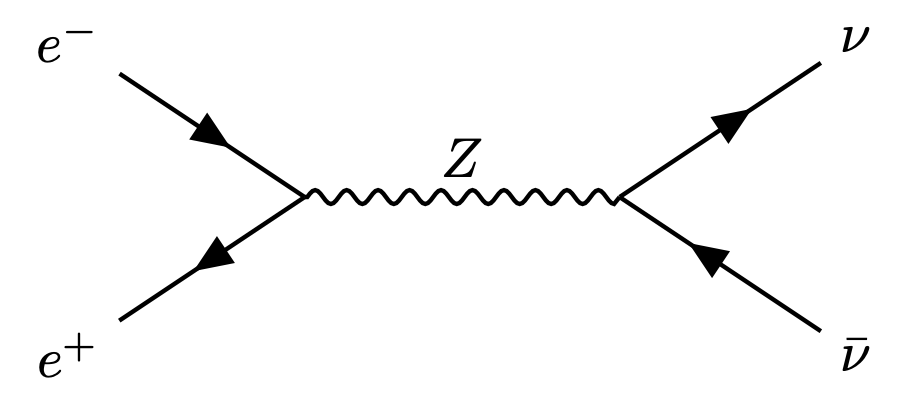}
$$
$$
\includegraphics[scale=0.18]{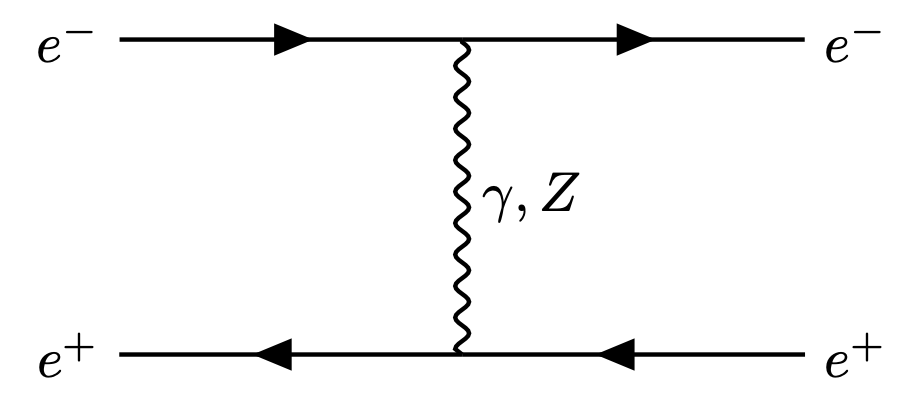}~~
\includegraphics[scale=0.18]{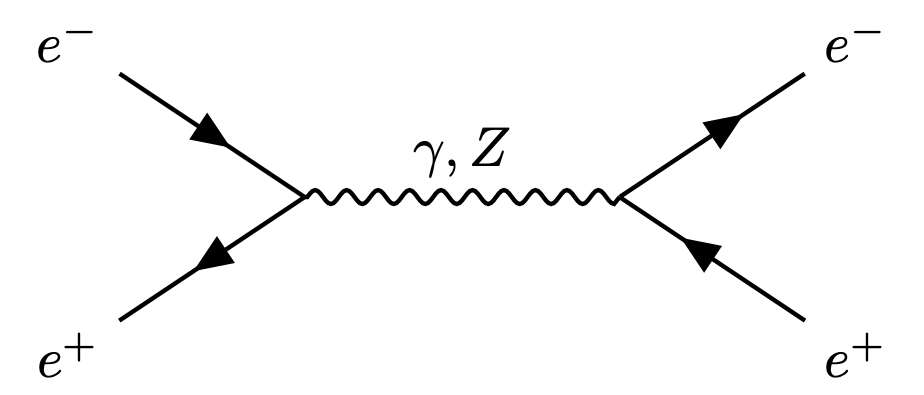}
$$
\caption{Top: SM background due to $e^+e^-\to\nu\bar{\nu}$ process. Bottom: Bhabha scattering contributing to the background for mono-photon with missing energy process where both $e^+,e^-$ in the final state are undetected in the collider. In all cases radiative photons may arise either from the initial or final state radiation or both.}
\label{fig:feyn-bckg}
\end{figure}

\begin{table}[htb!]
	\centering
	\begin{tabular}{|l|c|c|c|c|c|c|c|c|c|r|}
		\hline
		$\sqrt{s}$&$P_{e^-}$ & $P_{e^+}$ & $\sigma_{ee\to\nu\bar{\nu}}$ &$\sigma_{ee\to ee}$ \\
		(GeV)& &  & (pb)  & (pb) \\
		\hline
		&-0.8 & +0.3 & 102.0 & 618.8 \\
		250 &+0.8 & -0.3 & 14.72 &  628.3 \\
		&0.0 & 0.0 & 52.71   & 610.3 \\
		\hline
		&-0.8 & +0.3 & 130.93 & 622.06 \\
		1000&+0.8 & -0.3 & 7.88 & 603.33  \\
		&0.0 & 0.0 &  56.03   & 596.29 \\
		\hline
	\end{tabular}
	\caption{Production cross-section for dominant SM backgrounds at $\sqrt{s}=$ 250 and 1000 GeV.}
	\label{tab:bckg-pol}
\end{table}

Since we are typically interested in mono-photon signal with missing energy at the $e^+e^-$ collider, there are two main SM backgrounds which can mimic 
such a signal final state 

\begin{itemize}
\item $e^+e^-\to\n\bar{\nu}\gamma$ is an irreducible background~\cite{Liu:2019ogn} since neutrinos will also be missed inside the detector like the DM. Such a process is contributed by two Feynman diagrams: one involves the $t$-channel $W$ boson exchange and the other involves the $s$-channel $Z$ boson exchange as shown in the top panel of Fig.~\ref{fig:feyn-bckg}. 

\item Another possible background is the radiative Bhabha scattering $e^+e^-\to e^+e^-\gamma$ as shown in the bottom panel of Fig.~\ref{fig:feyn-bckg}, 
where both the final state leptons go undetected. Note here, that $e^+e^-\to f\bar{f} \gamma$ where $f$ is any SM fermion, lepton or jet forms a subset for this class 
of backgrounds, where $f$ has to be missed in the detector. 
\end{itemize}

We tabulate the production cross-section dominant SM background in table~\ref{tab:bckg-pol} for different choices 
of the beam polarizations which can be realized in ILC set-up. The SM background contribution varies quite differently with polarization. 
This can be understood easily. For $e^+e^-\to\nu\bar{\nu}$, the Feynman diagrams involving $t$-channel $W$ boson exchange only contributes 
to a non-zero $\sigma_\text{LR}$ since only left-handed fermions and right-handed anti-fermions take part in the weak charge current interaction. 
This indicates that the asymmetry parameter $\mathcal{A}_\text{LR}=1$. For the $s$-channel diagram for $e^+e^-\to\nu\bar{\nu}$, exchange of a vector 
boson ($Z$-boson) ensures non-zero $\sigma_\text{LR,RL}$ due to spin conservation. Since the coupling of $Z$-boson to $e_L^-e_R^+$ is stronger than $e_R^-e_L^+$, 
the diagrams involving $s$-channel $Z$-exchange has more contributions to $\sigma_\text{LR}$ than to $\sigma_\text{RL}$. 
Therefore, for $\sigma_{e^+e^-\to\nu\bar{\nu}}$ diminishes for polarization configuration $\{P_{e^-}:P_{e^+}\}=\{+0.8,-0.3\}$, 
contrary to the signal for operators with all positive Wilson coefficients, which provides the best way to probe leptophilic DM at ILC. 
Bhaba scattering ($e^+e^-\to e^+e^-\gamma$) cross-section although still looks quite 
substantial, but we will be able to tame it down significantly, which we discuss in context of the event selection criteria.

\subsection{Event distribution and analysis}

We generate the parton-level signal events for the benchmark points in Table~\ref{tab:BP} using the batch mode of {\tt CalcHEP-3.8.10}~\cite{Belyaev:2012qa}. 
The events are then fed to {\tt Pythia-6.4.28}~\cite{Sjostrand:2006za} for showering utilising in-built switches for initial and final state radiation (ISR/FSR). 
The SM background events are generated using {\tt MadGraph}~\cite{Alwall:2014hca} and then the event files are again analysed through {\tt Pythia}. In collider environment, we reconstruct the following objects and define a few observables as

\begin{itemize}
\item {\it Lepton ($l=e,\mu$):} Leptons are required to have a minimum transverse momentum $p_T>20$ GeV and pseudorapidity $|\eta|<2.5$. Two leptons are isolated objects if their mutual distance in the $\eta-\phi$ plane is $\Delta R=\sqrt{\left(\Delta\eta\right)^2+\left(\Delta\phi\right)^2}\ge 0.2$, while the separation between a lepton and a jet has to satisfy $\Delta R\ge 0.4$.
 
\item {\it Jets ($j$):} All the partons within $\Delta R=0.4$ from the jet initiator cell are included to form the jets using the cone jet algorithm {\tt PYCELL} built in {\tt Pythia}. We require $p_T>10$ GeV for a clustered object to be considered as jet. Jets are isolated from unclustered objects if $\Delta R>0.4$.  

\item {\it Photons ($\gamma$):} Photons are identified to register in the detector with minimum energy $E_\gamma>0.1$ GeV. 
 
\item {\it Unclustered Objects:}  All the final state objects which are neither clustered to form jets, nor identified as leptons, belong to this category. All particles with $0.5<p_T<20$ GeV and $|\eta|<5$, are considered as unclustered.
 
\item {\it Missing Transverse Energy or MET ($\slashed{E}_T$):} The transverse momentum of all the missing particles (those are not registered in the detector) 
can be estimated from the momentum imbalance in the transverse direction associated to the visible particles. Thus MET is defined as:
\bea
\slashed{E}_T = -\sqrt{(\sum_{\ell,j} p_x)^2+(\sum_{\ell,j} p_y)^2},
\eea
where the sum runs over all visible objects that include the leptons, jets and the unclustered components. 

\item{\it Missing Energy or ME ($\slashed{E}$)}: The energy which is carried away by the missing particles can be identified at lepton collider 
given the knowledge of CM energy of the reaction as 
\bea
\slashed{E}=\sqrt{s}-\sum_{\ell,j,\gamma} E; 
\eea

\item {\it Missing Mass ($m_{\rm miss}$)}: In a leptonic collider, owing to its clean kinematics, one can introduce the Lorentz invariant 
``missing mass'' of the system~\cite{Han:2020uak}. For the process $e^+ e^- \to \chi \bar\chi \gamma$, the missing mass is inferred to be
\bea
m_\text{miss}^2=(p_{e^+}+p_{e^-}-p_\gamma)^2 \,.
\eea
\noindent In the CM frame, $m_{\rm miss}^2=s-2\sqrt{s} E_\gamma$. Here, $p_{e^+}$ and $p_{e^-}$ are the four-momenta of incoming particle 
beams, and $p_\gamma$, $E_\gamma$ are the four-momentum and energy of the outgoing photon respectively.
\end{itemize}

We eventually put zero lepton and jet veto on the final state events of our interest. 
Following detector cuts are further used on the photons identified in {\tt Pythia}: 

\begin{itemize}

\item In order to ensure most of the events are localized around the central region of the detector we choose the pseduorapidity $\left|\eta_\gamma\right|\leq0.24$.

\item We also ensure the final state events contain at least one ``hard'' photon by choosing a cut on the transverse momentum of the photons: $p_T^\gamma>5$ GeV.

\end{itemize}

\begin{figure}[htb!]
$$
\includegraphics[scale=0.36]{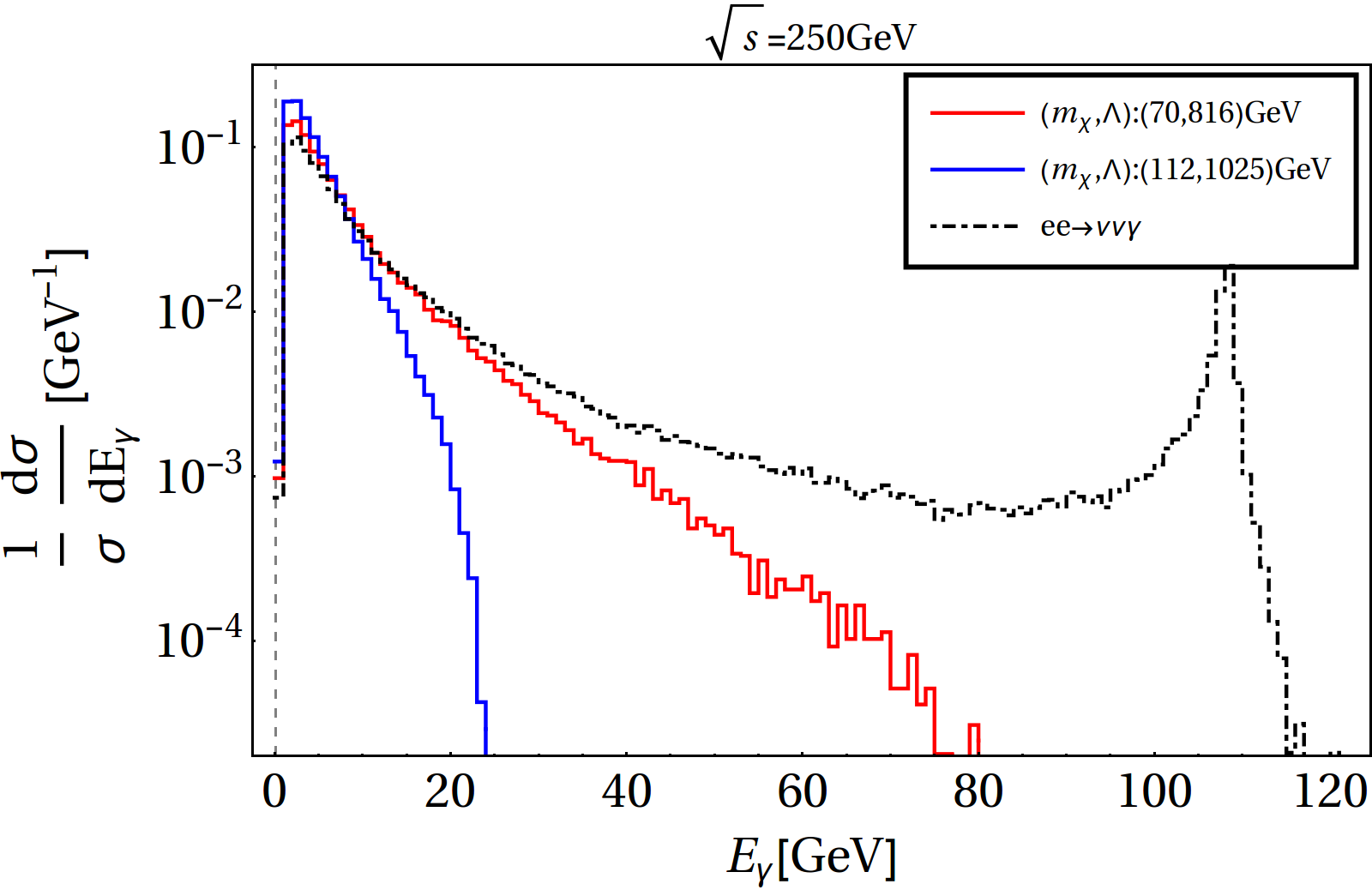}~~~~
\includegraphics[scale=0.36]{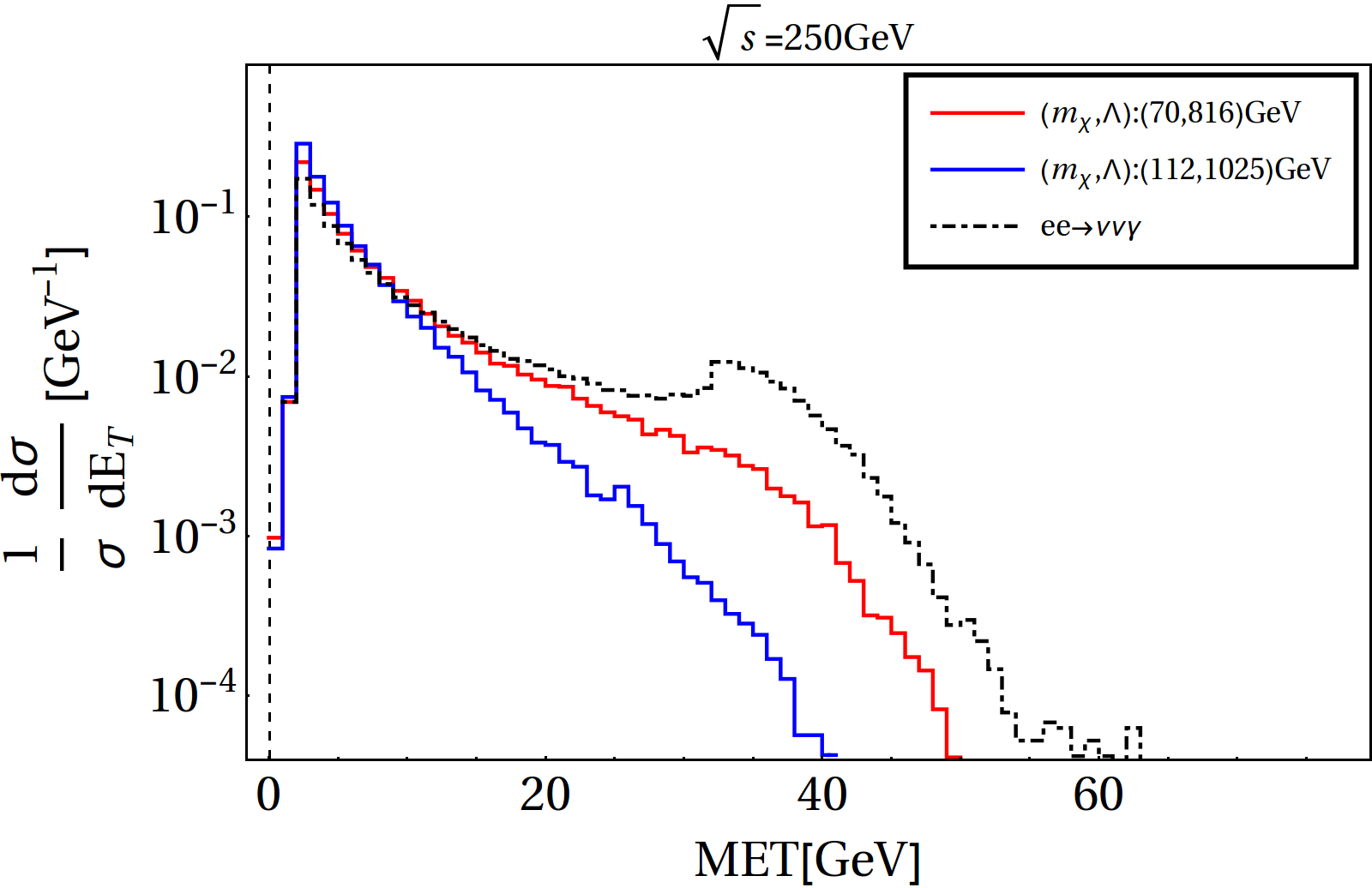}
$$
$$
\includegraphics[scale=0.36]{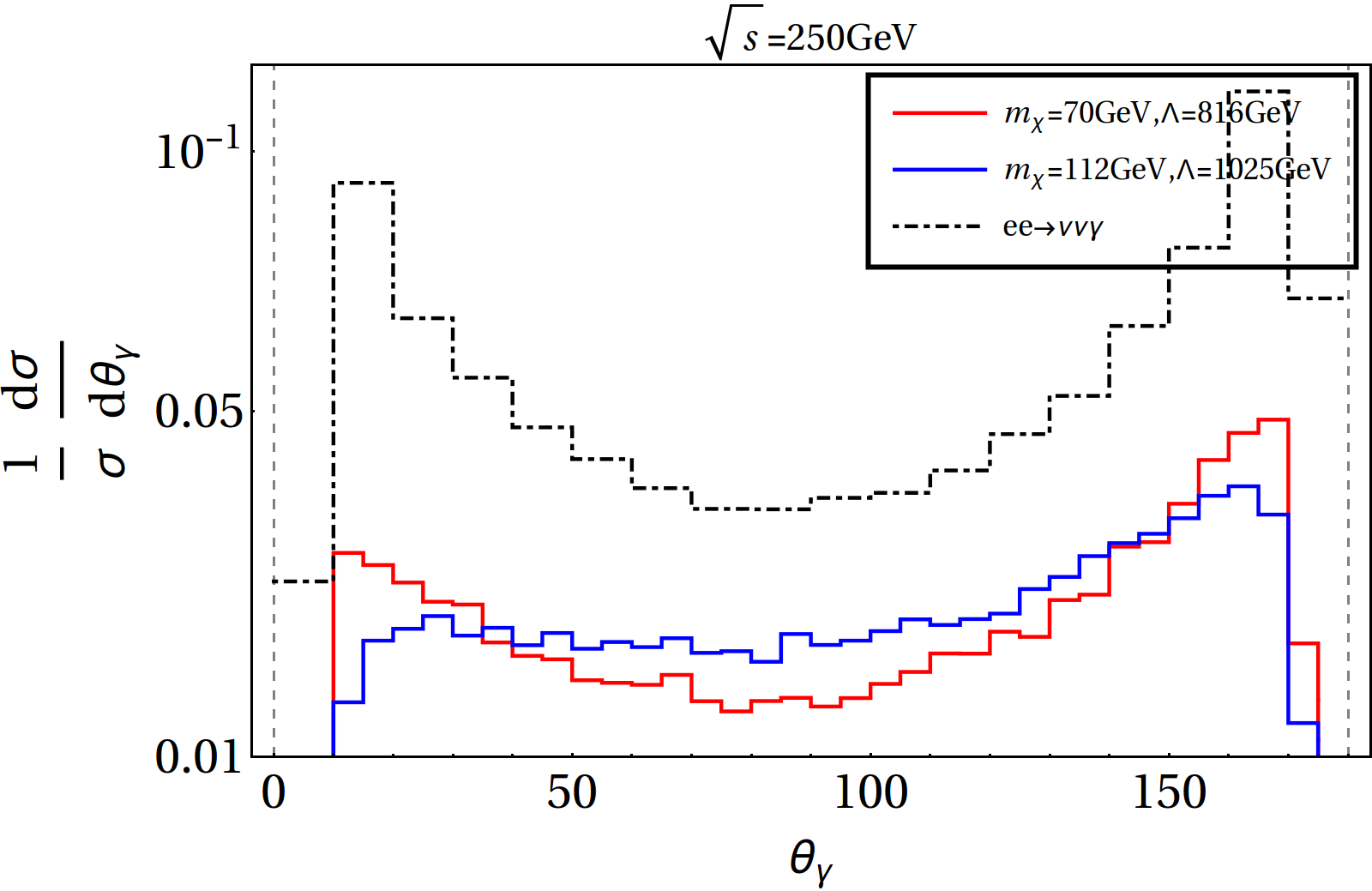}~~~~
\includegraphics[scale=0.36]{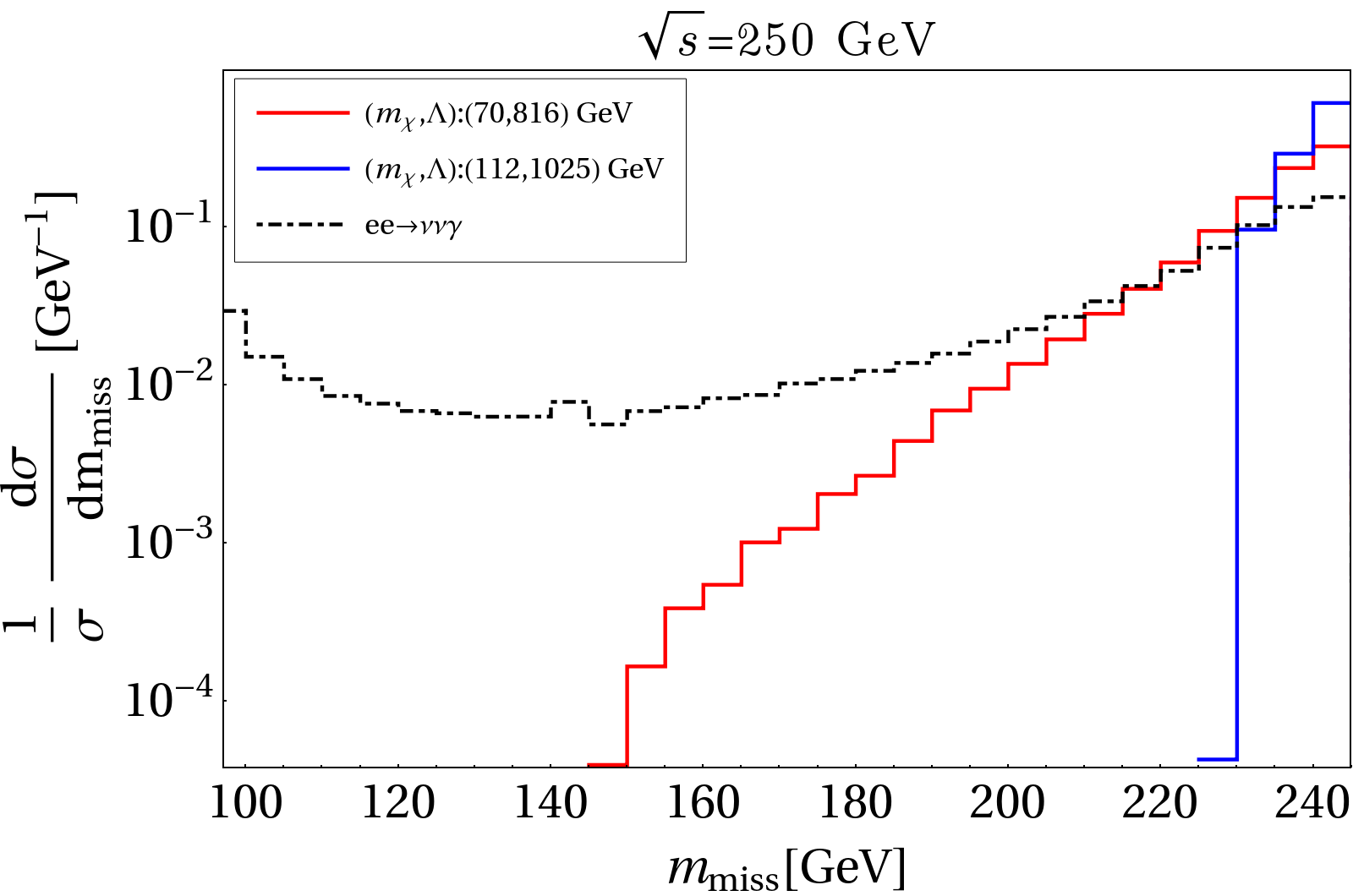}
$$
\caption{Top left: Normalized {\it inclusive} mono-photon event distribution for the BP1 and BP2 in table~\ref{tab:BP} and the SM backgrounds  as a 
function of the photon energy $E_\gamma$.  Top right: Normalized distribution of events for the signal and the SM backgrounds with missing energy 
(MET). Bottom left: Normalized signal and background distribution with $\theta_\gamma$. Bottom right: Normalized signal and SM background 
distribution with respect to $m_\text{miss}$. In all cases the signals are shown in red and blue curves, while the SM background is shown by the 
dashed black curve for $e^+e^-\to\nu\bar{\nu}$. We simulate the events at $\sqrt{s}=250$ GeV with $\{P_{e^-}:P_{e^+}\}=\{+0.8:-0.3\}$ beam polarization.}
\label{fig:distr}
\end{figure}

\begin{figure}[htb!]
	$$
	\includegraphics[scale=0.36]{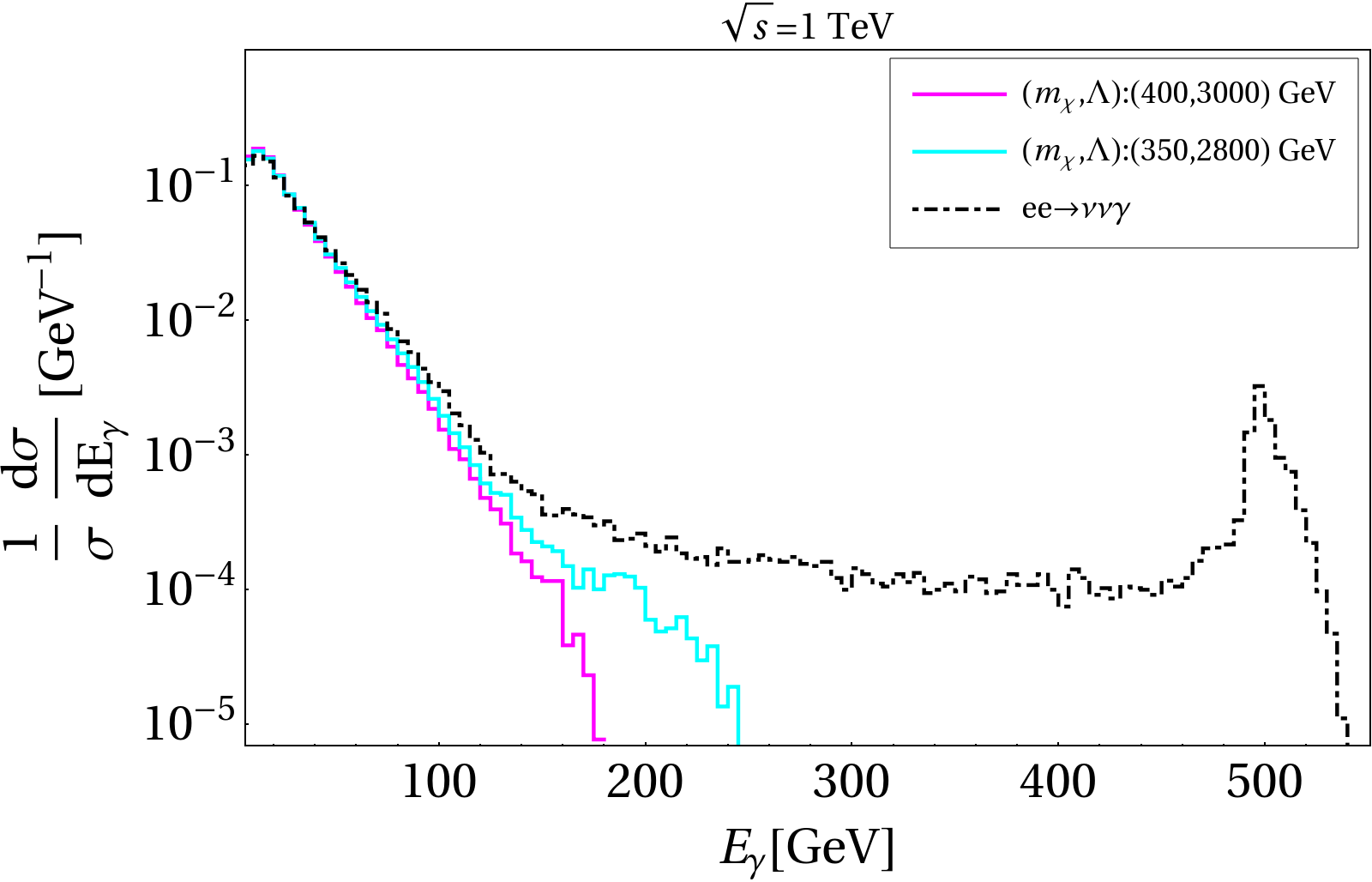}~~~~
	\includegraphics[scale=0.36]{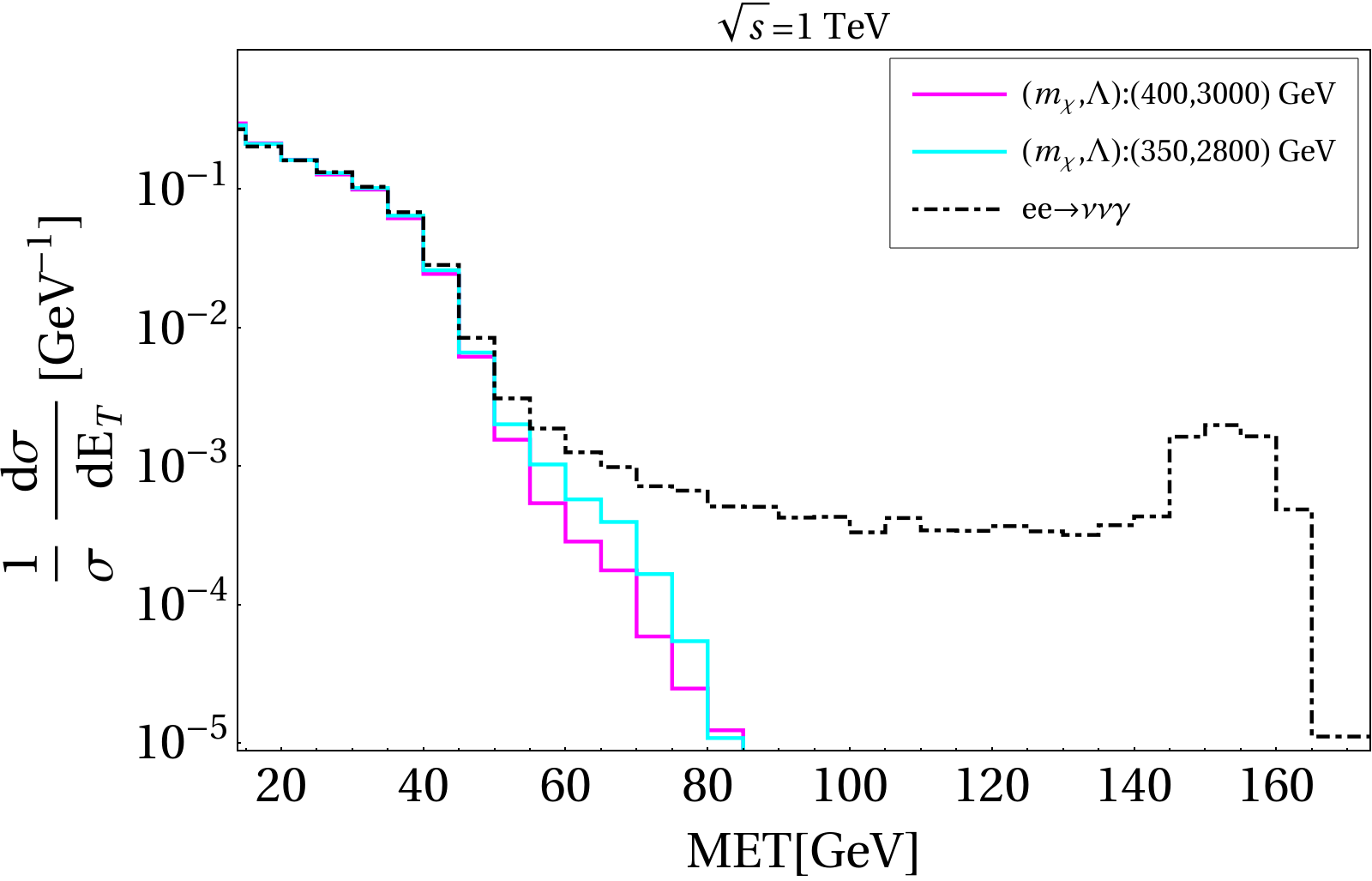}
	$$
	$$
	\includegraphics[scale=0.36]{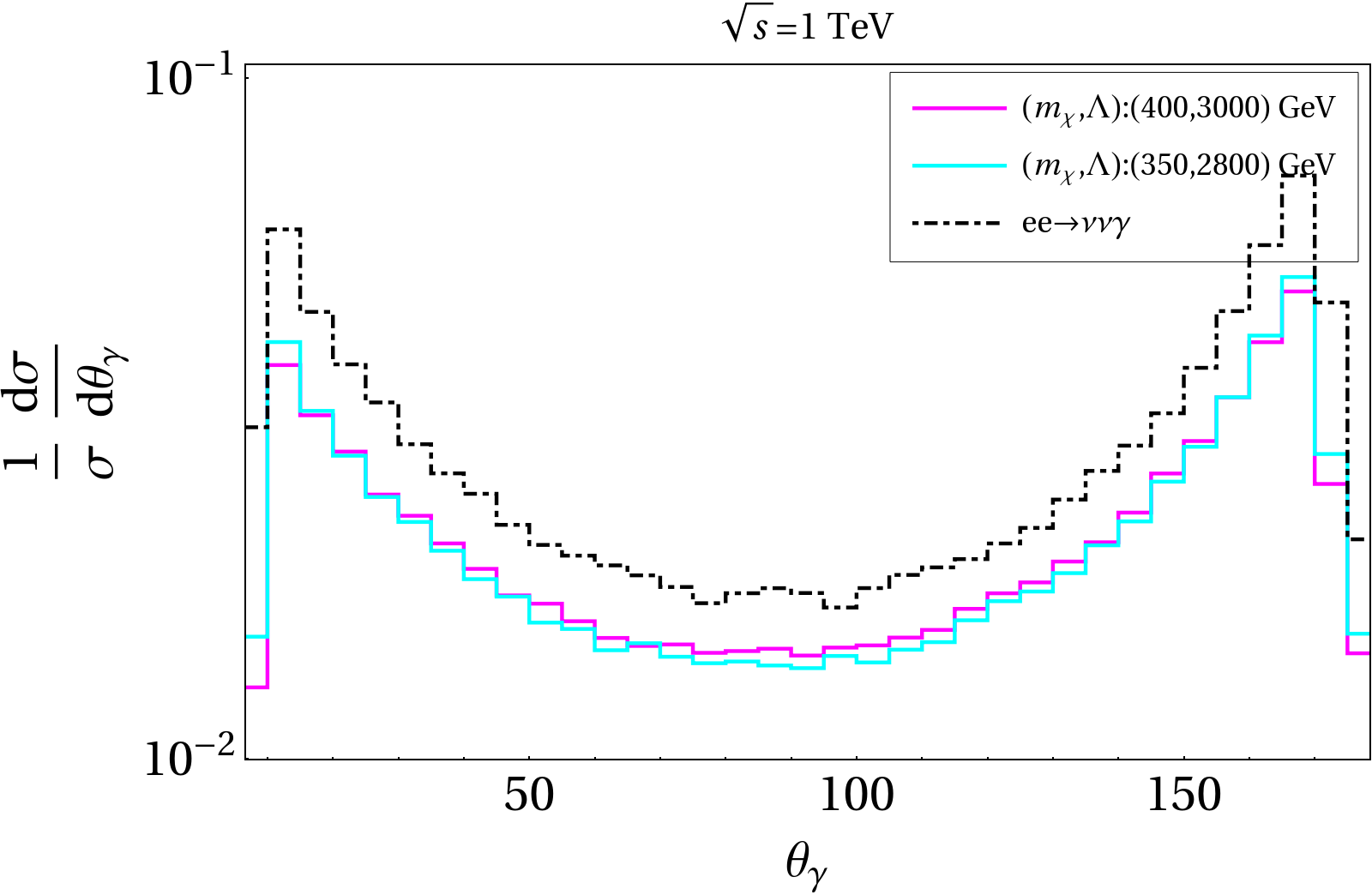}~~~~
	\includegraphics[scale=0.36]{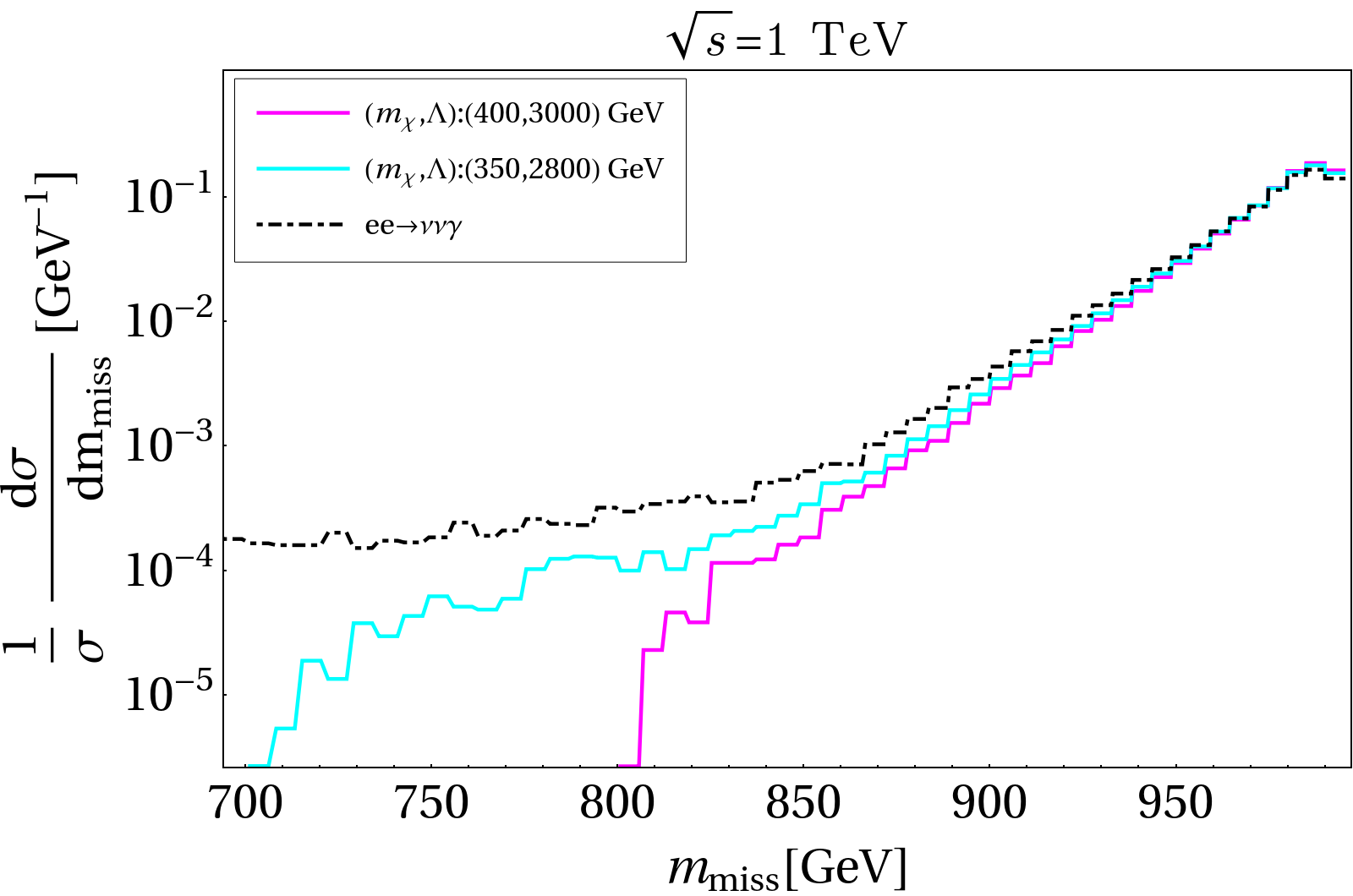}
	$$
	\caption{Same as Fig.~\ref{fig:distr} with BP4 and BP5, but with $\sqrt{s}=1$ TeV.}
	\label{fig:distr-dirac}
\end{figure}

Due to collinear singularity, the radiative Bhabha scattering process has large cross section when both final state electron and positron go along the beam 
directions. However, as shown in~\cite{Liu:2019ogn}, such backgrounds can be efficiently removed by adopting $E_\gamma>E_\text{B}^m(\theta_\gamma)$ 
on the final state mono photon with 

\bea
E_\text{B}^m(\theta_\gamma)=\sqrt{s}\Biggl[1+\frac{\sin\theta_\gamma}{\sin\theta_\text{bnd}}\Biggr]^{-1}
\label{eq:ebm-gamma}
\eea

\noindent where $|\cos\theta_\text{bnd}|=0.99$ corresponds to the boundary of the electromagnetic calorimeter (EMC). Thus, here onwards, we will omit 
SM background due to Bhabha scattering process and only consider the irreducible neutrino background. We would like to further mention that this particular 
choice of cut on the final state mono-photon energy (Eq.~\eqref{eq:ebm-gamma}) also efficiently eliminates reducible backgrounds due to processes like 
$e^+\,e^-\to f\,\bar{f}\,\gamma$ and $e^+\,e^-\to\gamma\,\gamma\,\gamma$ as shown in~\cite{Liu:2019ogn}. 

The distribution of normalized number of {\it inclusive} mono-photon events corresponding to the signal for Majorana DM 
as well as to the backgrounds, as a function of different observables, are illustrated in Fig.~\ref{fig:distr}. As seen from the top left panel, the mono-photon energy $E_\gamma$ spectrum does not exhibit any peak structure, whereas $e^+e^-\to\nu\bar{\nu} \gamma$ shows a peak around $E_\gamma = \frac{s-m_Z^2}{2\sqrt{s}}$. It is easy to show from the 2-body kinematics, the maximum photon energy due to the signal is given by~\cite{Liu:2019ogn}

\begin{equation}
E_\gamma<E_\chi^\text{max}=\frac{1}{2\sqrt{s}}\left(s-4m_\chi^2\right)\,,
\end{equation}

\noindent which exactly determines the end-point of the signal distribution. As a result, we see, for a heavier DM the radiated photons are less energetic. Note that the 
distribution for $E_\gamma$ is identical to ME ($\slashed{E}$) distribution given the presence of only one photon in the event. 
The normalized event distribution for MET (top right) are also similar to photon transverse momenta $p_T^\gamma$ for the same reason. 
Also a more massive DM results in a less-energetic radiated photon since most of the energy is carried by the DM itself. In the present scenario, 
since the photons come from the ISR, they tend to be soft and collinear. This makes their $\theta_\gamma$ distribution rather flat as shown in 
the bottom left panel. $m_\text{miss}$ distribution in the bottom right panel shows a clear distinction for DM signal from that of 
$\nu\bar{\nu}$ background, allowing to segregate them with a lower $m_\text{miss}$ cut, better for heavier DM mass.
A similar set of distributions is shown for Dirac DM in Fig.~\ref{fig:distr-dirac}, with a higher CM energy and the inferences broadly remain the same. 
In passing we note that the distributions as shown in Figs.~\ref{fig:distr}, \ref{fig:distr-dirac} primarily depend on the kinematics, therefore conclusions for 
other benchmark points as in Table~\ref{tab:BP} can be easily gauged.

\subsection{Cut-flow and signal significance}

Only for a small fraction of neutrino pair production events the radiated photon can be measured in the detector~\cite{Kalinowski:2020lhp}. 
Therefore, as mentioned earlier, we demand the final event selection to have photons with transverse momentum $p_T^\gamma>p_T^\text{min}=5$ GeV. 
On top of this we apply the following cuts in order to reduce the SM background as efficiently as possible, without harming the signal events. 

\begin{itemize}
 \item Cut1 ($\mathcal{C}_1$) : Events with zero lepton and jet-veto with exactly one photon in the final state. 
 
\item Cut2 ($\mathcal{C}_2$): We choose photons with energy lying within the window $0.1\leq E_\gamma< $ 60  (130) GeV or Majorana (Dirac) DM scenario. This helps to avoid the background events around the $Z$-mass window by retaining majority of the signal events.


\item Cut3 ($\mathcal{C}_3$): We apply a cut on the missing transverse energy $\slashed{E_T}\leq $ 33 (43) GeV for for Majorana (Dirac) DM scenario.

\item Cut4 ($\mathcal{C}_4$): Finally, we employ the missing mass cut $m_\text{miss} \geq$ 140 (220) GeV for BP1 (BP2), Majorana DM 
scenario and $m_\text{miss} \geq$ 700 (800) GeV for BP4 (BP5), Dirac DM scenario.
\end{itemize}
 \begin{table}[htb!]
 	\centering
 	\begin{tabular}{|l|c|c|c|c|c|c|c|c|c|r|}
 		\hline
 		Cuts &BP1 & BP2 & $\nu\bar{\nu}\gamma$ & $\epsilon^{\tt BP1}$ & $\epsilon^{\tt BP2}$ & $\sigma^{\tt BP1}=\frac{S}{\sqrt{S+B}}$ & $\sigma^{\tt BP2}=\frac{S}{\sqrt{S+B}}$ \\
 		& (fb) & (fb) & (fb) & & &$(\mathcal{L}=$ 100 $\text{fb}^{-1})$ & $(\mathcal{L}=$ 100 $\text{fb}^{-1})$ \\
 		\hline\hline
 		$\mathcal{C}_1$ & 72.53 & 1.32 & 303.62& 0.011 &0.004 & 37.40 & 0.75 \\
 		$\mathcal{C}_2$ & 71.99 & 1.31 & 208.91& 0.011 &0.004 & 42.95 & 0.90 \\
 		$\mathcal{C}_3$ & 67.16 & 1.23 & 176.64& 0.010 &0.004 & 43.01 & 0.96 \\
 		$\mathcal{C}_4$ & 67.16 & 1.23 & 133.73& 0.010 &0.004 & 47.38 & 1.14 \\
 		\hline
 	\end{tabular}
 	\caption{Event cross-section (fb), cut-efficiency ($\epsilon$) and signal significance $\sigma=\frac{S}{\sqrt{S+B}}$ for the benchmark points (BP1 and 
	BP2) corresponding to Majorana DM along with the irreducible SM background for CM energy ($\sqrt s$) = 250 GeV with 
	$\{P_{e^-}:P_{e^+}\}=\{+0.8:-0.3\}$ beam polarization.}
 	\label{tab:cut-flow}
 \end{table}
 \begin{table}[htb!]
 	\centering
 	\begin{tabular}{|l|c|c|c|c|c|c|c|c|c|c|r|}
 		\hline
 		Cuts &BP4 & BP5 & $\nu\bar{\nu}\gamma$ & $\epsilon^{\tt BP4}$ & $\epsilon^{\tt BP5}$ & $\sigma^{\tt BP4}=\frac{S}{\sqrt{S+B}}$ & $\sigma^{\tt BP5}=\frac{S}{\sqrt{S+B}}$\\
 		& (fb) & (fb) & (fb) & & &$(\mathcal{L}=$ 100 $\text{fb}^{-1})$ &$(\mathcal{L}=$ 100 $\text{fb}^{-1})$ \\
 		\hline\hline
 		$\mathcal{C}_1$ & 28.20 & 14.17 & 281.36 & 0.035 & 0.030 & 16.02 & 8.24\\
 		$\mathcal{C}_2$ & 28.10 & 14.16 & 275.05 & 0.035 & 0.030 & 16.13 & 8.32\\
 		$\mathcal{C}_3$ & 27.81 & 14.02 & 268.15 & 0.034 & 0.030 & 16.16 & 8.33\\
 		$\mathcal{C}_4$ & 27.81 & 14.02 & 265.42 & 0.034 & 0.030 &   16.24 & 8.40\\
 		\hline
 	\end{tabular}
 	\caption{Same as in Table~\ref{tab:cut-flow}, but for benchmark points (BP4, BP5) corresponding to Dirac DM scenario for CM energy 
	($\sqrt s$) = 1 TeV with $\{P_{e^-}:P_{e^+}\}=\{+0.8:-0.3\}$ beam polarization.}
 	\label{tab:cut-flow-dirac}
 \end{table}

\begin{figure}[htb!]
$$
\includegraphics[scale=0.5]{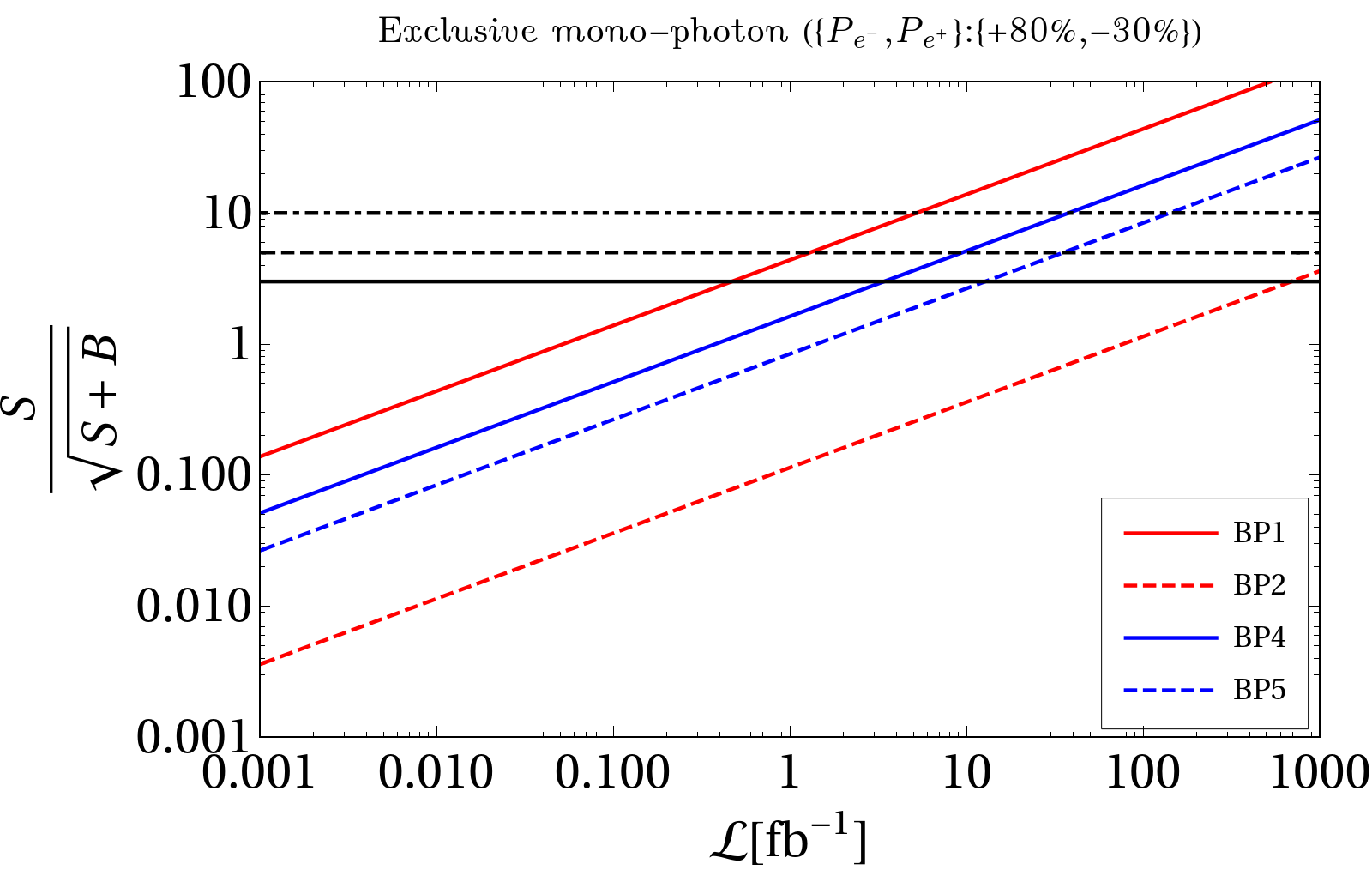}
$$
\caption{Signal significance for exclusive mono-photon with missing energy final state events, as a function of integrated luminosity for the chosen benchmark points in table~\ref{tab:BP}. The black solid, dashed and dotdashed horizontal straight lines denote respectively 3$\sigma$, 5$\sigma$ and 10$\sigma$ discovery limit. The red (blue) lines correspond to Majorana (Dirac) DM.}
\label{fig:signi}
\end{figure}

In Table~\ref{tab:cut-flow} we have tabulated the mono-photon signal for Majorana DM and background event cross-sections with the cuts employed 
following the sequence mentioned above. We also quote the efficiency factor $\epsilon=\frac{\sigma^{\tt sig}}{\sigma^{\tt prod}}$ 
(where $\sigma^{\tt prod(sig)}$ denotes the production (signal)-level cross-section), which depicts the loss of events in the process of employing 
selection cuts on the final state events. We also tabulate the significance of signal events with respect to SM backgrounds $\sigma=\frac{S}{\sqrt{S+B}}$
after each cuts. Here $S(=\sigma^{\tt sig}\times\mathcal{L})$ denotes the number of signal (final state) events for a given luminosity, while $B$ 
corresponds to the number of background events at the same luminosity. The main observations from Table~\ref{tab:cut-flow} is that with each 
cut the signal significance enhances, although mildly and for the low DM mass (BP1) the significance is larger than the discovery limit with luminosity 
$\mathcal{L}=$ 100 $\text{fb}^{-1}$. A similar observation is made for Dirac DM from table~\ref{tab:cut-flow-dirac}, although we require higher 
CM energy ($\sqrt s$ = 1 TeV) to produce them as dictated by relic density and direct search constraints, which results both BP4 and BP5 having 
discovery limit at $\mathcal{L}=$ 100 $\text{fb}^{-1}$. We also see that $\epsilon$ is quite suppressed after a photon tagging (more for Majorana case), 
and the cut flow do not alter them significantly, which testifies that the cuts employed here retains signal to a good extent. It may also be noted 
that the dramatic improvement with missing mass cut as reported in \cite{Han:2020uak}, is not observed in our case, owing to the limited CM 
energy and moderate DM mass for the chosen analysis.


The signal significance ($\sigma=\frac{S}{\sqrt{S+B}}$) is then plotted in terms of luminosity in Fig.~\ref{fig:signi}. We see that mass of the DM 
plays a crucial role; while for a luminosity of 1 $\text{fb}^{-1}$, BP1 (red line) with $\mchi=70$ GeV can reach 5$\sigma$ significance, 
BP2 (red dashed) with $\mchi=112$ GeV requires at least $\mathcal{L}> 10^3~\text{fb}^{-1}$ to be probed with 5$\sigma$ confidence. 
The signal cross-section for BP4 and BP5 (Dirac case) is smaller than BP1 and larger than BP2 (although with a different CM energy), so 
are the discovery reaches at those points, as shown by blue thick and dashed lines. 
\begin{figure}[htb!]
$$
\includegraphics[scale=0.42]{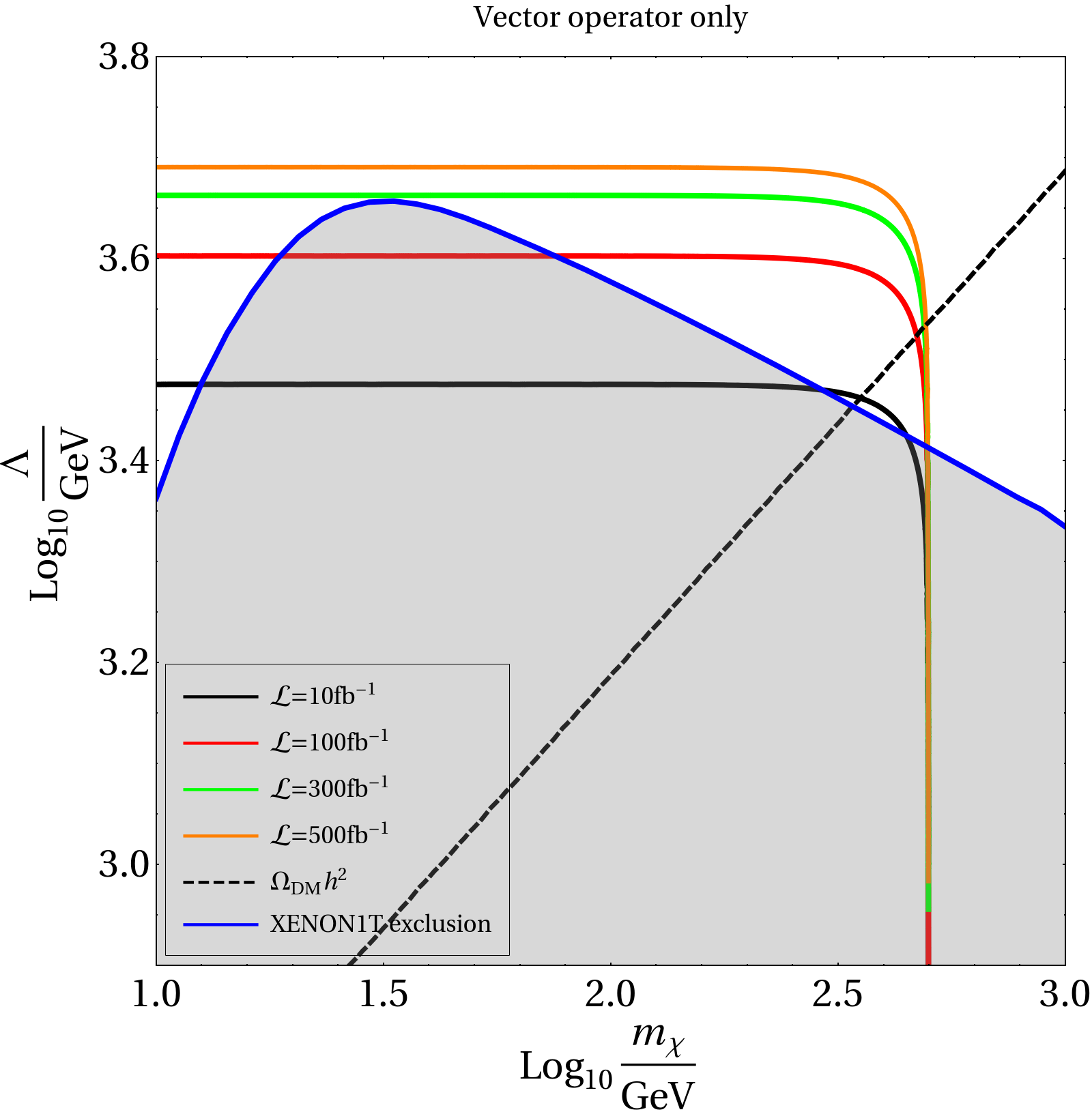}
$$
\caption{Parameter space in $\Lambda-m_\chi$ plane considering the vector operator $(\overline{\chi}\gamma^\mu\chi)(\overline{\ell}\gamma_\mu\ell)$, 
where the solid lines represent 3$\sigma$ significance contour for ILC at a CM energy of $\sqrt{s}=1$ TeV with integrated luminosity of $\{10, 100, 300, 500\}\text{fb}^{-1}$
shown in black, red, green, and orange respectively; the dashed diagonal line corresponds to the Planck observed relic density for DM and the solid blue curve shows the 
exclusion limit from spin-independent XENON1T data. }\label{fig:summary}
\end{figure}

It is useful to study the interplay of collider vs non-collider bounds in the context of effective operator as this provides information about the reach for different experiments in probing the DM parameter space. 
For hadrophilic DM such a study has been performed in many instances (for example, in~\cite{Belyaev:2018pqr}). Here we would like to illustrate such an example by considering a single operator in action in context of leptophlic 
DM scenario. In Fig.~\ref{fig:summary} we show the complementarity of direct search sensitivity to that of the collider search in mono-photon final state for the operator $\mathcal{O}^6_{DL}$ in DM mass ($\mchi$) 
versus NP scale ($\Lambda$) plane. The gray shaded region is discarded from the XENON1T exclusion limit, while along the black dashed straight diagonal line the Planck observed relic density is satisfied. 
The black, red, green and orange solid curves indicate 3$\sigma\equiv\frac{S}{\sqrt{S+B}}$ confidence limit at the ILC for a CM energy of $\sqrt{s}=1$ TeV, corresponding to luminosities 
$\mathcal{L}=\{10,100,300,500\}\,\text{fb}^{-1}$ respectively. Note that, all the collider confidence contours converge at $\mchi\simeq\sqrt{s}/2\sim 500$ GeV for a CM energy of 1 TeV. 
It is easy to understand that with larger luminosity the collider sensitivity overpowers the direct detection.

\section{Possible UV completion}\label{sec:uv}

In this section we sketch a few possible UV complete frameworks that are capable of generating leptophilic DM operators discussed so far in the draft. One of the simplest such possibilities can be found in~\cite{Fox:2008kb}, where the SM gauge sector is extended by a dark abelian gauge symmetry $U(1)_\text{DS}$. There is also a Dirac fermion $\chi$,  odd under a dark sector parity, that can be a potential DM candidate (all SM fields are even under the same parity). In order to ensure the dark gauge boson $Z'$ to be leptophilic, one has to ensure that it couples with equal and opposite charge to two generations of leptons for anomaly cancellation (for example, in a gauged $L_i-L_j$ model~\cite{He:1990pn,Ma:2001md,Baek:2001kca,Baek:2008nz,Heeck:2011wj,Das:2013jca,Biswas:2016yan,Duan:2017qwj,Foldenauer:2018zrz} with $i,j\in e,\mu,\tau$)\footnote{Leptophilic Majorana DM, discussed in the context of the anomaly-free theories can be found in Refs.~\cite{Duerr:2013dza, Schwaller:2013hqa, FileviezPerez:2019cyn}.}. The qurak interaction is prohibited since they do not carry any $L_i-L_j$ charge. At a scale $\mu\ll m_{Z'}$, the heavy dark gauge boson 
can be integrated out resulting in an operator of the form $\bar{\chi}\,\Gamma\,\chi\bar{\ell}\,\Gamma'\,\ell/m_{Z'}^2$, where $\Gamma^{(')}\equiv\{\gamma^\mu,\gamma^\mu\,\gamma^5\}$. 


Extra-dimensional model where the fermions have strong localizations at various points in the extra dimensions \cite{as2000, ms2000, Nussinov:Shrock02} furnishes another interesting possibility where the DM can couple to the leptons, while the coupling with the quarks is exponentially suppressed. Coupling fermions with appropriate kink (for one-extra dimension) or vortex (for two-extra dimensions) can localize the (zero-mode) fermions in the extra dimensions \cite{Rubakov:1983bb, Kaplan:1995pe, Dvali:2000ha}. Assuming that the fermions have support in the interval $[0, L]$ in the extra dimensions, constraints from precision electroweak searches, flavor-changing neutral current, and collider searches can be accommodated by choosing $L^{-1} \equiv \Lambda_L \gtrsim 100$ TeV \cite{Delgado:1999sv,  Dobrescu01}.  One appealing feature of this kind of extra-dimensional model is that the large hierarchy in SM fermion masses can be obtained by separating out the chiral parts in the extra dimension(s), without the requirement of large range of dimensionless Yukawa couplings.  Furthermore, proton decay can be exponentially suppressed to safety by separating out quark and lepton wavefunctions in the extra dimensions \cite{as2000, bvd}, although this does not hinder $n-\bar n$ oscillations \cite{nnb02, Nussinov:Shrock02,  Wise13, nnblrs}.  Set of solutions of SM fermion wavefunction centers exists that reproduce neutrino masses and mixing, and is consistent with experimental data \cite{nuled, barenboim01}.  Therefore, if the DM candidate $\chi$ is localized near the leptons in the extra dimensions, then the experimental bound from proton decay implies that $\chi$ is separated widely from the quark wavefunctions. Hence, the Wilson coefficients of $(\bar \chi \Gamma_\chi^\mu \chi) (\bar \ell \Gamma_{\ell, \mu} \ell)$ will be unsuppressed, while the Wilson coefficients for $(\bar \chi \Gamma_\chi^\mu \chi) (\bar \qr \Gamma_{\qr, \mu} \qr)$ will be exponentially suppressed by the separation distance between the corresponding fermion wavefunctions, thereby generating an effective leptophilic interaction as we study here \cite{dmled}.    

\section{Summary and Conclusions}
\label{sec:sum}

Dark matter (DM) in effective theory (EFT) formalism has been studied extensively due to the unknown nature of dark sector and economy of free parameters ($\mchi, \Lambda$) without loss of predictability of the theory itself. However, DM operators coupling to SM quarks face severe constraints from direct DM searches as well as from LHC, caveat to appropriate validation of EFT limit at hadron collider. Leptophilic DM, on the other hand, is motivated from the fact that 
it not only provides a model independent way to probe DM physics in EFT formalism, but presents an opportunity to study DM production at $e^+e^-$ collider abiding the EFT limit after addressing relic density. Dirac DM operators can additionally be probed in the future sensitivities of direct search experiment via one loop interaction with the SM quarks, while Majorana DM operators remain absent in direct detection due to the Lorentz structure of the current. 

We study the contribution from all possible DM operators of dimension six {\it simultaneously}, assuming the Wilson coefficients to contribute with 
equal strengths. Direct search for the Dirac DM is performed by taking into account the RG evolution of the Wilson coefficients from a high NP scale 
$\Lambda$ to low energy non-relativistic scale. We have seen that direct search constraints from latest XENON1T data constrains the Dirac DM 
model (with all Wilson coefficients chosen as one and having same sign) upto DM mass $\sim $400 GeV, with $\Lambda \simeq$ 3300 TeV.

Mono-photon arising from the initial state radiation (ISR), together with missing energy, turns out to be 
a potential signal to probe such operators at the ILC. For Dirac DM one requires higher CM energy ($\sqrt{s}=1$ TeV) satisfying the relic density and 
spin-independent direct search exclusion, while a  comparatively lower CM energy ($\sqrt{s}=500$ GeV) is suffice to probe Majorana DM due to 
the absence of any direct search limit. The SM backgrounds can be tamed down moderately with judicious choice of cuts on observables like 
missing energy, missing mass, photon transverse momentum etc.; although $\nu\bar{\nu}$ provides a substantial irreducible contribution to 
such signals. The key is to choose a maximally right polarised electron beam and left polarized positron beam ($P_{e^\pm} = ^{-30\%}_{+80\%}$), 
which helps to enhance the signal with positive Wilson coefficients and reduce the SM background significantly. A discovery reach of 5$\sigma$ can 
be achieved with luminosity $\mathcal{L}\simeq 1 \, \text{fb}^{-1}$ for low mass Majorana DM  ($\mchi \sim 70$ GeV), while that for Dirac DM with 
masses above 300 GeV requires larger luminosity $\mathcal{L}\gtrsim 50 \, \text{fb}^{-1}$. 


\acknowledgments

BB received funding from the Patrimonio Autónomo - Fondo Nacional de Financiamiento para la Ciencia, la Tecnología y 
la Innovación Francisco José de Caldas (MinCiencias - Colombia) grant 80740-465-2020. SG acknowledges support from the U.S. National Science  Foundation Grant NSF-PHY-1915093. This project has received funding /support from the European Union's Horizon 2020 research and innovation programme under the Marie Sklodowska-Curie grant agreement No 860881-HIDDeN. SB  would like to acknowledge DST-SERB grant CRG/2019/004078 from Govt. of India and Prof.\ Jose Wudka for useful discussions. SG would like to thank Prof.\ Robert Shrock for helpful discussions. The authors would like to thank Prof. Alexander Pukhov and Prof. M. C. Kumar for helping with the model implementation.

\appendix

\section{Annihilation cross-sections for leptophilic operators}
\label{app:cross}
In this appendix we list the annihilation cross-section times the relative velocity $v$,  expanded in powers of $v$ for different leptophilic operators.
\begin{shaded}
\begin{align}
	\big( \sigma v \big)_{{\cal O}_{DL}^6} &= \frac{c_{\ell_1}^2}{2 \pi \Lambda^4} m_\chi^2 \sum_\ell \sqrt{1-\delta_{\chi \ell}^2} \bigg[ \Big( 2+ \delta_{\chi \ell}^2 \Big) + \frac{v^2}{24 \Big( 1-\delta_{\chi \ell}^2 \Big)} \bigg( 8-4 \delta_{\chi \ell}^2 +5 \delta_{\chi \ell}^4 \bigg)  \bigg] + {\cal O}(v^4) ,  \\
	\big( \sigma v \big)_{{\cal O}_{L}^6} &=\kappa^2 \frac{c_{\ell_2}^2}{12 \pi \Lambda^4} m_\chi^2 \sum_\ell \sqrt{1-\delta_{\chi \ell}^2} \Big( 2+ \delta_{\chi \ell}^2 \Big) v^2 + {\cal O}(v^4) ,  \\
		\big( \sigma v \big)_{{\cal O}_{L1}^6} &= \kappa^2 \frac{c_{\ell_3}^2}{2 \pi \Lambda^4} m_\chi^2 \sum_\ell \sqrt{1-\delta_{\chi \ell}^2} \bigg[ \delta_{\chi \ell}^2 + \frac{v^2}{24 \Big( 1-\delta_{\chi \ell}^2 \Big)} \bigg( 8-22 \delta_{\chi \ell}^2 +17 \delta_{\chi \ell}^4 \bigg)  \bigg] + {\cal O}(v^4) ,  \\ 
\big( \sigma v \big)_{{\cal O}_{DL1}^6} &=  \frac{c_{\ell_4}^2}{ \pi \Lambda^4} m_\chi^2 \sum_\ell \bigg[  \Big( 1-\delta_{\chi \ell}^2 \Big)^{3/2}  + \frac{v^2}{24} \sqrt{1-\delta_{\chi \ell}^2} \Big( 4+5\delta_{\chi \ell}^2 \Big) \bigg] + {\cal O}(v^4) ,
\end{align}
\end{shaded}
where $\kappa=1 (2)$ for a Dirac (Majorana) DM $\chi$, and $\delta_{\chi \ell} \equiv \frac{m_\ell}{m_\chi}$.

\section{Models with negative Wilson coefficients}
\label{sec:negative}
The cases where Wilson coefficients are assumed negative (particularly for $c_{\ell_3}$ and $c_{\ell_4}$), offer different phenomenology. 
In this appendix, we elaborate such cases.
First, we show the relic density satisfying parameter space constrained from direct detection bound from XENON1T for ($1, 1, 1, -1$), ($1, 1, -1, 1$), 
and $(1, 1, -1, -1)$ choices of the $c_{\ell_i}$ for Dirac DM. To compare this with the BPs of the collider analyses in the text, we choose a few benchmark points as shown in Table~\ref{tab:BP_minus}.
    \begin{table}[htb!]
        \centering
        \begin{tabular}{|c|c|}
        \hline
        $c_{\ell_i}$ & ($m_\chi, \Lambda$) (GeV) \\
        \hline
        $(1, 1, 1, -1)$     &   ($350, 3300$) \\
        $(1, 1, -1, 1)$     &   ($350, 2800$)   \\
        $(1, 1, -1, -1)$     &   ($350, 3300$)\\
        \hline
        \end{tabular}
        \caption{Benchmark points for Dirac DM for negative choices of the Wilson coefficients $c_{\ell_3}$ and/or $c_{\ell_4}$.}
        \label{tab:BP_minus}
    \end{table}
    
\begin{figure}[htb!]
        \centering
        $$
        \includegraphics[width=0.8\textwidth]{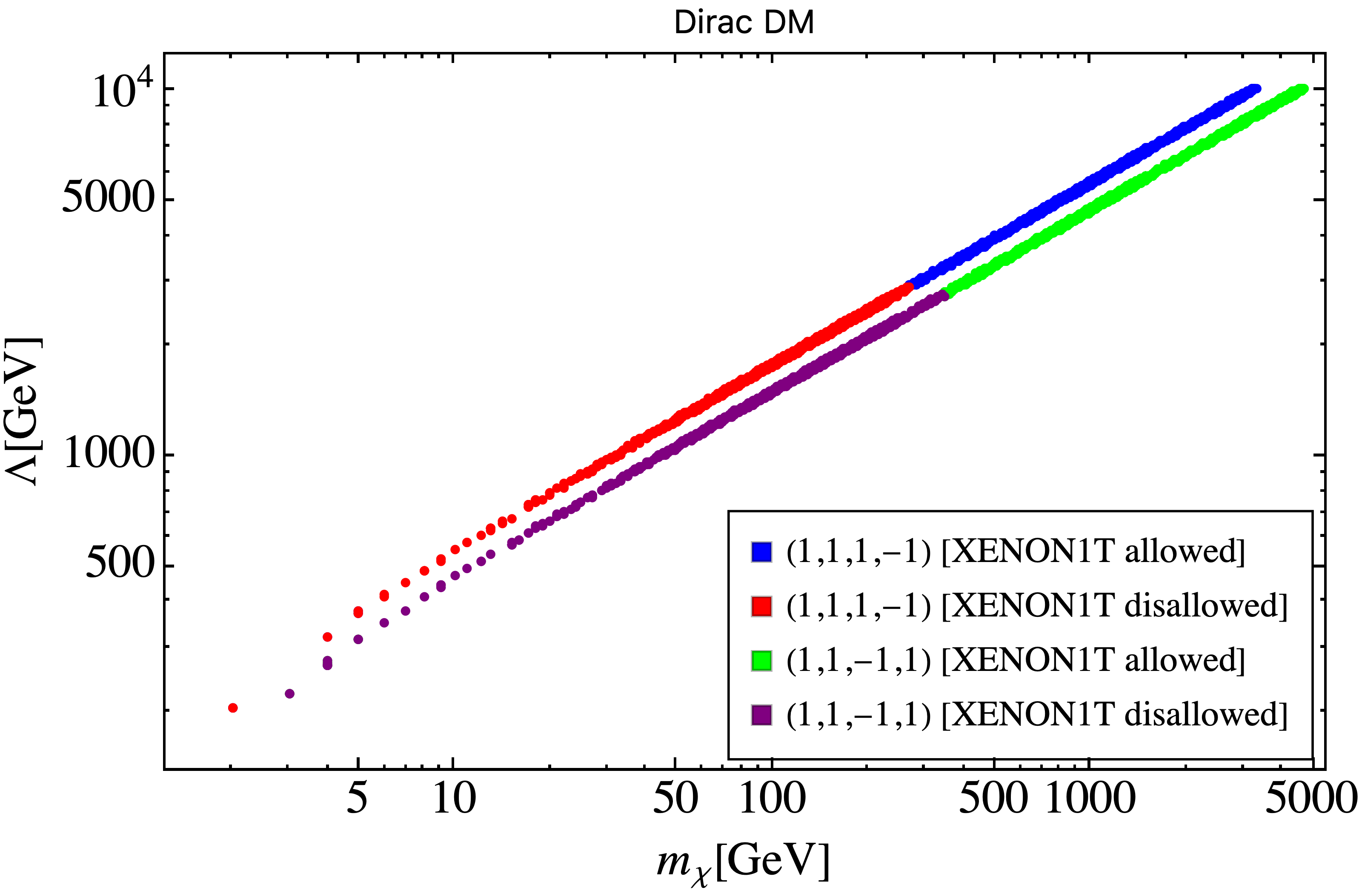}
        $$
        \caption{Relic density satisfying parameter space for different choices of the Wilson coefficients. The coupling values $(c_{\ell_1}, c_{\ell_2}, c_{\ell_3},  
        c_{\ell_4})$ is denoted as an ordered set. XENON1T disallowed parameter space is marked in separate colour (see figure inset). 
        The case $(1, 1, -1, -1)$ is not shown here as it is approximately degenerate with the $(1, 1, 1, -1)$ case. }
        \label{fig:relic_minus}
    \end{figure}

    The DM relic density and direct search allowed parameter space is shown in Fig.~\ref{fig:relic_minus}. The blue, and green points satisfy XENON1T bound on the spin-independent direct detection cross-section, while the red and purple points are ruled out from the same. The cross-over point from red to blue is $(275, 2950)$ GeV, and similar cross-over point for purple to green is $(345, 2800)$ GeV.

\begin{figure}[htb!]
$$
\includegraphics[scale=0.30]{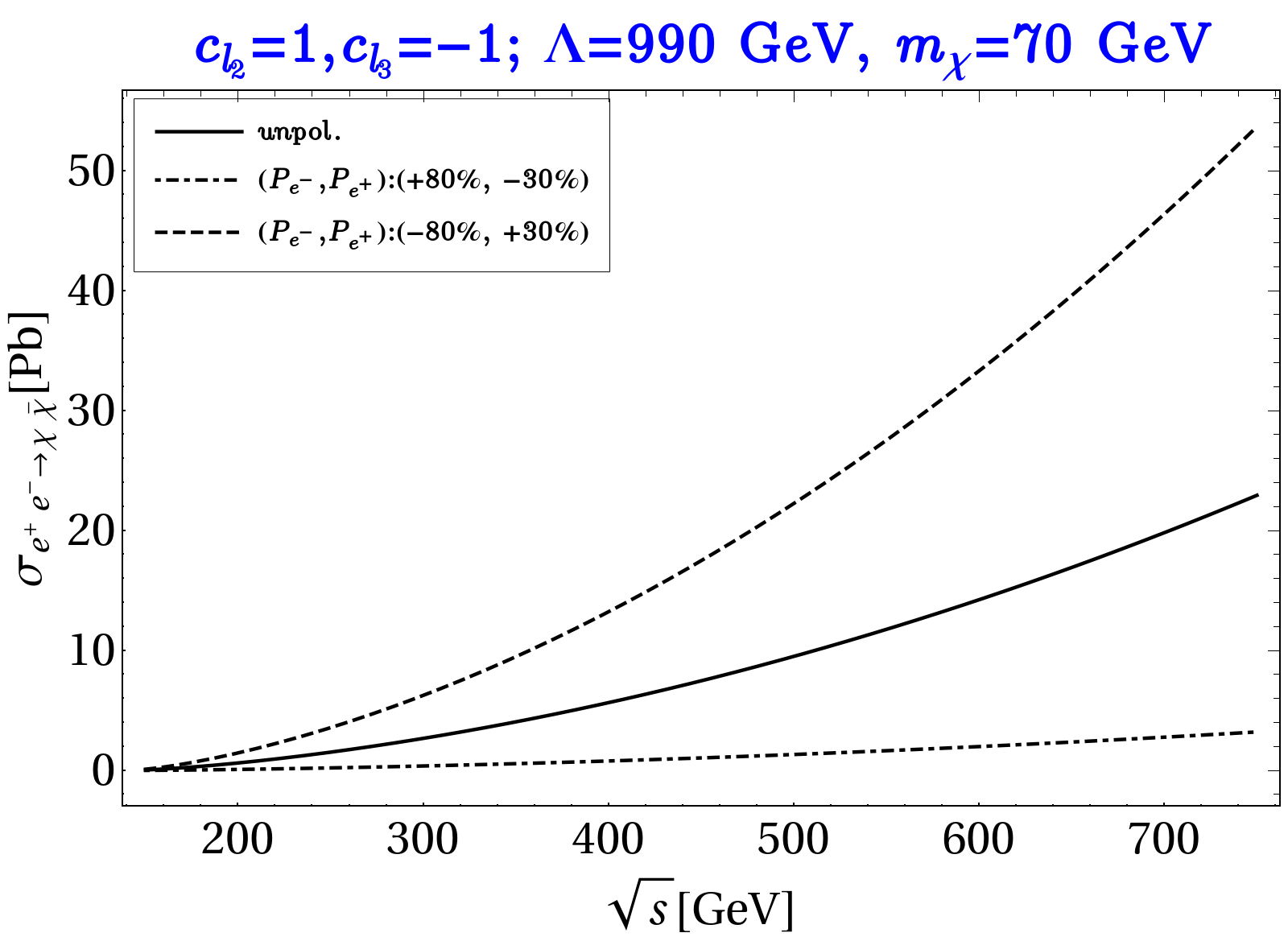}~~~~
\includegraphics[scale=0.30]{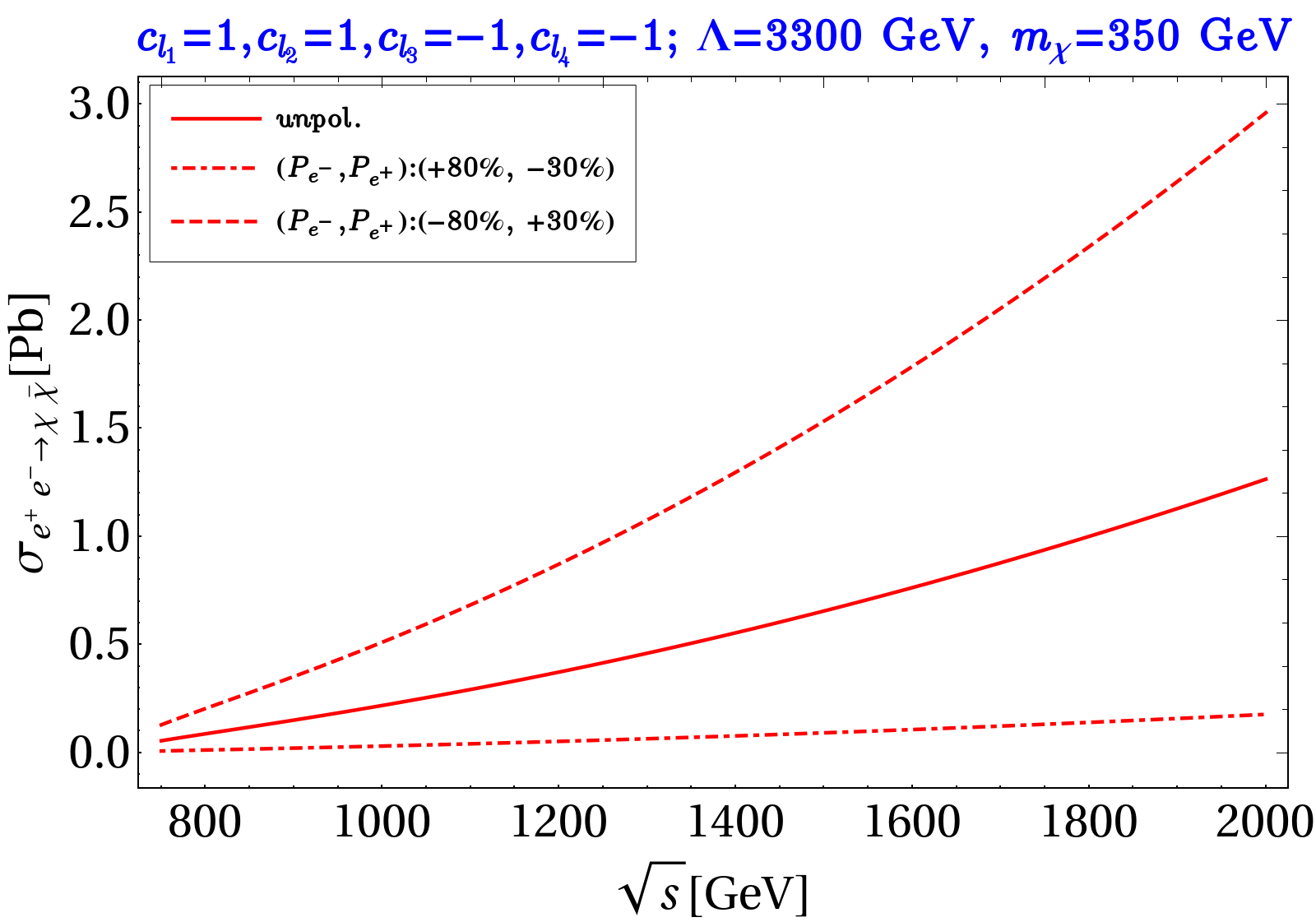}
$$
$$
\includegraphics[scale=0.30]{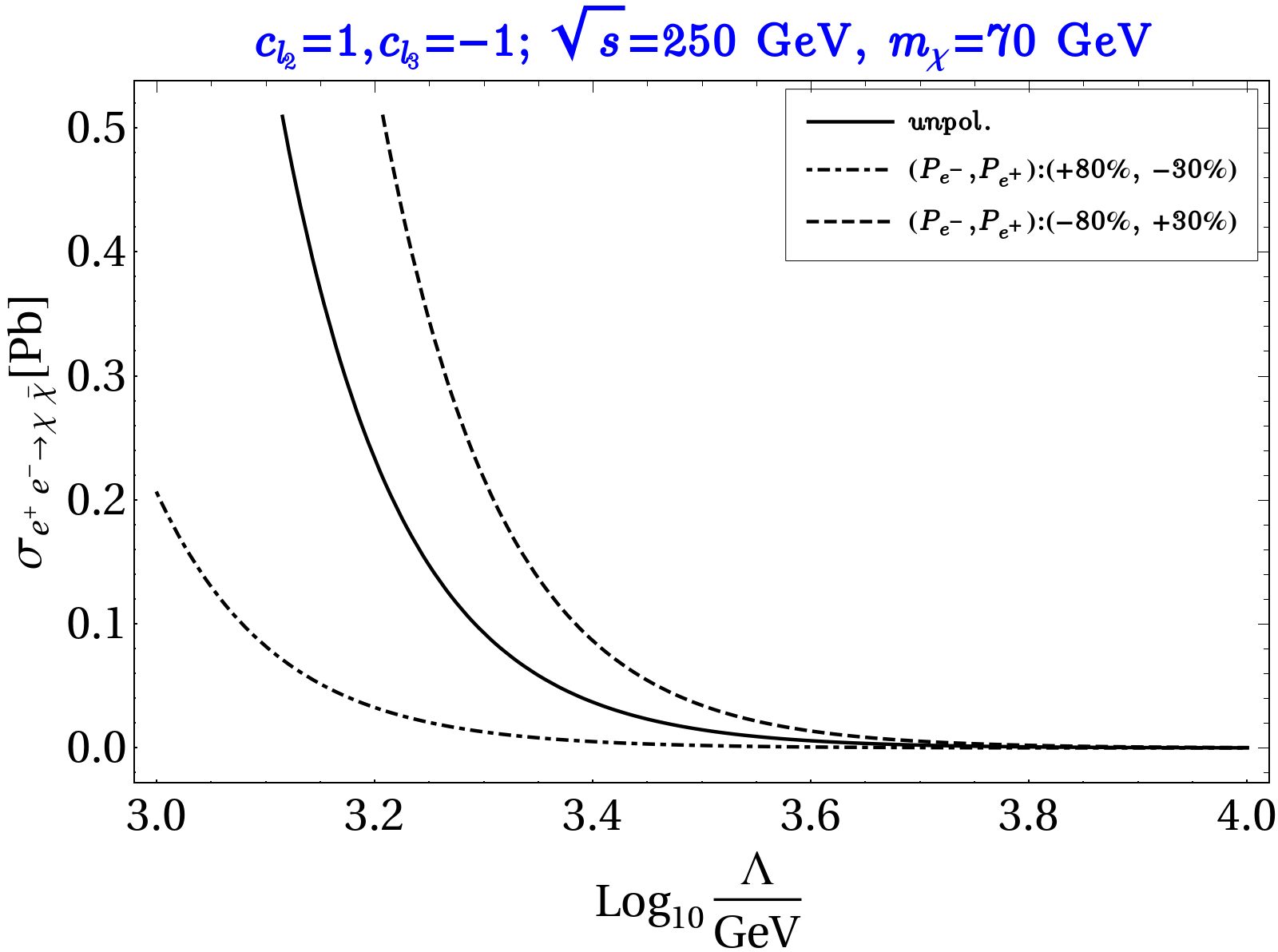}~~~~
\includegraphics[scale=0.29]{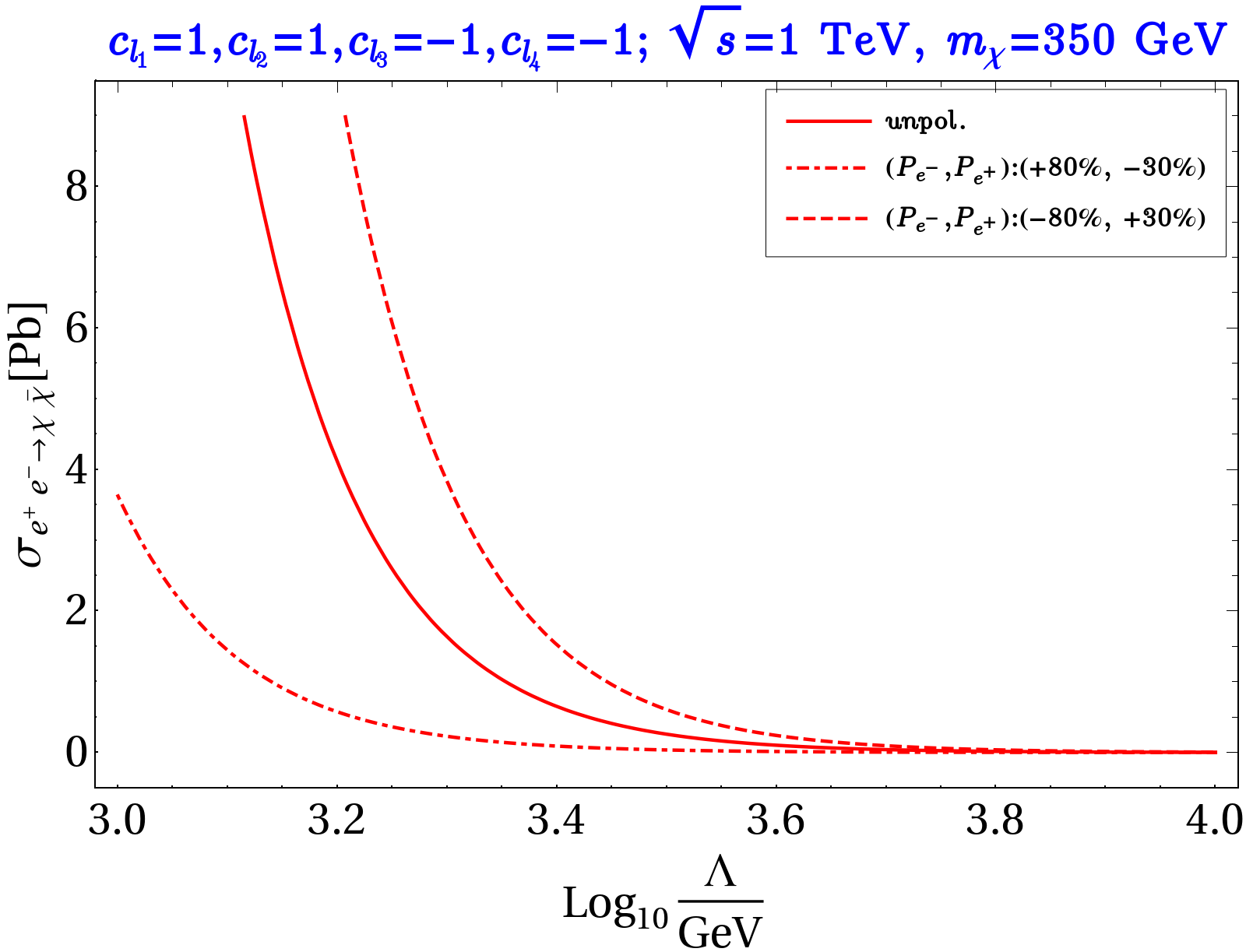}
$$
$$
\includegraphics[scale=0.30]{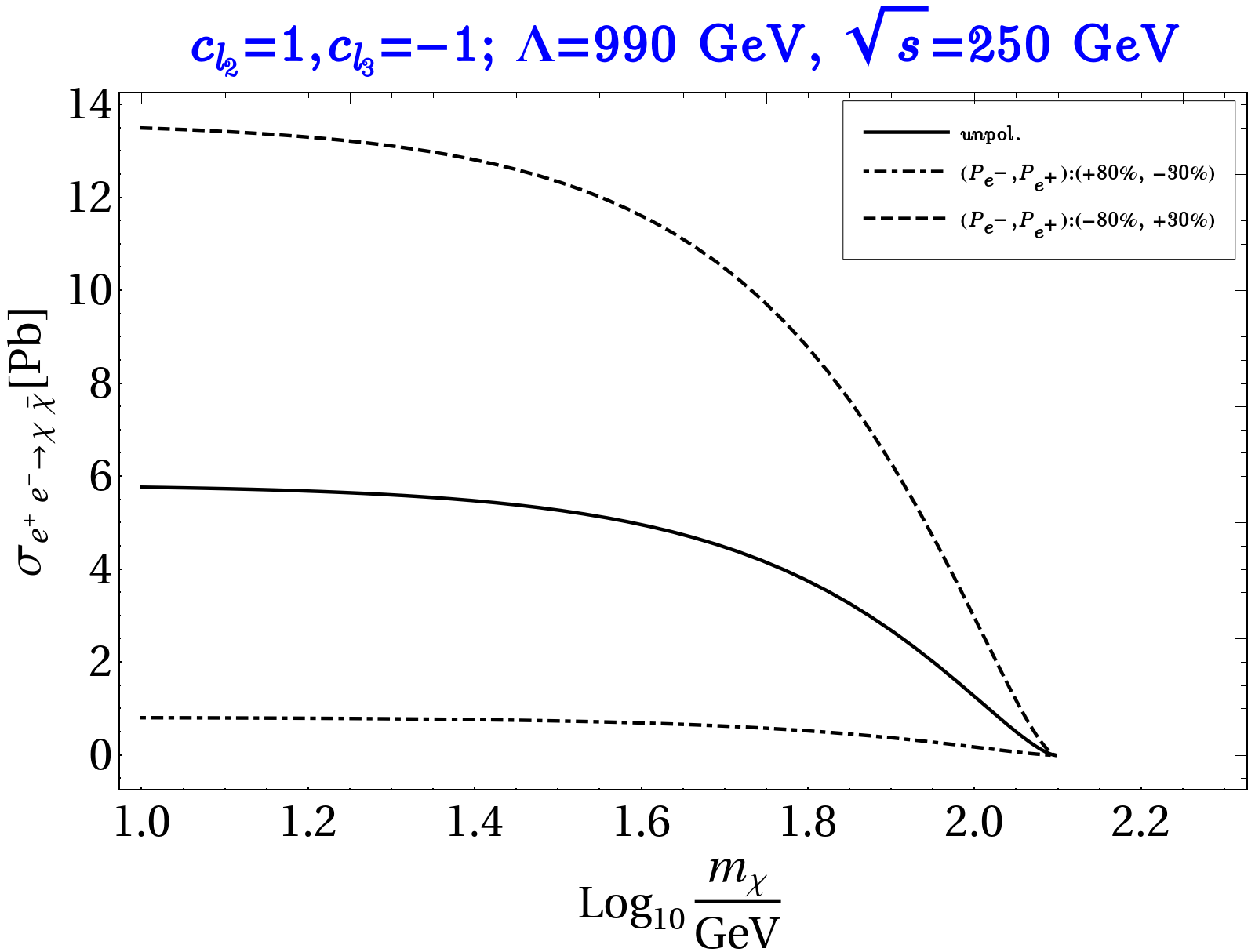}~~~~
\includegraphics[scale=0.30]{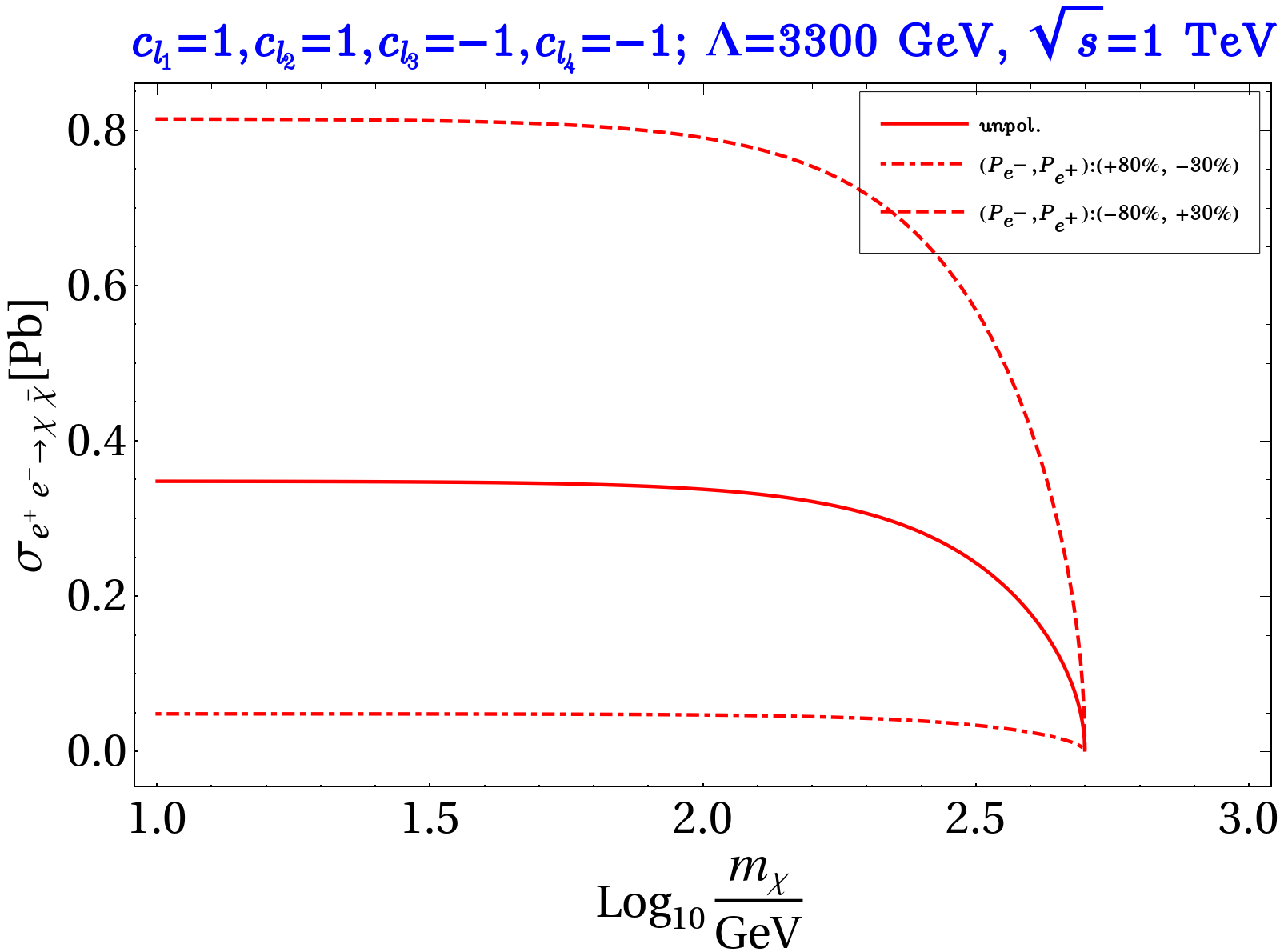}
$$
\caption{$\sigma_{e^+ e^- \to \chi \bar \chi}$ as function of $\sqrt{s}$ (top panel), $\Lambda$ (middle panel) and $\mchi$ (bottom panel) for Majorana DM (BP3) in 
the left column and Dirac DM (BP6) in the right column for three choices of polarisation: $\{P_{e^-}:P_{e^+}\}=\{0,0\};\{+0.8,-0.3\};\{-0.8,+0.3\}$ as mentioned in 
the figure inset. The other parameters kept fixed are mentioned in Figure heading.}
\label{fig:dm-prod-negcl}
\end{figure}

 We plot next the variation of DM production cross-section ($\sigma_{e^+ e^- \to \chi \bar \chi}$) for Majorana DM in the left column and Dirac DM 
 in the right column of Fig.~\ref{fig:dm-prod-negcl}. In the top, middle and bottom panel we show the variation with respect to $\sqrt{s}$, $\Lambda$ and $\mchi$ 
respectively for three different choices of the beam polarization: $\{P_{e^-}:P_{e^+}\}=\{0,0\};\{+0.8,-0.3\};\{-0.8,\\ +0.3\}$ shown by solid, 
dashed-dot and dashed black (red) curves for Majorana (Dirac) DM cases with operators having negative Wilson coefficients. See Figure caption for details. 
The features remain almost the same, excepting for the fact that cross-section for this case enhances for other polarization configuration, namely
$\{P_{e^-}:P_{e^+}\}=\{-0.8,+0.3\}$, unlike the cases with positive Wilson coefficients. Recall, that we could reduce the neutrino background significantly 
using the other polarisation $\{P_{e^-}:P_{e^+}\}=\{+0.8,-0.3\}$, which reduces the signal with this particular choices of Wilson coefficients. That is why, such a possibility is harder to probe at ILC.
    
\bibliographystyle{JHEP}
\bibliography{Bibliography}
\end{document}